\documentclass[sigconf, nonacm]{acmart}

\usepackage[linesnumbered,ruled,vlined]{algorithm2e}
\usepackage{algpseudocode}

\usepackage{amsthm,amssymb,amsmath,graphics,float,subfigure}

\usepackage{balance}
\usepackage{times}
\usepackage{bm}
\usepackage{multirow}
\newcommand{\stitle}[1]{\vspace{1ex} \noindent{\bf #1}}
\long\def\comment#1{}
\newcommand{\kw}[1]{{\ensuremath{\mathsf{#1}}}\xspace}

\newcommand{\hcs} {\kw{HCS}}
\newcommand{\hcss} {\kw{HCSs}}
\newcommand{\hcslist} {\kw{HCSList}}
\newcommand{\hcspivot} {\kw{HCSPivot}}
\newcommand{\listing} {\kw{Listing}}

\newcommand{\pivoter} {\kw{PIVOTER}}

\newcommand{\dlist} {\kw{Dlist}}
\newcommand{\plist} {\kw{Plist}}
\newcommand{\dpivot} {\kw{Dpivot}}
\newcommand{\ppivot} {\kw{Ppivot}}
\newcommand{\ndpivot} {\kw{Dpivot}-\kw{nup}}
\newcommand{\nppivot} {\kw{Ppivot}-\kw{nup}}

\newcommand{\dbl} {\kw{DBLP}}
\newcommand{\ama}{\kw{Amazon}}
\newcommand{\ema}{\kw{EmailEu}}
\newcommand{\epi}{\kw{Epinion}}
\newcommand{\cai}{\kw{Caida}}
\newcommand{\wik}{\kw{WikiV}}
\newcommand{\ski}{\kw{Skitter}}
\newcommand{\pok}{\kw{Pokec}}

\newcommand{\srp}{\kw{SRP}}
\newcommand{\srps}{\kw{SRPs}}
\newcommand{\gp}{\kw{GP}}
\newcommand{\gps}{\kw{GPs}}
\newcommand{\hgp}{\kw{HGP}}
\newcommand{\hgps}{\kw{HGPs}}

\newcommand\vldbdoi{XX.XX/XXX.XX}

\newcommand\vldbavailabilityurl{URL_TO_YOUR_ARTIFACTS}
\begin{document}

\title{Counting Cohesive Subgraphs with Hereditary Properties}
%\title{Efficient Algorithms for Counting Hereditary Cohesive Subgraphs on Large Graphs}
%\title{Efficient Counting of Cohesive Subgraphs with Hereditary Properties}
%\title{Hereditary Cohesive Subgraphs Counting On Large Graphs}
%\title{Efficient Counting of Hereditary Cohesive Subgraphs in Large Graphs}

\settopmatter{authorsperrow=3}

\author{ Rong-Hua Li}
\affiliation{%
	\institution{Beijing Institute of Technology}
	\city{Beijing}
	\country{China}
}
\email{ lironghuabit@126.com}

\author{Xiaowei Ye}
\affiliation{%
			\institution{Beijing Institute of Technology}
			\city{Beijing}
			\country{China}
		}
\email{yexiaowei@bit.edu.cn}

\author{Fusheng Jin}
\affiliation{%
		\institution{Beijing Institute of Technology}
	\city{Beijing}
	\country{China}
	}
\email{jfs21cn@bit.edu.cn}

\author{Yu-Ping Wang}
\affiliation{%
		\institution{Beijing Institute of Technology}
		\city{Beijing}
		\country{China}
	}
\email{ wyp_cs@bit.edu.cn}

\author{Ye Yuan }
\affiliation{%
	\institution{Beijing Institute of Technology}
	\city{Beijing}
	\country{China}
}
\email{ yuan-ye@bit.edu.cn}

\author{Guoren Wang}
\affiliation{%
		\institution{Beijing Institute of Technology}
		\city{Beijing}
		\country{China}
	}
\email{ wanggrbit@gmail.com}

\begin{abstract}

The clique model has properties of hereditariness and cohesiveness. Here hereditary property means a subgraph of a clique is still a clique. Counting small cliques in a graph is a fundamental operation of numerous applications. However, the clique model are often too restrictive for practical use, leading to the focus on other relaxed-cliques with properties of hereditariness and cohesiveness. To address this issue, we investigate a new problem of counting general hereditary cohesive subgraphs (\hcs). All subgraphs with properties of hereditariness and cohesiveness can be called a kind of \hcs. To count \hcs, we propose a general framework called \hcspivot, which can be applied to count all kinds of \hcs. \hcspivot can count most \hcs in a combinatorial manner without explicitly listing them.  Two additional noteworthy features of \hcspivot is its ability to (1) simultaneously count \hcss of any size and (2) simultaneously count \hcss for each vertex or each edge. The implementation of \hcspivot for clique counting is exactly the state-of-the-art clique counting algorithm \pivoter \cite{PIVOTER}. We focus specifically on two \hcs: $s$-defective clique and $s$-plex. We also propose several non-trivial pruning techniques to enhance the efficiency. We conduct extensive experiments on 8 large real-world graphs, and the results demonstrate the high efficiency and effectiveness of our solutions.

\end{abstract}

\maketitle

\section{Introduction} \label{sec:intro}
Counting small cohesive subgraphs in a graph is a fundamental problem in graph mining and has a wide range of applications in various fields, such as community detection, network analysis, and bioinformatics \cite{tutoralSeshadhriT19, surveyRibeiroPSAS21}. Among different types of motifs, clique is considered as one of the most important motifs due to the perfect cohesiveness and hereditary properties. Cohesiveness means great reach-ability between the vertices and hereditary property means that any induced subgraph of a clique is still a clique. In real-world, a group often has properties of cohesiveness and hereditary. For example, a group of friends should have abilities to reach each other and any subgroups are still friends. Therefore, counting cliques serves as a basic operator in many applications \cite{benson2016higher, lu2018community,15wwwnucleusdecomp,kdensWWW2015,motifClustering,17prehighorder}.

However, the constraint of clique is often very strict for real-world applications because the interaction between members of a group may not be direct, or there may be missing data in a real-world system. To overcome this limitation, many relaxed clique models have been proposed. Notable examples include $s$-defective clique \cite{Defective_bioinformatics_06,maximumDclique21,23sigmoddclique}, $s$-plex \cite{Plex_seidman1978graph, plex_AAAI21, DAi_kplex, d2k_kdd18}, $s$-clique \cite{luce1950connectivity}, $\gamma$-quasi-clique \cite{BrunatoHB07,quasiCliquePeiJZ05}, $k$-core \cite{seidman1983network}, $k$-truss \cite{12vldbtruss,14sigmodtruss}, and $k$-edge connected subgraph \cite{12edbtkecc,13sigmodkecc}. In this work, we focus mainly on the $s$-defective clique and $s$-plex models, since these two models are often close to a clique and also they maintain both the cohesiveness and hereditary properties.

Specifically, a graph is an $s$-defective clique if there exist at most $s$ missing edges compared to cliques. Since the maximum count of missing edges is restricted, $s$-defective clique is typically very cohesive. The subgraphs of an $s$-defective clique are still $s$-defective cliques because the deletion of nodes will not lead to the increasing of missing edges. Different from $s$-defective clique, an $s$-plex restricts each vertex has at most $s$ missing edges, i.e., has at most $s$ non-neighbors. Similar to $s$-defective clique, $s$-plex also has the property of hereditary because the deletion of nodes will not lead to increasing the count of non-neighbors of any vertex.

Instead of counting cliques in a graph, we study a new and more general problem of counting hereditary cohesive subgraphs (\hcss in short) in a graph. To address this problem, we develop two general counting frameworks which can be applied to counting both $s$-defective cliques and $s$-plexes. Similar to clique counts, the counts of \hcss have diverse applications in graph analysis. Below, we highlight two specific applications.

\vspace{0.2em}
\noindent \textit{\underline{Motif-based Graph Clustering.}} Motif-based graph clustering has been recognized as the state-of-the-art method for detecting real-world communities in a network \cite{benson2016higher,motifClustering}. The motif-based graph clustering method aims to minimize the so-called motif-based conductance, which is an important concept in network analysis whose definition is mainly based on the count of motifs \cite{benson2016higher,motifClustering}. Hence, the key for motif-based graph clustering methods is to compute the count of motifs. Our experiments show that compared to using traditional clique counts \cite{benson2016higher,motifClustering}, using the count of \hcss, such as $s$-defective clique and $s$-plex, can improve the quality of clustering. This highlights the practical importance of the problem of counting hereditary cohesive subgraphs.

\vspace{0.2em}
\noindent \textit{\underline{Network comparison.}} Comparing the counts of hereditary cohesive subgraphs across different graphs can aid in network comparison and similarity analysis. By quantifying the presence and abundance of these subgraphs, we can measure the structural similarities or differences between graphs. This enables us to identify graph properties or patterns that are shared or distinct, contributing to comparative network analysis and classification tasks. In this work, we define a new graph profile metric based on the count of \hcss. Our experiments show that such a new metric outperforms existing metrics in distinguishing different types of networks, suggesting that the counts of \hcss offer great potential as a novel structural feature for analyzing complex networks.

\stitle{Challenges.} Although counting \hcss has many practical applications, it is a challenging task. First, \hcss have a more complex structure than cliques (the cliques are a subset of \hcss), and the counts of \hcss are often much larger than that of cliques, making \hcss harder to enumerate and count. For example, on the \epi network, the count of $8$-cliques is $4.53\times 10^{8}$ while the count of $1$-defective clique with size $8$ is $4.02\times 10^{9}$ and the count of $1$-plex with size $8$ is $1.95\times 10^{10}$. %Moreover, the cohesive property of the subgraphs requires additional constraints to be satisfied, which further results in a significant increase in the search space and runtime of the algorithms.
Second, there is no previous work for counting $s$-defective clique and $s$-plex. Although there are several algorithms for counting cliques, they are not suitable for counting \hcss. This is because the complex structures of \hcss requires new search strategies, pruning techniques and optimizations that are not present in the clique counting algorithms. As a result, new algorithms and techniques need to be developed to tackle the challenges of counting \hcss.

\stitle{Contributions.} To overcome the computation challenges of counting \hcss, we first propose a listing-based backtracking framework, called \hcslist. This framework lists all \hcss only once, and thus it can also obtain the count. \hcslist utilizes the hereditary property to grow \hcss from small to large through backtracking. Based on the \hcslist framework, we devise two specific algorithms to count $s$-defective cliques and $s$-plexes respectively. To improve the efficiency of these algorithms, we also develop a $k$-core based pruning technique and an upper bounding technique which can significantly reduce the unpromising vertices and unnecessary search branches respectively.

The main limitation of \hcslist is that it needs to list all \hcss which is often intractable for counting relatively-large (e.g., size $>10$) \hcss in large graphs. To overcome this issue, we propose a novel pivot-based framework, called \hcspivot, which can count most \hcss in a combinatorial manner without explicitly listing them.  Instead of listing all \hcss with a given size $q$, \hcspivot lists \emph{large \hcss} (not necessarily maximal) and then counts \hcss with size $q$ in each large \hcss using a combinatorial method based on the hereditary property. To avoid repeatedly counting, \hcspivot makes use of a carefully-designed pivoting technique which can uniquely represent each \hcs using the large \hcss. Since listing the \emph{large \hcss} is much cheaper than listing all \hcss with size $q$, \hcspivot is often tractable to handle large graphs. Moreover, two important and useful features of \hcspivot are that (1) it can simultaneously count \hcss of any size, and (2) it can also obtain \emph{local count} of \hcss for each vertex or each edge, where the local count of a vertex (or an edge) means the number of \hcss containing that vertex (or edge). Based on the \hcspivot framework, we also develop two specific algorithms with two novel pivot-vertex selection strategies for counting $s$-defective cliques and $s$-plexes respectively. The results of comprehensive experiments demonstrate the high efficiency and effectiveness of our solutions. In summary, the main contributions of this work are as follows.

%Below, we briefly summarize the main contributions of this work.

%
\vspace{0.2em}
\noindent \textit{\underline{Novel \hcs counting frameworks.}} We propose two novel frameworks to count \hcss in a graph, namely \hcslist and \hcspivot respectively. \hcslist counts all \hcss by exhaustively enumerating them which is efficient when the number of \hcss is not very large. \hcspivot, however, counts most \hcss in a combinatorial manner without listing them, thus it is typically much more efficient than \hcslist when the number of \hcss is large. Both of these two frameworks are very general, and they can be easily applied to count any hereditary cohesive subgraph in a graph. To our knowledge, we are the first to study the \hcs counting problem and provide systematic and efficient approaches to solve this problem.

\vspace{0.2em}
\noindent \textit{\underline{New algorithms for $s$-defective clique and $s$-plex counting.}} We first propose two counting algorithms for the $s$-defective clique and $s$-plex counting problems based on the proposed \hcslist framework. We also present several non-trivial pruning techniques %and devise efficient data structures
to boost the efficiency of these algorithms. Then, based on the \hcspivot framework, we develop another two counting algorithms with two carefully-designed pivot-vertex selection strategies for counting $s$-defective cliques and $s$-plexes respectively. %To our knowledge, these four algorithms are the first

\vspace{0.2em}
\noindent \textit{\underline{Extensive experiments.}} We conduct comprehensive experiments using 8 real-world large graphs to evaluate the efficiency of our algorithms. The results show that (1) both \hcslist and \hcspivot are efficient when counting very small \hcss, and \hcspivot is several orders of magnitude faster than \hcslist for counting relatively-large \hcss. For example, on \dbl (425957 vertices and 2099732 edges), when $s=1$ and $q=9$ (the given size), \hcslist takes 162301.4 seconds and 193757.0 seconds to count all $s$-defective cliques and $s$-plexes respectively. However, with the same parameters, \hcspivot consumes only 0.1 seconds and 0.2 seconds to count all $s$-defective cliques and $s$-plexes respectively, which is around 7 orders of magnitude faster than \hcslist. (2) For both locally counting in all vertices (or edges) and simultaneously counting \hcss with size $q$ in a given range, the time costs of \hcspivot are within the same order of magnitude as those for computing the total count of \hcss in a graph with a given size $q$. These results further demonstrate the high efficiency of our pivot-based solutions.% which do not take much additional time costs for locally counting \hcss in all vertices (or edges) and simultaneously counting \hcss with different sizes.

We also perform extensive experiments to evaluate the effectiveness of our algorithms through two applications, including motif-based graph clustering and graph profile based network characterization. The results demonstrate that (1) the \hcs (including both $s$-defective clique and $s$-plex) count is very useful for motif-based graph clustering applications, which is often more effective than the clique counts, and it can also achieve the state-of-the-art performance for real-world community detection.  (2) The \hcs based graph profile is very effective to characterize different types of networks, while the state-of-the-art clique based method is often failed to distinguish various kinds of networks.

For reproducibility purpose, the source code of our work is available at \cite{fullversion}.%an anonymous link \url{https://anonymous.4open.science/r/HCS}.

\section{Preliminaries} \label{sec:preliminary}

\begin{figure}[t] %\vspace{-0.3cm}
	\begin{center}
		\subfigure[An example graph \label{sfig:graph_a}]{\includegraphics[width=0.33\linewidth]{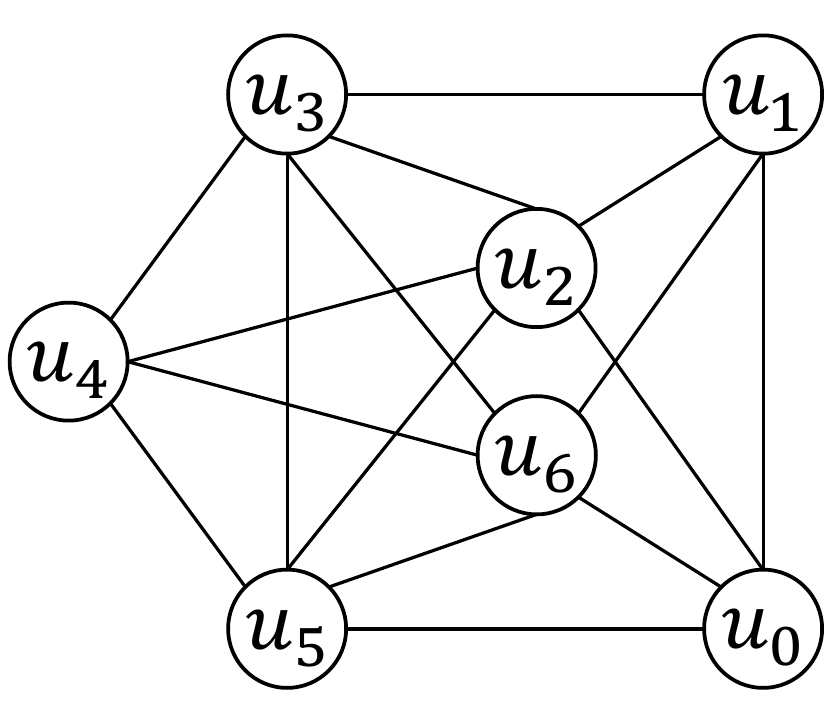}} %\hspace{-0.6cm}
		\subfigure[$1$-dclique/plex \label{sfig:graph_b}]{\includegraphics[width=0.33\linewidth]{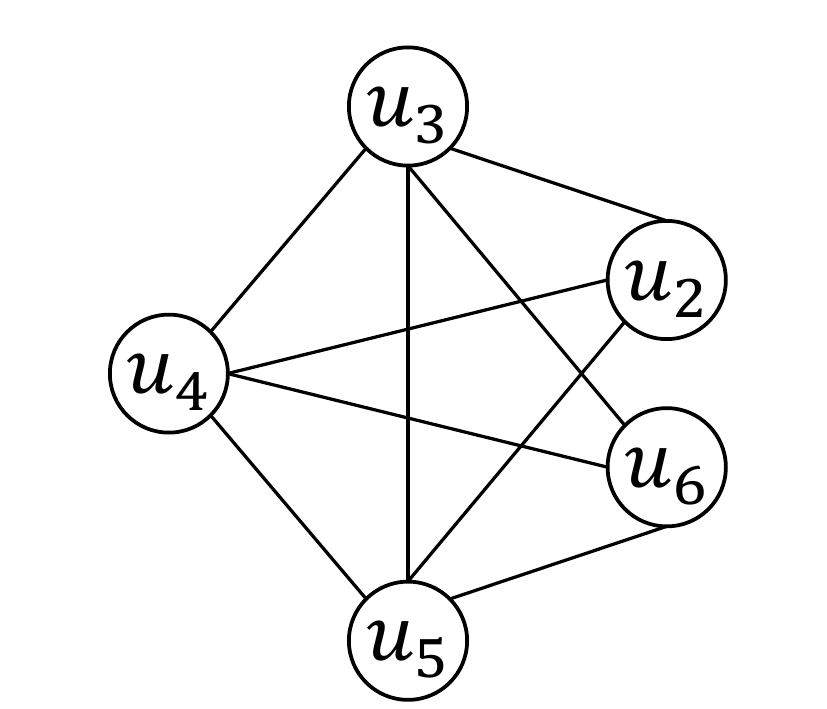}}%\hspace{-0.6cm}
		\subfigure[$1$-plex \label{sfig:graph_c}]{\includegraphics[width=0.33\linewidth]{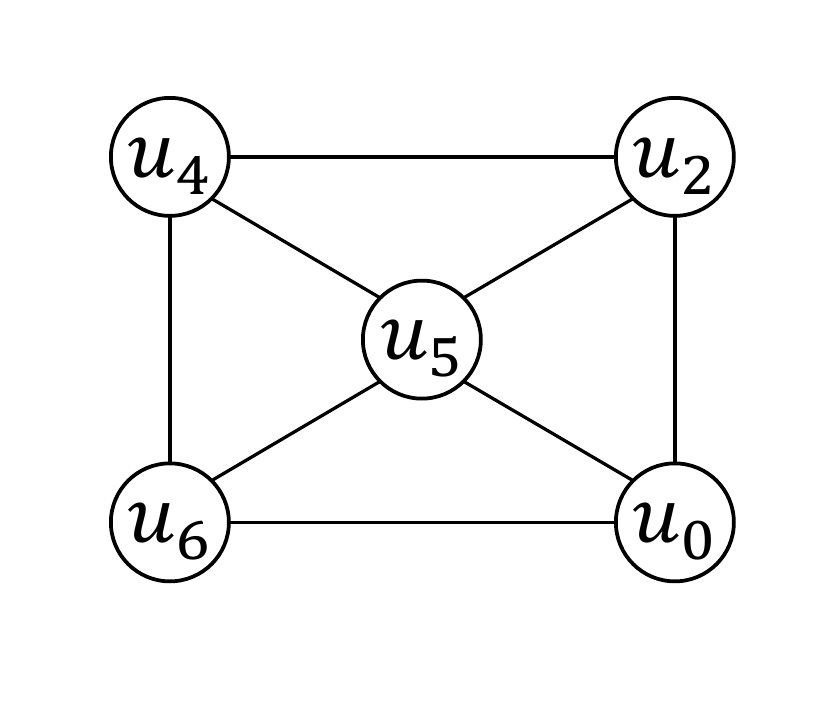}}
	\end{center}
\vspace{-0.4cm}
	\caption{Running example.}
	\label{fig:graph}
\vspace{-0.4cm}
\end{figure}
Denote by $G=(V,E)$ an undirected graph where $V$ is the set of vertices and $E\subseteq V\times V$ is the set of edges. The neighbors of each vertex $u\in V$ is $N(u)\triangleq \{v|(u,v)\in E\}$. The $2$-hop neighbors of each vertex $u$ is $N_2(u)\triangleq \{w|(u,v)\in E, (w,v)\in E,  (u,w)\notin E\}$. Given a graph $G(V, E)$ and a vertices set $Q\subseteq V$, define the edges induced by $Q$ as $E(Q)\triangleq\{(u,v)|(u,v)\in E, u\in Q, v\in Q\}$. We use $G(Q) = (Q, E(Q))$ to denote the subgraph induced by $Q$. For representation simplicity, we replace the size $|Q|$ with $q$. We define the total missing edges of $G(Q)$ as $\overline{m}(Q) = {q\choose 2} - |E(Q)|$, and define the missing edges of a specific vertex $u$ as $\overline{m}(u, Q) = |Q\setminus \{u\}|-|N(u)\cap Q|$. Note that $|Q\setminus \{u\}|=q-1$ if $u\in Q$, $|Q\setminus \{u\}|=q$ otherwise.

A $k$-core of $G$ is a maximal subgraph in which every vertex has a degree no less than $k$ within the subgraph \cite{seidman1983network}. The core number of a vertex $u$ denotes the maximum $k$ such that there is a $k$-core containing $u$. The degeneracy ordering of $V$ is an ordering $\{v_1,v_2,...\}$ such that $v_i$ has the minimum degree in the subgraph $G(\{v_i,v_{i+1},...\})$ \cite{degeneracyOrder}. A nice property of degeneracy ordering is that $\forall i, {|N(v_i)\cap \{v_{i+1},v_{i+2},..\}|} \le \delta$, where $\delta$ is the value of degeneracy. Note that the degeneracy value $\delta$ is equal to the maximum core number of the vertices in $G$, which is often very small in real-world networks \cite{13maximalclique, subgraphcountingbook}. In the ordered graph, we define the (2-hop) out-going neighbors of a vertex $v_i$ as $\vec{N}(v_i)=N(v_i)\cap \{v_{i+1},v_{i+2},...\}$ and $\vec{N_2}(v_i)=N_2(v_i)\cap \{v_{i+1},v_{i+2},...\}$.

\begin{definition}[Hereditary graph]\label{def:hereditary}
	A graph with property $\mathcal{P}$  is hereditary if all induced subgraphs also meet $\mathcal{P}$.
\end{definition}

Definition~\ref{def:hereditary} gives the concept of hereditary graphs. Among the hereditary subgraphs, we are interested in \textit{Hereditary Cohesive Subgraphs} (\hcs in short), because cohesive subgraphs are often with many important practical applications in network analysis \cite{subgraphcountingbook}. It is easy to verify that the classic clique subgraph (completed subgraph) is a kind of \hcs, as clique is cohesive and any subgraph of a clique is also a clique. However, since the strict constraint of clique model, many relaxed clique models are often used in practical applications. In this work, we focus mainly on two widely-used relaxed clique models, $s$-defective clique ($s$-dclique in short)  \cite{Defective_bioinformatics_06} and $s$-plex \cite{Plex_seidman1978graph}, which also satisfy the hereditary property.

\begin{definition}[$s$-dclique \cite{Defective_bioinformatics_06}]\label{def:dclique}
	Given a graph $G$ and a vertices set $Q\subseteq V$, $G(Q)$ is a $s$-dclique if $\overline{m}(Q) \le s$.
\end{definition}

\begin{definition}[$s$-plex \cite{Plex_seidman1978graph}]\label{def:plex}
	Given a graph $G$ and a vertices set $Q\subseteq V$, $G(Q)$ is a $s$-plex if $\forall u\in Q, \overline{m}(u, Q) \le s$.
\end{definition}

It is worth mentioning that Definition~\ref{def:plex} is slightly different from the traditional $k$-plex definition \cite{Plex_seidman1978graph}, as we exclude $u$ when defining $ \overline{m}(u, Q)$. In essence, the $s$-plex is equivalent to the $(k-1)$-plex based on the traditional definition \cite{Plex_seidman1978graph}.

By Definition~\ref{def:dclique} and Definition~\ref{def:plex}, it is easy to verify that both $s$-dclique and $s$-plex satisfy the hereditary property. Note that clique is a special case of $s$-dclique and $s$-plex (clique is a $0$-dclique and $0$-plex). Both $s$-dclique and $s$-plex are not naturally cohesive as clique \cite{Plex_seidman1978graph}. However, as suggested in \cite{Plex_seidman1978graph}, we can easily make them cohesive by restricting their diameters no larger than 2. Fortunately, as shown in Lemma~\ref{lem:cohesive_dclique} and Lemma~\ref{lem:cohesive_plex}, very mild conditions can achieve this goal. Due to the space limits, all missing proofs can be found in the full version of this paper \cite{fullversion}.

\begin{lemma}\label{lem:cohesive_dclique}
	A $s$-dclique with size $q$ such that $q-2 \ge s$ has diameter at most $2$.
\end{lemma}
%\begin{proof}
%Let $G^\prime$ be a $s$-dclique with size $q$ such that $q-2 \ge s$. Assume to the contrary that the length of the shortest path between $u$ and $v$ in $G^\prime$ is $3$. Clearly, it has $N(u)\cap N(v) = \emptyset$, i.e. there are no edge between $N(u)$ and $v$, and between $N(v)$ and $u$. Suppose that $u$ and $v$ have $t$ non-neighbors in $G^\prime$ in total. Then, the number of missing edges in $G^\prime$ is at least $|N(u)|+|N(v)|+1+t$, i.e., $s\ge |N(u)|+|N(v)|+1+t$. Note that the total number of vertices in $G^\prime$ is $q=|N(u)|+|N(v)|+2+t$. Thus, we have $s\ge q-1$, which is a contradiction.
%\end{proof}

\begin{lemma}[\cite{Plex_seidman1978graph}]\label{lem:cohesive_plex}
A $s$-plex with size $q$ that $q \ge 2s+1$ has diameter at most $2$.
\end{lemma}

For practical applications, the value of $s$ is often not very large (e.g., $s\ge 4$) \cite{Defective_bioinformatics_06,dcliqueAAAI2022,maximumDclique21,plex_AAAI21,DAi_kplex}. When $s$ is large, the $s$-dclique and $s$-plex are likely to loose their cohesiveness. Thus, the conditions in Lemma~\ref{lem:cohesive_dclique} and Lemma~\ref{lem:cohesive_plex} are easy to meet. Note that although our frameworks and algorithms are designed for the $s$-dcliques and $s$-plexes that have diameter at most $2$, they can be easily extended to the $s$-dcliques and $s$-plexes with arbitrary diameters.

For representation simplicity, we use $(q,s)$-dclique ($(q,s)$-plex) to represent an $s$-dclique ($s$-plex) with size $q$. Clearly, by our definition, both $(q,s)$-dclique and $(q,s)$-plex are \hcss. In this paper, we focus mainly on the problem of counting these two hereditary cohesive subgraphs in a graph.

\begin{example}
	Fig.~\ref{sfig:graph_a} illustrates an example graph $G$. $G$ is a $(7,2)$-plex and any subgraph of $G$ is a $2$-plex. In Fig.~\ref{sfig:graph_b}, the subgraph induced by $\{u_2,u_3,u_4,u_5,u_6\}$ is both a $(5,1)$-dclique and $(5,1)$-plex. The subgraph induced by $\{u_0,u_2,u_4,u_5,u_6\}$ (Fig.~\ref{sfig:graph_c}) is a  $(5,1)$-plex. It is easy to derive that $G$ contains $1$ $(5,1)$-dclique and $9$ $(5,1)$-plexes.
\end{example}

\stitle{Problem Statement.} Given a graph $G(V, E)$, parameters $q, s$ and a kind of \hcs ($(q,s)$-dclique or $(q,s)$-plex), our goal is to exactly count the number of \hcss in $G$.

\begin{figure}[t] %\vspace{-0.3cm}
	\begin{center}
	\includegraphics[width=0.8\linewidth]{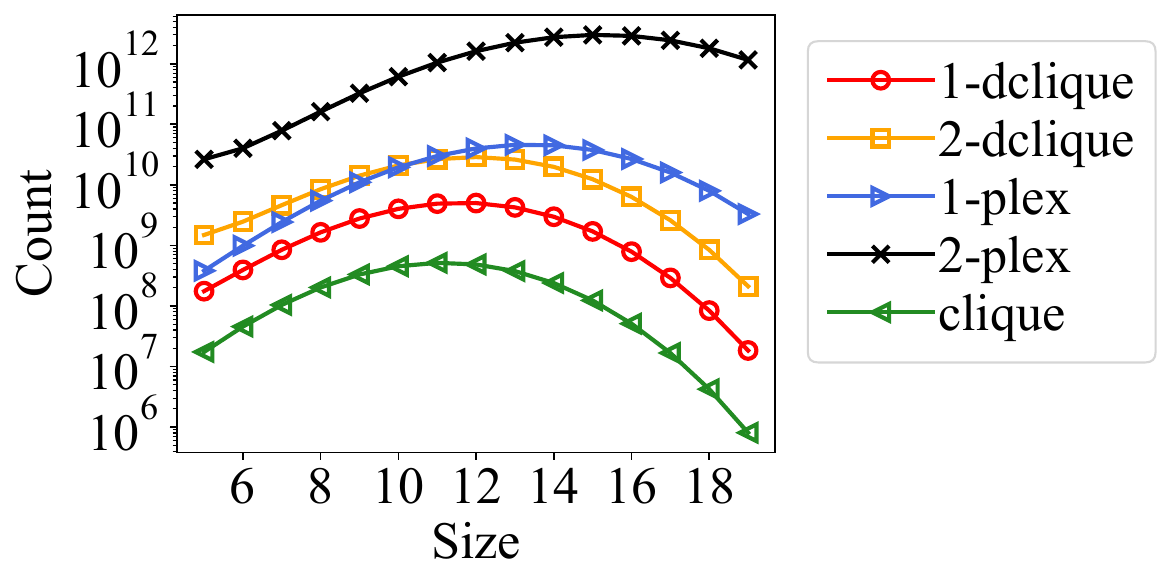}
	\end{center}\vspace*{-0.4cm}
	\caption{The counts of different kinds of \hcs.}\vspace*{-0.4cm}
	\label{fig:cnt}
\end{figure}

\stitle{Challenges.} As discussed in \cite{TuranShadow,PIVOTER,ccpath}, counting the number of $q$-cliques in a graph is an immensely challenging problem due to the exponential explosion in the number of $q$-cliques. Compared to the $q$-clique counting problem, the problem of counting the number of \hcss studied in this work is much more difficult, because the count of \hcs is often orders of magnitude larger than the number of cliques. For example, the graph in Fig.~\ref{sfig:graph_a} contains only two $4$-cliques, while it contains twenty $(4,1)$-dcliques and twenty-five $(4,1)$-plexes. Fig.~\ref{fig:cnt} plots the count of $1$-dcliques, $1$-plexes, $2$-plexes and cliques on the \epi network ($|V|=75879, |E|=811480$). The counts of $s$-dcliques and $s$-plexes are orders of magnitude larger than that of cliques. Moreover, existing approaches for counting $q$-cliques cannot be directly applied to \hcs counting due to the increased complexity and additional constraints associated with \hcss. %Therefore, it is imperative to develop new and efficient algorithms to address this problem. In the following sections, we will propose two distinct and novel solutions to solve the \hcs counting problem.

\section{The Listing-based Solutions} \label{sec:baseline}
In this section, we first propose a general listing-based framework to count the \hcss. Then, based on this framework, we design specific algorithms for counting $s$-dcliques and $s$-plexes, along with some important carefully-designed pruning techniques.

\subsection{A general listing framework \hcslist} \label{subsec:hcs-list}

\begin{algorithm} [t]
\caption{\hcslist}
\label{alg:hcslist}
%\small
\scriptsize
\KwIn{$G=(V, E)$, two integers $q$ and $s$}
\KwOut{The count of hereditary cohesive subgraphs.}
\SetKwProg{Fn}{Procedure}{}{}
%$G'(V',E') \gets $ the $(q-s-1)$-core of $G$\;
Let $V'$ be ordered by degeneracy ordering $\{v_1,v_2,...\}$\;
\For{$i = 1$ to $|V'|$} {
	$C \gets \vec{N}(v_i)\cup \vec{N_2}(v_i)$\; \tcc{Construct the candidate set} %\tcp{Construct the candidate set} %\algorithmiccomment{Construct the candidate set}
	$\listing(C, \{v_i\})$\;
}

\Fn{$\listing(C, R)$}{
\If{$|R|=q-1$}{
	$answer\gets answer + |C|$\;
	\Return\;
}

\For{$u\in C$}{
	$C\gets C\setminus \{u\}$\;
	$C'\gets \{v\in C|R\cup \{u\} \cup \{v\} is\ an\ \hcs\}$\; \tcc{Update the candidate set} %\algorithmiccomment{Update the candidate set}
	$\listing(C', R\cup \{u\})$\;
}
}
\end{algorithm}

A simple approach to count \hcss in a graph is to list them. However, listing all \hcss in a graph can be challenging due to potential overlap and a potentially large number of HCSs. To address this problem, we propose a general backtracking algorithm, called \hcslist, which can enumerates each \hcs in a graph exactly once. The key idea of \hcslist is described as follows. Since \hcs satisfies the hereditary property, we can enumerate an \hcs starting from a single vertex and maintaining a set $R$ that represents the current sub-\hcs. We then iteratively grow $R$ to the desired size, and use a backtracking technique to list all \hcss.

Algorithm~\ref{alg:hcslist} gives the details of \hcslist. The algorithm first orders the graph into a degeneracy ordering (line~1), which can be computed in $O(|E|)$ time using the core decomposition algorithm \cite{corePeel}. Based on the degeneracy ordering, the algorithm lists each \hcs on the lowest-rank vertex of \hcs using the \listing procedure  (lines~2-4), which can guarantee that each \hcs is only enumerated once. The \listing procedure maintains a candidate set $C$ and a sub-\hcs $R$, where each vertex in $C$ can be added into $R$ to form a larger sub-\hcs. Initially, for each vertex $v$, \listing sets $R=\{v\}$ and constructs the candidates set $C$ for $v$. Note that in this paper, we focus mainly on the \hcs with a diameter less than 3, thus the initial candidate set is a set of all one-hop and two-hop outgoing neighbors of $v$ (line~3). Then, for each $v$, the \listing procedure lists all \hcss that contains $v$ as the lowest-rank vertex in the next backtracking call (lines~5-12). Line~11 utilizes the hereditary of \hcs that $R\cup \{u\} \cup\{v\}$ are contained in an \hcs which means that $R\cup \{u\} \cup \{v\}$ is an \hcs itself. Finally, it is not necessary to enumerate $C$ when $|R|=q-1$ because $R\cup \{u\}$ is an \hcs for all $u\in C$ (lines~6-8). It is easy to show that \hcslist only enumerates each \hcs once, thus the correctness of the algorithm can be trivially guarantee.

\stitle{Complexity analysis of \hcslist.} Theorem~\ref{the:listing_complexity} gives the time and space complexity analysis of Algorithm~\ref{alg:hcslist}. Note that for the time complexity, we mainly analyze the size of the recursion tree of \hcslist, as its time complexity is mainly dominated by the recursion tree size.

\begin{theorem}\label{the:listing_complexity}
The time complexity of Algorithm~\ref{alg:hcslist} is $O(|V| \frac{\delta^{2}!}{(\delta^{2} - q)!} \tau)$, and the space complexity of Algorithm~\ref{alg:hcslist} is $O(|V|+|E|+q\delta^2)$, where $\delta$ is the degeneracy of the input graph and $\tau$ is the time consumed in each tree node of the recursion tree.
\end{theorem}
%\begin{proof}
%Since the graph is ordered by degeneracy ordering, the size of the candidate set $C$ in line 3 is $O(\delta^2)$. Clearly, for each vertex $v$, the depth of the recursion tree in the \listing procedure is exactly $q-1$ (line~6). For each $v$, the size of the 2-level of the recursion tree is $O(|C|)\le O(\delta^2)$, and it is easy to derive that the size of the $(q-1)$-level is $O(\frac{\delta^{2}!}{(\delta^{2} - q)!})$. Therefore, the total time complexity of Algorithm~\ref{alg:hcslist} is $O(|V| \frac{\delta^{2}!}{(\delta^{2} - q)!}\times \tau)$. For the space complexity, the main cost is the $O(\delta^2)$ space for storing the candidate set $C$ at each recursion tree node in the \listing procedure. Clearly, the total space needed for this is $O(q\delta^2)$, and the space required for storing the input graph is $O(|V|+|E|)$. As a result, the overall space complexity of the algorithm is $O(|V|+|E|+q\delta^2)$.
%\end{proof}

Note that \hcslist is a general framework. The parameter $\tau$ in Theorem~\ref{the:listing_complexity} depends on specific \hcss, and we will give detailed analyses for two \hcss, including $s$-dclique and $s$-plex, in the following sections. Additionally, when counting $s$-dclique and $s$-plex specifically, we need to take into account of the details of design and implementation. The key is how to design effective prune strategies to reduce the search space. Below, we propose several effective and carefully-designed pruning techniques for both $s$-dclique and $s$-plex counting based on the \hcslist framework.

\subsection{Listing-based $s$-dclique counting} \label{sec:list-dclique}
%In order to transform the \hcslist framework into an efficient algorithm for counting $s$-dcliques, we still need to address two problems. First, we need to develop effective pruning techniques that can enhance efficiency by leveraging the intrinsic properties of $s$-dcliques. Second, we need to design efficient data structures that enable fast implementation. Below, we give the details of our solutions for these problems.
In order to transform the \hcslist framework into an efficient algorithm for counting $s$-dcliques, we still need to develop effective pruning techniques that can enhance efficiency by leveraging the intrinsic properties of $s$-dcliques.

\stitle{Pruning techniques for $s$-dclique counting.} Our first pruning technique is designed to reduce the candidate set $C$ before enumeration. Specifically, we aims to reduce the size of $\vec{N}(v_i) \cup \vec{N_2}(v_i)$. Below, we presents two useful lemmas which will be used to reduce $\vec{N}(v_i)$ and  $\vec{N_2}(v_i)$ respectively.

\begin{lemma}\label{lem:dclique_core}
For each vertex $u\in \vec{N}(v_i)$ ($v_i$ is in lines~2-4 of Algorithm~\ref{alg:hcslist}), if $u$ and $v_i$ are contained in a $(q,s)$-dclique, then we have
$|\vec{N}(u)\cap \vec{N}(v_i)| \ge q-s-2$.
\end{lemma}
%\begin{proof}
%Let $Q$ be a subset of $C\cup \{v_i\}$ and has size $q$. Assume $G(Q)$ is a $s$-dclique that contains $u$ and $v_i$. Let $n_{2hop} = |Q \cap \vec{N_2}(v_i)|$, then we have $|Q \cap \vec{N}(v_i)| = q - n_{2hop} - 1$. Since the maximum number of missing edges is $s$, and $v_i$ already has $n_{2hop}$ missing edges, the maximum number of missing edges of $u$ in $\vec{N}(v_i)$ is $s-n_{2hop}$, i.e., $\overline{m}(u, Q \cap \vec{N}(v_i)) \le s-n_{2hop}$. Therefore, $|\vec{N}(u) \cap \vec{N}(v_i)| \ge |\vec{N}(u) \cap (Q \cap \vec{N}(v_i))| = |Q \cap \vec{N}(v_i)| - 1 - \overline{m}(u, Q \cap \vec{N}(v_i)) \geq q - s - 2$.
%\end{proof}

Lemma~\ref{lem:dclique_core} shows that for an $s$-dclique with size $q$ that contains $v_i$, the subgraph induced by all the vertices in $\vec{N}(v_i)$ must be a $(q-s-2)$-core. As a result, for each $v_i$ in line~3 of Algorithm~\ref{alg:hcslist}, we can reduce $\vec{N}(v_i)$ by computing the $(q-s-2)$-core on the subgraph induced by $\vec{N}(v_i)$.

\begin{lemma}\label{lem:n2-size-reduce}
For each vertex $u \in \vec{N_2}(v_i)$, if $u$ and $v_i$ is contained in a $s$-dclique with size $q$, then $u$ has at least $q-s-1$ neighbors in $\vec{N}(v_i)$.
\end{lemma}
%\begin{proof}
%  Let $Q$ be a $s$-dclique with size $q$ that contains $u$ and $v_i$. Since $u \in \vec{N_2}(v_i)$, $u$ and $v_i$ is disconnected in $Q $ by definition. We claim that $u$ and $v_i$ have at least $q-s-1$ common neighbors. We can prove this claim by a contradiction. Suppose to the contrary that $u$ and $v_i$ have  $q-s-2$ common neighbors. Then, there are $s$ non-common neighbors in total for $u$ and $v_i$, which means that $s$ edges will be missed for $u$ and $v_i$. Since $u$ and $v_i$ is disconnected, $Q$ misses at least $s+1$ edges, which is a contradiction. Based on this claim, the lemma is clearly followed.
%\end{proof}

By Lemma~\ref{lem:n2-size-reduce}, we can further prune the candidate set $C$ by removing the vertices that have less than $q-s-1$ neighbors in the  $(q-s-2)$-core of $\vec{N}(v_i)$  from $\vec{N_2}(v_i)$.

Our second pruning technique is to eliminate unnecessary branches of the procedure \listing in advance. However, how do we determine if a branch of \listing is unnecessary without actually accessing it? Below, we propose an upper bounding technique to solve this issue.

Assuming the procedure \listing is in the state between line~10 and line~11 of Algorithm~\ref{alg:hcslist}, where $u$ is already removed from $C$. We need to determine whether the following branch (line~12) can be pruned based on the values of $u$, $R$, and $C$. Let the vertices in $N(u)\cap C$ be ordered according to the count of non-neighbors in $R$, from smallest to largest, as $\{v_1,v_2,...\}$. It follows that $\overline{m}(v_i, R)\le \overline{m}(v_{i+1}, R)$. We can define $\omega(u, R, C)=\max_i\{\sum_{j\le i}{\overline{m}(v_j, R)} \le s-\overline{m}(R\cup \{u\})\}$. Since $\{v_1,v_2,...\}$ is ordered, $\omega(u,R,C)$ is an upper bound on the number of vertices in $N(u)\cap C$ that can be added into $R\cup \{u\}$. With this information, we can calculate the total upper bound $\gamma_d(u,R,C)$ which can be used to prune the branch in advance if $\gamma_d(u,R,C)< q$. Lemma~\ref{lem:upper_dclique} provides further details on this process.
\begin{lemma} \label{lem:upper_dclique}
Let $\gamma_d(u,R,C)=|R\cup \{u\}| + \min\{s-\overline{m}(R\cup \{u\}), \overline{m}(u, C)\} + \omega(u, R, C)$. Then, $\gamma_d(u,R,C)$ is an upper bound on the size of the $s$-dclique that the enumeration branch of Algorithm~\ref{alg:hcslist} can potentially reach.
\end{lemma}
%\begin{proof}
%As previously explained, $\omega(u, R, C)$ is the upper bound of the number of vertices in $N(u)\cap C$ that can be added into $R\cup \{u\}$. Similarly, it is easy to see that $\min\{s-\overline{m}(R\cup \{u\}), \overline{m}(u, C)\}$ is the upper bound of the number of vertices in $C\setminus N(u)$ that can be added into $R\cup \{u\}$. By adding these upper bounds together, we get the total upper bound represented by $\gamma_d(u,R,C)$.
%\end{proof}

It is worth remarking that the work which focused on the maximum $s$-dclique problem \cite{dcliqueAAAI2022} also defines an upper bound. However, their upper bound uses $|N(u)\cap C|$ directly instead of our $\omega(u,R,C)$. Thus, our upper bound can be deemed as a tighter bound which may also be more effective than the one defined in \cite{dcliqueAAAI2022} for solving the maximum $s$-dclique problem.

\stitle{Implementations.} In the implementation of the $s$-dclique counting algorithm, we maintain an array $A$ of size $|V|$ that $A[v]$ stores the value of $\overline{m}(v, R)$. When $u$ is added into $R$, for each $v \in C'\setminus N(u)$, $A[v]$ should increase $1$. Correspondingly, $A[v]$ should decrease $1$ after the removal of $u$ (line~12 of Algorithm~\ref{alg:hcslist}). Clearly, the maintenance of $A$ costs linear time. With the help of the array $A$, we can get the count of non-neighbors of each vertex in constant time.

The computation of the upper bound $\omega(u, R, C)$ does not really sort the set $N(u)\cap C$. Instead, we construct a bucket array $B$ where $B[i]$ is the count of vertices in $N(u)\cap C$ that have $i$ non-neighbors in $R$. Based on $B$, it is easy to derive that the time complexity of the computation of $\omega(u,R,C)$ is only $O(s)$.

\stitle{Time complexity}. In the worst case, the number of recursion tree nodes of the listing-based $s$-dclique counting algorithm is consistent with Theorem~\ref{the:listing_complexity}. Theorem~\ref{the:sdclique_listing_complexity} gives the time complexity of a specific recursion tree node with candidate set $C$.

\begin{theorem}\label{the:sdclique_listing_complexity}
The time complexity in each recursion node of the listing-based $s$-dclique counting algorithm is $O(|C|^2)$.
\end{theorem}
%\begin{proof}
%The key computation cost is checking whether $R\cup \{u\} \cup \{v\}$ is a $s$-dclique (line~11 of Algorithm~\ref{alg:hcslist}). As described above, the maintenance of $A$ takes linear time for each vertex $u\in C$. Thus, the total cost for maintaining $A$ is $O(|C|^2)$. With the help of the array $A$, computing $\overline{m}(v,R\cup \{u\})$ only needs constant time for each $v\in C$. Therefore, the total cost is also $O(|C|^2)$. Putting it all together, the total time complexity on each recursion node is $O(|C|^2)$.
%\end{proof}

In addition, the time cost taken by the candidate set pruning techniques can be dominated by $O(\delta |E|)$. This is because in our pruning technique, we need to construct a subgraph induced by $\vec N(v_i)$ for each $v_i$, which takes at most $O(\delta |E|)$ in total using a technique developed in \cite{15vldbegonet}. Since the total size of all the neighborhood subgraphs can be bounded by $O(\delta |E|)$, the total time complexity to compute the $(q-s-2)$-cores for all $v_i$ is $O(\delta |E|)$.

\subsection{Listing-based $s$-plex counting}\label{sec:listing_plex}
%Similar to $s$-dclique, designing an efficient $s$-plex counting algorithm based on Algorithm~\ref{alg:hcslist} also needs to solve two issues: (1) how to design prune techniques, (2) how to devise efficient data structures for fast implementation.
Similar to $s$-dclique, designing an efficient $s$-plex counting algorithm based on Algorithm~\ref{alg:hcslist} also needs to  design prune techniques.

\stitle{Pruning techniques for $s$-plex counting.} Like $s$-dclique, we can also use $k$-core based pruning technique to reduce the candidate size for $s$-plex counting. Specifically, we can first compute a $(q-s-2)$-core on the subgraph induced by $\vec N(v_i)$ without missing any $s$-plex, based on the results established in \cite{d2k_kdd18}.

\begin{lemma}[\cite{d2k_kdd18}]\label{lem:plex_prune}
Assume that $u$ and $v$ are two vertices in a $(q,s)$-plex, then $u$ and $v$ have at least $q-2s-2$ common neighbors if $(u,v)$ is an edge in the graph, and have at least $q-2s$ common neighbors otherwise.
\end{lemma}

With Lemma~\ref{lem:plex_prune}, we can remove unpromising vertices from the candidate set $C$ in line~3 of Algorithm~\ref{alg:hcslist}. Specifically, we first reduce $\vec{N}(v_i)$ to a $(q-s-2)$-core by removing the vertices $\{u | u\in \vec{N}(v_i), \ |\vec{N}(u)\cap \vec{N}(v_i)| < q-s-2\}$. Then, we delete the vertices in $\vec{N_2}(v_i)$ that have degree less than $q-2s$ in the $(q-s-2)$-core. Finally, we combine the remaining vertices to get the updated candidate set without losing any $(q,s)$-plex.

To reduce the enumeration branches, we also devise an upper bound on the size of the $s$-plex that the current branch can access. Our upper bound is inspired by \cite{DAi_kplex}. Specifically,
when the \listing procedure is in the state between line~10 and line~11 of Algorithm~\ref{alg:hcslist}, we calculate an upper bound which is defined as $\gamma_{p}(u,R,C)=|R\cup \{u\}| +\max_i\{\sum_{j\le i}{\overline{m}(v_j,R)\le \sum_{v\in R}(s-\overline{m}(v,R))}\}+ \min(s-\overline{m}(u,R),\overline{m}(u,C))$, where  $\{v_1,v_2,...\}$ is the set $N(u)\cap C$ ordered according to $\overline{m}(v_i,R)$. This upper bound contains three parts: (1) $|R\cup \{u\}|$, the listed result, (2) $\max_i\{...\}$, the maximum count of neighbors of $u$, and (3) $\min(s-\overline{m}(u,R),\overline{m}(u,C))$, the maximum count of non-neighbors of $u$. It is easy to derive that $\gamma_{p}(u,R,C)$ is valid upper bound. If this upper bound is less than $q$, we can prune the current branch as it will not produce a valid $s$-plex.

\stitle{Implementations.} In the implementation of the listing-based $s$-plex counting algorithm, we maintain an array $As$ with size $|V|$. For $v\in C$, $As[v]$ is exactly the set of vertices $R\setminus N(v)$. Since the count of the non-neighbors of $v$ is at most $s$, $As$ takes at most $O(s|V|)$ space. Similar to $s$-dclique, the maintenance of $As$ takes $O(|C|)$ time for each $u\in C$. Specifically, $u$ should be inserted into $As[v]$ for $v\in C\setminus N(u)$ if $u$ is added into $R$ and removed otherwise. With the array $As$, we can get the non-neighbors of each vertex efficiently.

Note that an efficient implementation of checking whether $R\cup \{u\} \cup \{v\}$ is a $s$-plex (line~11 of Algorithm~\ref{alg:hcslist}) is nontrivial. A straightforward implementation is that for each $v\in C$ and for each $w\in R\cup \{u\} \setminus N(v)$, we verify whether $\overline{m}(w, R\cup \{u\} \cup \{v\})\le s$. The drawback of this implementation is that it needs to check the non-neighbors of each vertex in $C$. To improve this, we observe that for each $w\in R\cup \{u\} \setminus N(v)$, only the vertices that $\overline{m}(w,R\cup \{u\}) = s$ do not meet the condition because $w$ and $v$ are non-adjacent. We can claim that $w$ is also a non-neighbor of $u$. To see this, we assume to the contrary that $w\in N(u)$, then we have $\overline{m}(w,R) = s$. This is impossible because $w$ is the non-neighbor of $v\in C$ and $\overline{m}(w,R\cup \{v\})=s+1$ which is contradictory to the fact that $R\cup \{v\}$ is a $s$-plex. Therefore, our improved implementation is that for each $w\in R\setminus N(u)$, if $\overline{m}(w,R\cup \{u\}) = s$, we remove the non-neighbors of $w$ in $C$. Note that in this improved implementation, it is no need to check the non-neighbors of the vertices in $C$.

\stitle{Time complexity.} For the $s$-plex counting algorithm, the number of the recursion-tree nodes is also consistent with Theorem~\ref{the:listing_complexity}. However, in each tree node, the time consumed for $s$-plex counting is higher than that for $s$-dclique counting. Theorem~\ref{the:splex_listing_complexity} shows the time complexity taken in each tree node for $s$-plex counting.
\begin{theorem}\label{the:splex_listing_complexity}
	The time complexity in each recursion-tree node of the listing-based $s$-plex counting algorithm is $O(s|C|^2)$.
\end{theorem}
%\begin{proof}
%According to the analysis of the implementation above, checking whether $R\cup \{u\} \cup \{v\}$ is a $s$-plex (line~11 of Algorithm~\ref{alg:hcslist}) takes at most $O(s|C|^2)$ time. The maintenance of $As$ consumes linear time for each vertex $u\in C$. Thus, the total cost of maintaining $As$ is $O(|C|^2)$. Putting it all together, the total time complexity in each recursion-tree node of the listing-based $s$-plex counting algorithm is $O(s|C|^2)$.
%\end{proof}

By Theorem~\ref{the:sdclique_listing_complexity} and \ref{the:splex_listing_complexity}, the listing-based $s$-plex counting algorithm is more sensitive to the parameter $s$, compared to the listing-based $s$-dclique counting algorithm. As confirmed in our experiments (see Table~\ref{tab:running_time}), with the increase of $s$, the running time of the $s$-plex counting algorithm increases more significantly compared to the $s$-dclique counting algorithm.

\section{The pivot-based solutions} \label{sec:pivot-algs}
Although we develop many pruning techniques, the \hcslist framework is still not very efficient for a relatively large $q$ due to the exponential explosion of the number of \hcs in real-world graphs. For example, the \dbl network has $2.8\times 10^8$ $(5,1)$-dcliques, but it has $9.0\times 10^{13}$ $(10,1)$-dcliques. The sheer quantity of \hcs makes listing-based algorithms infeasible. Another disadvantage of the \hcslist framework is that it has redundant calculations. For example, two branches of the \listing procedure may have large overlapping sets of vertices, resulting in unnecessary duplicated calculations. A natural question that arises is whether it is possible to merge different branches of the \listing procedure to reduce redundant calculations. Additionally, is it possible to obtain the counts of \hcs in a combinatorial way, rather than listing each individual one? We propose a novel pivot-based framework that addresses these questions.

To aid understanding, we will first introduce a new listing strategy for the pivot-based framework. Then, we explain the key pivoting technique for implementing combinatorial counting. Finally, based on the pivoting technique, we devise specific algorithms for counting $s$-dcliques and $s$-plexes with various optimization techniques.

\subsection{Warm up: a new listing strategy} \label{subsec:new-listing}
\begin{algorithm}[t]
\caption{The new listing strategy}
\label{alg:listing_strategy}
%\small
\scriptsize
\SetKwProg{Fn}{Procedure}{}{}

\Fn{$\listing(C, R)$}{
\If{$|R|=q-1$}{
	$answer\gets answer + |C|$\;
	\Return\;
}

Split $C$ into $C_1$ and $C_2$\;
$\listing(C_1, R)$\;
\For{$u\in C_2$}{
	$C\gets C\setminus \{u\}$\;
	$C'\gets \{v\in C|R\cup \{u\} \cup \{v\} is\ an\ \hcs\}$\;
	$\listing(C', R\cup \{u\})$\;
}
}
\end{algorithm}

The new listing strategy utilize the idea of divide-and-conquer. Unlike the \listing procedure in Algorithm~\ref{alg:hcslist}, it divides the candidate set into two parts: one part is used directly as the candidate set for the next level of recursion, while the other part is used for listing, similar to the approach used in \hcslist. Algorithm~\ref{alg:listing_strategy} describes such a new listing strategy.

In Algorithm~\ref{alg:listing_strategy}, the candidate set is split into two sets, $C_1$ and $C_2$ (line~5). $C_1$ is the candidate set of the next recursive call directly (line~6) and $C_2$ is to list each vertex (lines~7-10). The following theorem shows that such a new listing strategy can also correctly count all \hcss.

\begin{theorem}\label{the:new_strategy_correct}
Algorithm~\ref{alg:listing_strategy} does not miss any \hcs.
\end{theorem}
%\begin{proof}
%It is easy to check that if an \hcs is in $R\cup C_1 $, it will be accessed in the recursive call in line~6. If an \hcs is in $R\cup C_2$ or $R\cup C_2\cup C_1$, it will be accessed in the recursive call in line~10. As a result, no \hcs will be missed by Algorithm~\ref{alg:listing_strategy}.
%\end{proof}

Note that the new listing strategy does not combine similar branches or eliminate redundant computation compared to the \listing procedure in \hcslist. It still requires listing each \hcs. Indeed, the following theorem shows that the worst-case time and space complexity of Algorithm~\ref{alg:listing_strategy} is equal to those of the \listing procedure in Algorithm~\ref{alg:hcslist}.

\begin{theorem}\label{the:pivot_time}
The worst-case time and space complexity of Algorithm~\ref{alg:listing_strategy} is the same as those of the \listing procedure in Algorithm~\ref{alg:hcslist}.
\end{theorem}

Compared to the \listing procedure in Algorithm~\ref{alg:hcslist}, a useful feature of Algorithm~\ref{alg:listing_strategy} is that it classifies all \hcss into three distinct categories (except the vertices contained in $R$):
\begin{itemize}
	 \item[(I)] the \hcs only containing vertices in $C_1$;
	 \item[(II)] the \hcs only containing vertices in $C_2$;
	 \item[(III)] the \hcs containing vertices both in $C_1$ and $ C_2$.
\end{itemize}
%With this classification, we develop a pivot-vertex technique that can reduce redundant computation and can get the count of the \hcs of class~(3) in a combinational way. %The details is in the next subsection.
With this classification, we are able to develop a pivot-based technique that can significantly reduce redundant calculations by computing the count of the class-III \hcss in a combinatorial manner.

\stitle{Remark.} It is worth remarking that the listing procedure of Algorithm~\ref{alg:hcslist} can be considered as a special case of Algorithm~\ref{alg:listing_strategy}. When $C_1$ in Algorithm~\ref{alg:listing_strategy} is always empty during the recursion procedure, Algorithm~\ref{alg:listing_strategy} degenerates to the listing procedure of Algorithm~\ref{alg:hcslist}.

\subsection{A general pivot-based counting framework} \label{subsec:pivot-framework}
%As described above, the \hcs of class~(3) has a sub-part in $C_1$.  Since the hereditary, the sub-part in $C_1$ is also an \hcs. Clearly, the sub-\hcs should be class~(1) and it should be accessed by the recursion call in line~6 of Algorithm~\ref{alg:listing_strategy}. Based on this observation, we invented the pivot-vertex technique to accelerate counting a portion of \hcs of class~(3). When the \hcs of class~(1) is listed, the pivot-vertex can get a portion of \hcs of class~(3) in a combinational way.

As described previously, the \hcs of class-III contains a sub-part in $C_1$. By virtue of being hereditary, this sub-part is also an \hcs. The sub-\hcs must be of class-I because it only contains vertices in $C_1$. Note that the recursive call in line~6 of Algorithm~\ref{alg:listing_strategy} has already searched all class-I \hcss, thus it is no need to repeatedly list such sub-\hcss when counting the class-III \hcss. Inspired by this, we propose a novel pivoting technique to improve the efficiency of counting the \hcss of class-III.

%To enhance the efficiency of counting the \hcss of class-III, we develop a novel pivoting technique. When listing the \hcss of class-I, the pivoting technique can count a portion of \hcss of class-III simultaneously.

\begin{definition}[pivot vertex]\label{def:pivot}
Let $\mathbb{H}$ be the set of \hcss in $C_1$ that $\forall H\in \mathbb{H}, R\cup H\ is\ an\ \hcs$. A vertex $u\in C_2$ is called a pivot vertex if $\forall H\in \mathbb{H}, R\cup H\cup \{u\} \text{ is an \hcs}$.
\end{definition}

Let $u_p$ denote a pivot vertex. By utilizing $\mathbb{H}$, we can obtain the set of \hcss of class-III $\{H\cup \{u_p\} | H\in \mathbb{H}\}$ without actually listing them. It is straightforward to deduce that the number of \hcs of size $q+1$ in $\{H\cup \{u_p\} | H\in \mathbb{H}\}$ is equivalent to the number of \hcs of size $q$ in $\mathbb{H}$. Thus, by combining these calculations, we can improve the computational efficiency.

Based on this idea, we propose a new counting framework, which is detailed in Algorithm~\ref{alg:pivot}. In addition to the candidate set $C$ and the sub-\hcs $R$, the \listing procedure in Algorithm~\ref{alg:pivot} also maintains a vertex set $D$. The set $D$ contains all the pivot vertices selected so far. Initially, $D=\emptyset$ and the candidate set $C$ is divided into two sets $C_1$ and $C_2$ (line~8). If there exists a vertex $u_p$ that can serve as a pivot vertex, it is removed from $C_2$ and added to $D$ (lines~9-11). If there is no pivot vertex, the process proceeds in the same manner as Algorithm~\ref{alg:listing_strategy} (line~13). Once the size of $R$ reaches $q-1$, each vertex in $D$ and $C$ can be added to $R$ to generate a new \hcs (lines~2-3). When $C$ is empty, every $q-|R|$ vertices in $D$ combined with $R$ form a new \hcs (lines~5-6).

By replacing the \listing procedure in \hcslist with the \listing procedure outlined in Algorithm~\ref{alg:pivot}, we obtain a new general framework, called \hcspivot. Note that the pruning techniques proposed in Section~\ref{sec:baseline} can be directly applied to \hcspivot.
Below, we analyze the correctness of \hcspivot.

\begin{algorithm}[t]
	\caption{The pivot-based counting framework}
	\label{alg:pivot}
%\small
\scriptsize
	\SetKwProg{Fn}{Procedure}{}{}
	
	\Fn{$\listing(C, R, D)$}{
		\If{$|R|=q-1$}{
			$answer\gets answer+|D|+|C|$\;
			\Return\;
		}
		\If{$|C|=0$}{
			$answer\gets answer + {|D|\choose q-|R|}$\;
			\Return\;
		}
%		\If{$R\cup C$ is an \hcs} {
%			$D\gets D\cup C$\;
%			$answer\gets answer + {|D|\choose q-|R|}$\;
%			\Return\;
%		}
		
		Split $C$ into $C_1$ and $C_2$\;
		\If{there exists a pivot vertex $u_p$}{
			$C_2\gets C_2\setminus \{u_p\}$\;
			$\listing(C_1, R, D\cup \{u_p\})$\;
		}
		\Else{
			$\listing(C_1, R, D)$\;
		}
		
		\For{$u\in C_2$}{
			$C\gets C\setminus \{u\}$\;
			$C'\gets \{v\in C|R\cup \{u\} \cup \{v\} is\ an\ \hcs\}$\;
			$\listing(C', R\cup \{u\}, D)$\;
		}
	}
\end{algorithm}

\stitle{Correctness of the \hcspivot framework.} Clearly, the backtracking process of Algorithm~\ref{alg:pivot} can be represented as a recursion tree, where the root node has $D=\emptyset$ and the leaves have either $|R|=q-1$ or $C|=0$. It is important to note that each \hcs lies on exactly one path of the recursion tree. Let us label the recursive calls of \listing in line~11, line~13, and line~17 as $L_1$, $L_2$, and $L_3$, respectively. Each node in the recursion tree either has a child node from $L_1$ or $L_2$, and multiple child nodes from $L_3$. If an \hcs consists only of vertices in $C_1$, it will be in the path down to either $L_1$ or $L_2$. If an \hcs contains a pivot vertex $u_p$ and all other vertices are in $C_1$, it will be in the path down to $L_1$. Finally, if an \hcs contains vertices in $C_2$ (excluding $u_p$ if it exists), it will be in the path down to $L_3$. These include all cases, and it is clear that each \hcs can occur in only one of them. Thus, we can claim that \hcspivot is capable of accurately computing the count of \hcss. Theorem~\ref{the:pivot_corrrect} formally proves that \hcspivot is correct.

\begin{theorem}\label{the:pivot_corrrect}
 \hcspivot correctly counts the \hcss.
\end{theorem}

Note that if no pivot vertex is selected in each recursion of Algorithm~\ref{alg:pivot}, Algorithm~\ref{alg:pivot} degenerates to Algorithm~\ref{alg:listing_strategy}. Thus, the time complexity of \hcspivot is the same as that of \hcslist in the worst case. However, for the two specific \hcss considered in this paper, i.e., $s$-dclique and $s$-plex, we can always select valid pivot vertices in most recursions (for $s$-dclique counting, we can also guarantee that the pivot vertex always exists in each recursion), which can substantivally boost the performance of \hcspivot. Indeed, as shown in our experiments, \hcspivot can be up to 7 orders of magnitude faster than \hcslist based on such a powerful pivoting technique.

\begin{figure*}[t] \vspace{-0.3cm}
	\subfigure[$(4,1)$-dcliques containing $u_0$]{\includegraphics[width=0.49\linewidth]{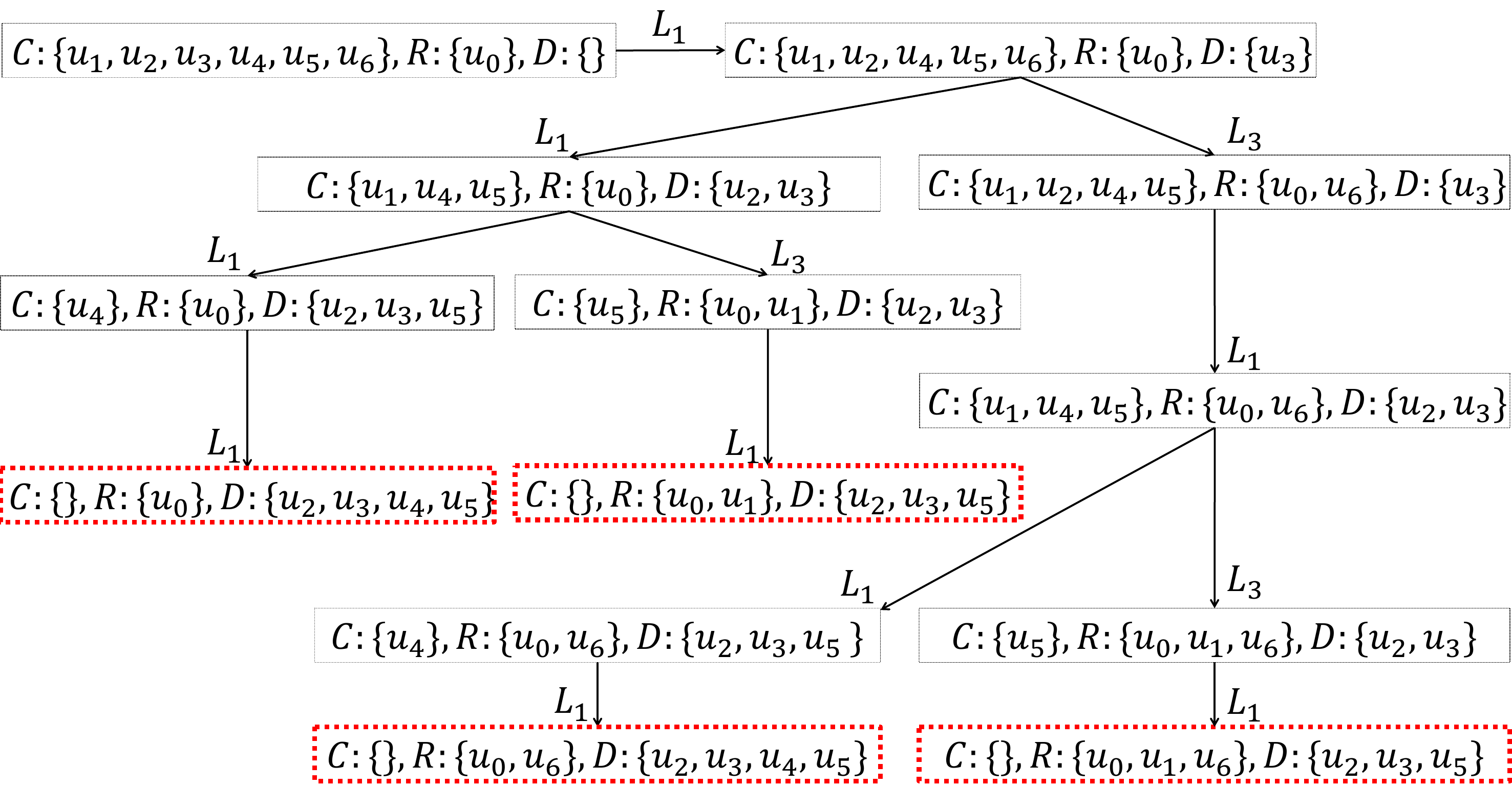} \label{fig:tree_a}}
	\subfigure[$(4,1)$-plex  containing $u_0$]{\includegraphics[width=0.49\linewidth]{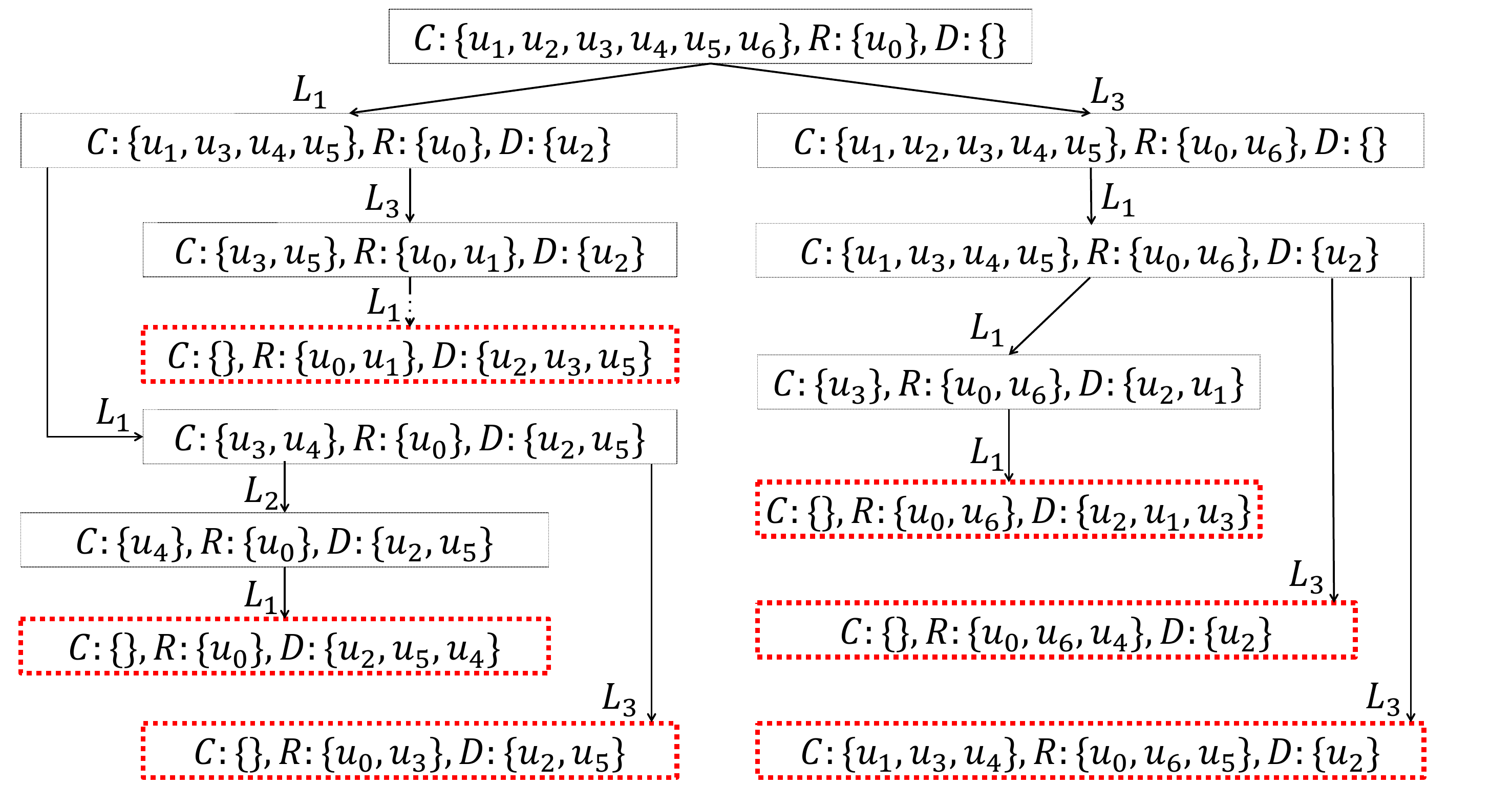} \label{fig:tree_b}}
	\vspace*{-0.3cm} \caption{Illustration of the recursion tree of Algorithm~\ref{alg:pivot} on the graph in Fig.~\ref{sfig:graph_a}.} \vspace*{-0.3cm}
	\label{fig:tree}
\end{figure*}

\subsection{Pivot-based $s$-dclique counting}\label{sub:dpivot}
Note that \hcspivot is a general framework for counting \hcss. When using it to count $s$-dcliques, it is need to solve two issues: (1) how to split $C$ into $C_1$ and $C_2$, and (2) how to choose a pivot vertex.

Lemma~\ref{lem:dclique_split}, derived from Definition~\ref{def:pivot}, provides the criteria for a vertex $u$ to be a pivot vertex for $s$-dclique counting.

\begin{lemma}\label{lem:dclique_split}
Let $\mathbb{H}$ be the set of $s$-dcliques in $C_1$ such that $\overline{m}(R\cup H) \le s$ for each $H\in \mathbb{H}$, and $u$ be a vertex in $C_2$. If $\overline{m}(R\cup H \cup \{u\})\le s$ for each $H\subseteq \mathbb{H}$, then $u$ is a pivot vertex.
\end{lemma}

\stitle{A basic pivoting technique.}
By Lemma~\ref{lem:dclique_split}, it is easy to see that if $\overline{m}(u, R\cup H) = 0$ for all  $H\in \mathbb{H}$, the vertex $u$ must be a pivot vertex because $\overline{m}(R\cup H \cup \{u\}) =\overline{m}(R\cup H) + \overline{m}(u, R\cup H) \le s$. Based on this observation, a basic pivoting method is to select a vertex $u$ that $\overline{m}(u, R) = 0$ and $\overline{m}(u, H) = 0$. The pivot vertex $u$ can be selected from the common neighbors of $R$ and $C_1$ is set as $N(u)\cap C$. Here we take the maximum $|N(u)\cap C|$ over all $u$ to ensure that the set $C_1$ is as large as possible. This is because a large $C_1$ will result in a small $C_2$ which can reduce the number recursive branches in the recursion tree of Algorithm~\ref{alg:pivot}, thus improving the efficiency of the algorithm.  However, such a straightforward method has two limitations. First, it requires the pivot must be a common neighbor of $R$ which is very restrictive and such a pivot may not exist (the common neighbor of $R$ may be empty). Second, %computing the set of common neighbors of $R$ is often very expensive in each recursion. Moreover,
even though the pivot vertex has the maximum degree among the common neighbors of $R$, $|C_2|$ may still be large, meaning that there will be many recursive branches generated by the vertices in $C_2$, which may also slow down the algorithm.

%By Lemma~\ref{lem:dclique_split}, it is easy to see that if $\overline{m}(u, R\cup H) = 0$ for all  $H\in \mathbb{H}$, the vertex $u$ must be a pivot vertex. Based on this observation, a basic pivoting method is to select a vertex from the common neighbors of $R \cup C_1$ as the pivot vertex. More specifically, let $C(R)=\cap_{v\in R}N(v)$ be the set of common neighbors of $R$. The pivot vertex can be selected as $u=\arg\max_{u\in C(R)}{|N(u)\cap C|}$ and $C_1$ is set as $N(u)\cap C$. Here we take the maximum $|N(u)\cap C|$ over all $u$ to ensure that the set $C_1$ is as large as possible. This is because a large $C_1$ will result in a small $C_2$ which can reduce the number recursive branches in the recursion tree of Algorithm~\ref{alg:pivot}, thus improving the efficiency of the algorithm. However, such a straightforward method has two limitations. First, it requires the pivot must be a common neighbor of $R$ which is very restrictive and such a pivot may not exist (the common neighbor of $R$ may be empty). Second, computing the set of common neighbors of $R$ is often very expensive in each recursion. Moreover, even though the pivot vertex has the maximum degree among the common neighbors of $R$, $|C_2|$ may still be large, meaning that there will be many recursive branches generated by the vertices in $C_2$, which may also slow down the algorithm.

\stitle{An improved pivoting technique.} To overcome the limitations of the basic pivoting method, we propose an improved pivoting technique which can ensure that the pivot vertex always exists in each recursion. Specifically, we relax the restriction $\forall H\subseteq \mathbb{H}, \overline{m}(R\cup H \cup \{u\})\le s$ in Lemma~\ref{lem:dclique_split} to $\forall H\subseteq \mathbb{H}, \overline{m}(H \cup \{u\})\le s$. This change means that we do not consider the non-neighbors of $u$ in $R$. Instead of choosing vertices in the common neighbors of $R$, we directly select the vertex with maximum degree in $C$ as the pivot vertex and let $C_1=N(u)\cap C$. This may result in the problem of $\exists u\in D, \overline{m}(u, R) > 0$ at the leaf of the recursion tree (lines~5-7 of Algorithm~\ref{alg:pivot}). Recall that for the basic pivoting technique, it has $\forall u\in D, \overline{m}(u, R)=0$ since $u$ is chosen from the common neighbors of $R$. Thus, we can choose arbitrary $q-|R|$ vertices from $D$. However, now it may occurs the case that $\overline{m}(R\cup H) > s$ where $|H|=q-|R|, H\subseteq D$. We need to compute how many subsets with size $q-|R|$ in $D$ are valid answers. Fortunately, this is not a complicated issue, since $G(D)$ is always a clique as described in the following Theorem~\ref{the:DIsclique}.

\begin{theorem}\label{the:DIsclique}
If the vertex $u$ is selected as the pivot vertex and let $C_1=N(u)\cap C$, $G(D)$ must be a clique for the parameter $D$ in each recursion node of Algorithm~\ref{alg:pivot}.
\end{theorem}
%\begin{proof}
%Since $C_1=N(u)\cap C$, the set of candidates $C$ is made up of the common neighbors of $D$ at all recursion nodes. When $D$ is empty, it is clearly that $G(D\cup \{u\})$ is a clique. When $D$ is not empty and $G(D)$ is already a clique, $G(D\cup \{u\})$ remains a clique due to $u$ being a common neighbor of $D$. Therefore, $G(D)$ must always be a clique.
%\end{proof}

Note that with our improved pivoting technique, the task of choosing arbitrary $q-|R|$ vertices from $D$ when $|C|=\emptyset$ (as stated in line~6 of Algorithm\ref{alg:pivot}) transforms into computing the number of subsets $H$ with $H\subseteq D$ and $|H|=q-|R|$ such that $H\cup R$ is a valid $(q,s)$-dclique.
%\textcolor{blue}{
\begin{theorem}
	 If $\sum_{u\in H}\overline{m}(u, R)\le s-\overline{m}(R)$ holds, $H\cup R$ is a valid $(q,s)$-dclique for $H\subseteq D, |H|=q-|R|$.
\end{theorem}
\comment{
\begin{proof}
	Since $D$ is a clique by Theorem~\ref{the:DIsclique}, i.e. $\overline{m}(H)=0$ and $R$ has already missed $\bar m(R)$ edges, the missing edges in $R\cup H$ is $\overline{m}(R\cup H) = \overline{m}(R) + \overline{m}(H) +\sum_{u\in H}\overline{m}(u, R)\le s$.
\end{proof}
}
%}

As a consequence, the problem of counting all valid $(q,s)$-dcliques is equivalent to a classic variant of \textit{0-1 Knapsack Problem}, where the knapsack size is $s-\overline{m}(R)$ and the item weight is $\overline{m}(u, R)$ for each $u\in D$. This can be solved in $O(s|D|^2)$ time and $O(s|D|)$ space using \textit{Dynamic Programming} \cite{Knapsack}.

The correctness of the pivot-based $s$-dclique counting algorithm can be guaranteed by Theorem~\ref{the:pivot_corrrect} and Theorem~\ref{the:DIsclique}. In each recursion, the time taken for selecting a pivot vertex is at most $O(|C|^2)$, the time consumed for computing the $(q,s)$-dclique counts when $|C=\emptyset|$ is $O(s|D|^2)$, and the time spent on lines~14-17 of Algorithm~\ref{alg:pivot} is $O(|C^2|)$ (Theorem~\ref{the:sdclique_listing_complexity}). Thus, the worst-case time complexity in each recursion is $O(\max(|C|^2,s|D|^2))$. Note that the practical performance of our pivot-based algorithm is extremely faster than the worst-case bound thanks to the benefits of the new pivot vertex selection strategy: (1) By selecting the vertex with the maximum degree in $C$ as the pivot vertex, the size of $C_1=N(u_p)\cap C$ is maximized and the number of vertices to list is minimized; (2)  Pivot-vertices always exist at each node of the full recursion tree, and thus more $s$-dcliques are counted in a combinational way. The following example illustrates how our algorithm works.

\begin{example}
Fig.~\ref{fig:tree_a} depicts a sub-tree of the total recursion tree generated by Algorithm~\ref{alg:pivot} on counting the $(4,1)$-dcliques that include vertex $u_0$. The calls to \listing in line~11, line~13, and line~17 of Algorithm~\ref{alg:pivot} are labeled as $L_1$, $L_2$, and $L_3$, respectively. The root node has a candidate set of $C=\{u_1, u_2, \ldots, u_6\}$, a sub-dclique of $R=\{u_0\}$, and an empty set $D=\{\}$. The vertex $u_3$ is selected as the pivot vertex because it has the maximum degree. The pivot vertex is then added to the set $D$ and the candidate set $C$ becomes $\{u_1,u_2,u_4,u_5,u_6\}$. The vertex $u_2$ has the maximum degree in the updated candidate set, so it is processed through $L_1$ and the vertex $u_6$ is processed through $L_3$.  The last step is to solve a \textit{0-1 Knapsack Problem} at the leaf node. Let us consider the example of the leaf node $R=\{u_0\}$, $D=\{u_2,u_3,u_4,u_5\}$. The problem is to choose $q-|R|=3$ vertices from $D$. Since $G(D)$ must be a clique, we need to ensure that the three chosen vertices have at most $s-\overline{m}(R)=1$ non-neighbors in $R$. Clearly, the two correct answers are $\{u_2,u_3,u_5\}$ and $\{u_2,u_4,u_5\}$.
\end{example}

\subsection{Pivot-based $s$-plex counting} \label{subsec:splex-pivot}
Compared to counting $s$-dcliques, counting $s$-plexes is a more complicated task as each vertex is allowed to miss at most $s$ edges.
Lemma~\ref{lem:plex_split}, which derives from Definition~\ref{def:pivot}, provides the criteria for a vertex $u$ to be considered as a pivot vertex.

\begin{lemma}\label{lem:plex_split}
Let $\mathbb{H}$ be the set of $s$-plex in $C_1$ that $\forall H\in \mathbb{H}, \forall v\in R\cup H, \overline{m}(v, R\cup H) \le s$. Let $u$ be a vertex that $u\in C_2$ and $H'$ represents $R\cup H \cup \{u\}$ in short. If $\forall H\subseteq \mathbb{H}, \forall v\in H', \overline{m}(v,H')\le s$, $u$ is a pivot vertex.
\end{lemma}

The condition $\forall H\subseteq \mathbb{H}, \forall v\in H', \overline{m}(v,H')\le s$ means that for any $H\in \mathbb{H}$, adding the vertex $u$ into the set $R\cup H$ always results in an $s$-plex. It encompasses both (1) the requirement that $\overline{m}(u, H')\le s$, meaning that $u$ itself has at most $s$ non-neighbors, and (2) the requirement that $\forall v\in (R\cup H)\setminus N(u), \overline{m}(v, H')\le s$, meaning that the non-neighbors of $u$ still have at most $s$ non-neighbors after the addition of $u$. Based on these observations,
we propose a pivot-selection technique for counting $s$-plexes based on Theorem~\ref{the:plex_split}.

\begin{theorem}\label{the:plex_split}
Let $u\in C$, and let $C_1=N(u)\cap C$. The vertex $u$ can serve as a pivot vertex if $\forall v\in R\setminus N(u), \overline{m}(v,R\cup C_1)\le s-1$.
\end{theorem}
%\begin{proof}
%Note that $H'=R\cup H \cup \{u\}$ can be considered as $R\cup H$ with the addition of $\{u\}$. For every $v \in R\cap N(u)$, it still has a maximum of $s$ missing edges after $u$ is added to $R\cup H$. For $v\in R\setminus N(u)$, $\overline{m}(v, H')\le s$ is ensured because $\overline{m}(v,R\cup C_1)\le s-1$. And for $v\in H$, since $u$ and $v$ are neighbors, $\overline{m}(v, R\cup H) = \overline{m}(v, H')$. Putting all it together, all vertices in $H'$ have at most $s$ missing edges, and thus $u$ is a valid pivot vertex.
%\end{proof}

%According to the split strategy in Theorem~\ref{the:plex_split}, we choose the vertex $u$ with the maximum degree and let $C_1= N(u)\cap C$ in Algorithm~\ref{alg:pivot}. Then we check whether all the non-neighbors of $u$ in $R$ have at most $s-1$ non-neighbors in $R\cup C$. If yes, $u$ is the pivot vertex. Example~\ref{exa:plex} gives an example about how the split strategy woks on the graph in Figure~\ref{fig:graph}.

According to Theorem~\ref{the:plex_split}, the steps of choosing the pivot vertex is as follows. First, for each $v\in C$, we evaluate whether all vertices in $R\setminus N(u)$ have at most $s-1$ non-neighbors in $R\cup (C\cap N(v))$. Then, we obtain a set of vertices that can serve as the pivot vertex. Among them, we choose the one that has maximum degree in $C$ as the pivot vertex. If no vertex in $C$ can serve as the pivot vertex, we set $C_1 = C\cap N(v)$ that $v$ has maximum degree in $C$.

However, for each $v\in C$ and for each $w\in R\setminus N(v)$, computing the count of non-neighbors of $w$ in $R\cup (C\cap N(v))$ is quite a heavy work. To enhance the efficiency, we improve the strategy of choosing the pivot vertex according to Corollary~\ref{cor:plex_choose_pivot}, which comes from Theorem~\ref{the:plex_split}.

\begin{corollary}\label{cor:plex_choose_pivot}
Let $u$ be a vertex in $C$ and let $C_1=N(u)\cap C$. $u$ can serve as a pivot vertex if $\forall v\in R\setminus N(u), \overline{m}(v,R\cup C)\le s-1$.
\end{corollary}

The correctness of Corollary~\ref{cor:plex_choose_pivot} can be ensured by Theorem~\ref{the:plex_split} because $C_1$ is a subset of $C$. According to Corollary~\ref{cor:plex_choose_pivot}, we just need to compute the count of non-neighbors in $R\cup C$ for each vertex in $R$, which can be done in linear time. Similarly, among the set of vertices that can serve as the pivot vertex, we choose the one that has maximum degree in $C$ as the pivot vertex, and let $C_1$ be the set of neighbors of the pivot vertex. It is important to note that if the condition ``$\forall v\in R\setminus N(u), \overline{m}(v,R\cup C)\le s-1$" does not hold, there is no pivot vertex can be selected. In this case, we just pick the maximum-degree vertex $u$ in $C$ and set $C_1=N(u)\cap C$.% to partition $C$.

Based on this improved pivoting technique, the time taken for selecting a pivot vertex is $O(|C|^2)$ in each recursion. Clearly, computing the set $C_1$ consumes at most $O(|C|^2)$ time, and the time spent on lines~14-17 of Algorithm~\ref{alg:pivot} is $O(s|C|^2)$ (Theorem~\ref{the:splex_listing_complexity}). Thus, the total time cost in each recursion is $O(s|C|^2)$. Similar to the pivot-based $s$-dclique counting algorithm, the practical performance of the pivot-based $s$-plex counting algorithm is also much better than the worst-case time bound, as confirmed in our experiments.

%First, we select the vertex with the maximum degree in $C$, and denote it as $u$. Then, we let $C_1=N(u)\cap C$. Next, we evaluate whether all vertices in $R\setminus N(u)$ have at most $s-1$ non-neighbors in $R\cup C_1$. If this condition holds true, we consider $u$ to be the pivot vertex. If not, determine that the pivot vertex does not exist. As demonstrated in Example~\ref{exa:plex}, this split strategy is applied to the graph in Fig.~\ref{fig:graph}.

%The correctness of the pivot-based $s$-plex counting algorithm is proved by Theorem~\ref{the:pivot_corrrect}. The worst case time complexity of the algorithm is $O(2^n)$, as confirmed by Theorem~\ref{the:pivot_time}. Similar to the pivot-based $s$-dclique counting algorithm, the practical performance of the $s$-plex counting algorithm is much better than $O(2^n)$. However, counting $s$-plex is slower than $s$-dclique because the count of pivot-vertices is smaller and the vertices to list are larger.  The experiment results in Section~\ref{sec:exp} confirm the analysis.

\begin{example}\label{exa:plex}
Fig.~\ref{fig:tree_b} is the recursion tree of Algorithm~\ref{alg:pivot} on counting $(4,1)$-plex. We label the calls of \listing in line~11, line~13, line~17 of Algorithm~\ref{alg:pivot} as $L_1, L_2, L_3$, respectively. The root node has $C=\{u_1,u_2,...,u_6\}$ and $R=\{u_0\}$. Among the root node, $u_2,u_5$ and $u_6$ can serve as the pivot vertex and they all have $4$ neighbors in $C$. The vertices $u_1, u_3$ and $u_4$ cannot serve as the pivot vertex. This is because they have non-neighbors in $C\cup R$ and the count of the non-neighbor in $C\cup R$ is at most $s-1=0$ according to Corollary~\ref{cor:plex_choose_pivot}. Suppose that choose $u_2$ as the pivot vertex, and  then $C_1=N(u_2)=\{u_1,u_3,u_4,u_5\}, C_2=\{u_6\}$. $C_1$ is the candidate set of the child node through $L_1$. The vertex $u_6$ is inserted into $R$ in the child node through $L_3$. Note that on the recursion node with $C=\{u_3,u_4\}$, $R=\{u_0\}$, and $D=\{u_2,u_5\}$, no vertex can serve as the pivot vertex. In this case, we set $C_1=C\cap N(u_3)=\{u_4\}$ and $C_2=\{u_3\}$. Here a child node through $L_2$ with candidate set $C_1$ occurs, and $u_3$ is inserted into $R$ in the child node through $L_3$.

%The search process would have gone down to a child node through $L_3$, but it was cut off by the pruning technique described in subsection~\ref{sec:listing_plex}. The pruning process used the upper bound $|R\cup \{u\}| + s-\overline{m}(u,R)+\max_i\{...\}$ from subsection~\ref{sec:listing_plex}, where it has $|R\cup \{u\}|=2, s-\overline{m}(u,R) =1$, and $\max_i\{...\} = 0$. As a result, the upper bound was less than $5$, and the branch was deemed unnecessary to explore further.

In Fig.~\ref{fig:tree_b}, the red dotted leaves are the answers. The recursion node with $C=\{u_1,u_3,u_4\}$, $R=\{u_0,u_6,u_5\}$, and $D=\{u_2\}$ is the leaf node because $|R|=q-1$ (line~2 of Algorithm~\ref{alg:pivot}). In the red dotted leaves, the combination of any $q-|R|$ vertices in $D$ with $R$ forms a $(4,1)$-plex. For instance, in the leaf node with $R=\{u_0,u_6\}$ and $D=\{u_2,u_1,u_3\}$, every two vertices of $D$ combined with $R$ will result in a $(4,1)$-plex.
%Since $u_3$ has the maximum degree, $C_1=N(u_3)=\{u_1,u_2,u_4,u_5,u_6\}$. The non-neighbor of $u_3$ in $R$ is $u_0$ and $u_0$ has a non-neighbor in $C_1$, so the condition in Theorem~\ref{the:plex_split} does not meet. $u_3$ is not a pivot vertex. Thus the root node has two child node, one goes down through $L_2$ and another goes down through $L_3$. On the search node with $C=\{u_1,u_4,u_5\}, R=\{u_0\}$ and $D=\{u_2\}$, the pivot vertex is $u_5$, $C_1=\{u_4\}$ and $C_2=\{u_1\}$. It was necessary to go down to a child node through $L_3$, but this branch was cut off by the pruning technique in subsection~\ref{subsec:prune}. Specifically, in the upper bound  $|R\cup \{u\}| + s-\overline{m}(u,R)+max_i\{...\}$ in subsection~\ref{subsec:prune}, it has $|R\cup \{u\}|=2,s-\overline{m}(u,R) =1$ and $max_i\{...\} = 0$.  Thus the upper bound is less than $5$ and the branch should be pruned. At each leaf node, any $q-|R|$ vertices in $D$ combined with $R$ is a $(5,1)$-plex. For example, in the leaf node with $R=\{u_0,u_3\}, D=\{u_1,u_2,u_5,u_6\}$, both $\{u_1,u_2,u_5\}, \{u_1,u_2,u_6\},\{u_1,u_5,u_6\}$ and $\{u_2,u_5,u_6\}$ combined with $\{u_0,u_3\}$ are $(5,1)$-plexes.
\end{example}

\subsection{Discussions}\label{ssec:discussion}
%\stitle{Counting \hcs with size in a range $[q_l,q_r]$ simultaneously.} Algorithm~\ref{alg:pivot} is versatile in counting \hcs of different in a range $[q_l,q_r]$ simultaneously. This is possible because the search tree of Algorithm~\ref{alg:pivot} is almost not affected by the parameter $q$. Thus we can count \hcs with size in a range $[q_l,q_r]$ simultaneously by making a few modifications to the algorithm: (1) change line~2 into $|R|=q_r-1$; (2) change line~6 into $answer_q\gets answer_q + {|D|\choose q-|R|}$ for $q\in [q_l, q_r]$.

\stitle{Counting \hcs with size in a range $[q_l,q_r]$ simultaneously.} A striking feature of Algorithm~\ref{alg:pivot} is that it is highly adaptable and capable of simultaneously computing the counts of \hcss of different sizes in a range of $[q_l, q_r]$. This is achievable due to the fact that the recursion tree of Algorithm~\ref{alg:pivot} is only slightly impacted by the parameter $q$. By making a few modifications to the algorithm, we can count \hcss with sizes in the range $[q_l, q_r]$ simultaneously. These modifications include: (1) modifying $|R|=q-1$ in line~2 to $|R|=q_r-1$, and (2) modifying $answer\gets answer + {|D|\choose q-|R|}$ in line 6 to $answer_q\gets answer_q + {|D|\choose q-|R|}$ for every $q\in [q_l, q_r]$. Note that the listing-based framework, i.e., Algorithm~\ref{alg:hcslist}, cannot simultaneously compute the counts of \hcss of different sizes in a range. In the experiments, we will use the counts of \hcss with various size to build a \textit{graph profile} \cite{xu2008superfamily,guimera2007classes} for a given network. The experimental results show the ability of the proposed \emph{graph profile} to characterize different types of networks.

%\stitle{Local counting.} Local counting means the count of \hcs that containing a specific vertex or edge. Given a vertex $u$ or a edge $(u,v)$, define the local count as $c_u$ or $c_{(u,v)}$. The vertices can be split into two categories, $u\in R$ and $u\in D$. For $u\in R$, it has $c_u\gets c_u+{|D|\choose q-|R|}$. However, for $u\in D$, it has $c_u\gets c_u+{|D|-1\choose q-|R|-1}$. The edges can be split into three categories, (1) $(u\in R,v\in R), (2) (u\in R, v\in D)$ and (3) $(u\in D, v\in D)$. For class (1), it has $c_{(u,v)}\gets c_{(u,v)} + {|D|\choose q-|R|}$. For class (2), it has $c_{(u,v)}\gets c_{(u,v)} + {|D|-1\choose q-|R|-1}$. For class (3), it has $c_{(u,v)}\gets c_{(u,v)} + {|D|-2\choose q-|R|-2}$.

\stitle{Local counting.} Another important feature of Algorithm~\ref{alg:pivot} is that it can obtain the local count of \hcs for each vertex or edge. Here local count refers to the number of \hcss that contain a specific vertex or edge. For a given vertex $u$ or edge $(u,v)$, we define its local count as $c_u$ or $c_{(u,v)}$, respectively. To implement locally counting using Algorithm~\ref{alg:pivot}, we only need to add a counter for each vertex (or each edge) in line~6. More specifically, to compute the local counts for vertices, we partition the vertices into two categories: $u\in R$ and $u\in D$. If $u\in R$, then $c_u$ is updated as $c_u\gets c_u + {|D|\choose q-|R|}$. On the other hand, if $u\in D$, then $c_u$ is updated as $c_u\gets c_u + {|D|-1\choose q-|R|-1}$. To calculate the local counts for edges, we divide the edges into three categories: (1) $(u\in R,v\in R)$, (2) $(u\in R, v\in D)$, and (3) $(u\in D, v\in D)$. For category (1), $c_{(u,v)}$ is updated as $c_{(u,v)}\gets c_{(u,v)} + {|D|\choose q-|R|}$. For category (2), $c_{(u,v)}$ is updated as $c_{(u,v)}\gets c_{(u,v)} + {|D|-1\choose q-|R|-1}$. Finally, for category (3), $c_{(u,v)}$ is updated as $c_{(u,v)}\gets c_{(u,v)} + {|D|-2\choose q-|R|-2}$. Note that the listing-based framework, i.e., Algorithm~\ref{alg:hcslist}, cannot compute the local counts for all vertices (or edges) in such a combinatorial manner. In the experiments, we will apply the local counts to construct a matrix $M$, where $M_{ij}$ is the count of \hcs containing the edge $(i,j)$. Such a matrix $M$ is then used for graph clustering applications which achieves much better performance compared to the state-of-the-art graph clustering methods, as shown in our experiments.

%In the experiment, we apply the similarity matrix on an application of graph clustering and get great results.

\stitle{Difference with the maximal $s$-dclique or $s$-plex enumeration.} %Our algorithm is quite different from the maximal $s$-plex or $s$-dclique enumeration algorithms. Firstly, the \hcs $R\cup D$ at the leaf search node is a large \hcs but not necessary to be maximal. Secondly, all of the large \hcs $R\cup D$ at the leaf search nodes do not have overlapped sub-\hcs to make sure each \hcs is counted only once according to Theorem~\ref{the:pivot_corrrect} while the maximal $s$-dcliques or $s$-plexes may have overlapped sub-structure.
Note that existing maximal $s$-dclique or $s$-plex  enumeration algorithms \cite{dcliqueAAAI2022,23sigmoddclique, plex_AAAI21, 22wwwkplex, DAi_kplex} also use a pivoting technique to reduce the enumeration branches. Our pivot-based algorithm (i.e., Algorithm~\ref{alg:pivot}), however, is significantly different from those existing algorithms. First, the recursion tree of Algorithm~\ref{alg:pivot} is completely different from those of the maximal $s$-dclique or $s$-plex  enumeration algorithms. Specifically, the leaf nodes of our recursion tree may not be maximal results. However, for the maximal $s$-plex or $s$-dclique enumeration algorithms \cite{dcliqueAAAI2022,23sigmoddclique, plex_AAAI21, 22wwwkplex, DAi_kplex}, the leaf nodes of the recursion tree are maximal $s$-plexes or $s$-dcliques. Second, to ensure that each \hcs is counted exactly once, our algorithm guarantees that there are no overlapping sub-\hcs among the \hcss at the leaf nodes. This differs from the maximal $s$-dcliques or $s$-plexes enumeration algorithms, which may contain overlapping sub-structures at the leaf nodes. Third, our definition of pivot vertex (Definition~\ref{def:pivot}) is specifically designed for the \hcs counting problem, which distinguishes it from the maximal $s$-dcliques or $s$-plexes enumeration problems. Moreover, to improve the efficiency, we have devised unique and novel pivoting techniques to prune unnecessary candidate sets and branches (Theorem~\ref{the:DIsclique} and \ref{the:plex_split}), which significantly differ from the pivoting techniques utilized in the enumeration of maximal $s$-dcliques or $s$-plexes. %So our problem is harder and our algorithm is more complex.

%\stitle{Relation to \kw{PIVOTER} \cite{PIVOTER}.} \kw{PIVOTER} is a clique counting algorithm by adapting the maximal clique enumeration algorithm Bron-Kerbosch \cite{BK}. Since clique is special kind of \hcs by setting $s$ to zero, our proposed Algorithm~\ref{alg:pivot} can also be applied to count cliques. Specifically, Algorithm~\ref{alg:pivot} is exactly the \kw{PIVOTER} algorithm after removing the lines~12-13 and fixing pivot vertex as the maximum degree vertex, which means that the \kw{PIVOTER} algorithm is a special case of our algorithm. There are also many difference between Algorithm~\ref{alg:pivot} and \kw{PIVOTER}. Firstly, The candidate set of Algorithm~\ref{alg:pivot} is greatly larger than \kw{PIVOTER}. Because The candidate set of Algorithm~\ref{alg:pivot} includes both the 1-hop and 2-hop neighbors (line~3 of Algorithm~\ref{alg:hcslist}), while the candidate set of \kw{PIVOTER} only includes the 1-hop neighbors. Secondly, the idea of splitting the candidate set into $C_1$ and $C_2$ is not available to \kw{PIVOTER}. This is a general strategy that can be adapted to all \hcs counting problem. Thirdly, we devise many pruning techniques according to the complex restrictions  in Definition~\ref{def:dclique} and \ref{def:plex}.

%Algorithm~\ref{alg:pivot} in our proposal can also be applied to count cliques, as cliques are a special type of hypercliques with $s=0$.  Thus the \kw{PIVOTER} algorithm
\stitle{Relation to the pivot-based $q$-clique counting algorithm.} \kw{PIVOTER} is the state-of-the-art pivot-based $q$-clique counting algorithm \cite{PIVOTER} which builds upon the classic Bron-Kerbosch algorithm for maximal clique enumeration \cite{BK}. Since $q$-clique is a type of \hcs, our pivot-based algorithm (i.e., Algorithm~\ref{alg:pivot}) can also be applied to count $q$-cliques. We note that \kw{PIVOTER} is a special case of Algorithm~\ref{alg:pivot}. Specifically, we remove lines~12-13, select the maximum-degree vertex in $C$ as the pivot vertex $v_p$, and set $C_1=N(v_p)\cap C$, then Algorithm~\ref{alg:pivot} is exactly equivalent to the \kw{PIVOTER} algorithm. Compared to \kw{PIVOTER}, which can only be applied to count $q$-cliques, Algorithm~\ref{alg:pivot} is a general framework which is capable of processing all \hcs counting problems. Moreover, the idea of splitting the candidate set into $C_1$ and $C_2$ in Algorithm~\ref{alg:pivot} is not used in \kw{PIVOTER} \cite{PIVOTER}, which is critical to develop a general approach to all \hcs counting problems. Based on Algorithm~\ref{alg:pivot}, we need to design different candidate set partition and pivot vertex selection strategies when handling various \hcs counting problems. %In this paper, we develop several non-trivial candidate set partition and pivot vertex selection strategies for two \hcs counting problems (i.e., $s$-dclique and $s$-plex counting).
Additionally, our algorithm incorporates several carefully-designed pruning techniques that are tailored to the specific restrictions defined in Definitions~\ref{def:dclique} and \ref{def:plex}, which are also not shown in \kw{PIVOTER} \cite{PIVOTER}.

%It is a specific case of our algorithm. To obtain the \kw{PIVOTER} algorithm from Algorithm~\ref{alg:pivot}, we can simply set the parameter $s$ as $0$ and  remove lines~12-13. However, there are several key differences between Algorithm~\ref{alg:pivot} and the \kw{PIVOTER} algorithm. Firstly, the candidate set used in Algorithm~\ref{alg:pivot} is larger, as it includes both 1-hop and 2-hop neighbors, while the PIVOTER algorithm only considers 1-hop neighbors. Secondly, the splitting of the candidate set into $C_1$ and $C_2$ and the way to select the pivot-vertices is unique to Algorithm~\ref{alg:pivot}, and serves as a general strategy for solving all \hcs counting problems. Finally, our algorithm incorporates several pruning techniques that are tailored to the specific restrictions defined in Definitions~\ref{def:dclique} and \ref{def:plex}, which is not shown in the clique counting algorithm.

%This property suggests that we are using hereditary from a large \hcs ($R\cup D$ in line~6) to compute small \hcs.

%Before the split strategy, we introduce the way to judge whether $u$ is a pivot vertex of $C_1$.

%\begin{theorem}\label{the:plex_judge}
%Given $R,C_1$ and $u$, $u$ is a pivot vertex if $\forall v\in C_1, |(R\setminus N(u)) \cap (R\setminus N(v))| =0$.
%\end{theorem}
%\begin{proof}
%The equation $|(R\setminus N(u)) \cap (R\setminus N(v))| =0$ means $u$ and $v$ does not have common non-neighbors in $R$. Thus $H\cup R\cup \{u\}$ must be a legal $s$-plex where $H$ is an \hcs with all vertices in $C_1$.
%\end{proof}

\stitle{Relation between listing and pivot based solutions.} Note that our listing-based solutions are both essential and nontrivial, and they are not always inferior to pivot-based solutions. In cases where $q$ is a small constant, the time complexity of listing-based solutions is polynomial, while pivot-based solutions always have exponential time complexity for any value of $q$. Our experiments have also shown that listing-based solutions outperform pivot-based solutions in certain scenarios (refer to Table~\ref{tab:running_time}). In Section~\ref{subsec:new-listing}, the new listing strategy serves as a precursor to the subsequent pivot-based solutions. Its main purpose is to improve the understandability of the pivot-based solutions.

\stitle{Extension to counting \hcss with a larger diameter.} To apply Algorithm~\ref{alg:hcslist} and Algorithm~\ref{alg:pivot} to count \hcss with diameter larger than $2$, we need to modify line~3 of Algorithm~\ref{alg:hcslist} to include vertices in longer hops. As for Algorithm~\ref{alg:pivot}, we can make use of the basic pivot technique in Definition~\ref{def:pivot}.

\vspace{-0.1cm}
\section{ Experiments}\label{sec:exp}

\begin{table}[t]
	%\small
\scriptsize
	\centering
	\caption{Datasets}
	%\vspace*{-0.3cm}
	\begin{tabular}{c|c|c|c|c}
		\toprule
		\textbf{Networks} & $\bm{|V|}$ & $\bm{|E|}$ & $\bm{\delta}$  & \textbf{Type}\\
		\midrule
		
		\textbf{\wik} & 7115 & 201524 & 53 & Social network\\
		\textbf{\cai} & 26475 & 106762 & 22 & Autonomous system network\\
		\textbf{\epi} & 75879 & 811480 & 67 & Social network\\
		\textbf{\ema} & 265009 & 728960 & 37 & Communication network\\
		\textbf{\ama} & 403394 & 4886816 & 10 & Communication network\\
		\textbf{\dbl} & 425957 & 2099732 & 113 & Co-authorship network\\
		\textbf{\pok} & 1632803 & 44603928 & 47 & Social network\\
		\textbf{\ski} & 1696415 & 22190596 & 111 & Autonomous system network\\
		\bottomrule
	\end{tabular}
	
	\label{tab:datasets}\vspace{-0.3cm}
\end{table}

\begin{table*}[t!]\vspace*{-0.1cm}
	%\small
	\scriptsize
	\centering
	\caption{Running time (sec) of different algorithms with various $q$ and $s$} \vspace*{-0.3cm}
	\begin{tabular}{c| c|cc|c|cc|  c|cc|c|cc|  c|cc|c|cc}
		\toprule
		\multirow{4}{*}{\textbf{Networks}} & \multicolumn{18}{c}{\textbf{Running time (sec)}}  \\
		\cline{2-19}
		&   \multicolumn{6}{c|}{$\bm{s=1}$} &
		 \multicolumn{6}{c|}{$\bm{s=2}$} &
		  \multicolumn{6}{c}{$\bm{s=3}$} \\
		\cline{2-19}
		& \multicolumn{3}{c|}{$s$-dclique}
		& \multicolumn{3}{c|}{$s$-plex}
		& \multicolumn{3}{c|}{$s$-dclique}
		& \multicolumn{3}{c|}{$s$-plex}
		& \multicolumn{3}{c|}{$s$-dclique}
		& \multicolumn{3}{c}{$s$-plex}\\
		\cline{2-19}
		& $\bm{q}$ & \dlist & \dpivot & $\bm{q}$ & \plist & \ppivot
		&$\bm{q}$ & \dlist& \dpivot& $\bm{q}$ & \plist  & \ppivot
		&$\bm{q}$ & \dlist & \dpivot& $\bm{q}$ & \plist & \ppivot \\
		\midrule

\multirow{3}{*}{\wik} & 5 & 10.2 & \textbf{9.6} & 5 & 18.3 & \textbf{16.7} & 7 & \textbf{126.4} & 134.9 & 7 & 4315.5 & \textbf{3139.3} & 10 & 3083.8 & \textbf{720.3} & 15 & - & \textbf{70049.5} \\
& 10 & 13.1 & \textbf{5.7} & 10 & 80.9 & \textbf{14.6} & 12 & 60.1 & \textbf{43.5} & 12 & 5059.8 & \textbf{1618.3} & 15 & \textbf{95.3} & 115.4 & 20 & 3612.8 &\textbf{ 1548.9} \\
& 15 & \textbf{1.4} & 1.8 & 15 & 8.1 & \textbf{2.4} & 17 & \textbf{3.7} & 7.8 & 17 & - & \textbf{88.9 }& 20 & \textbf{3.9} & 9.4 & 25 & \textbf{1.9} & 2.1 \\
\midrule
\multirow{3}{*}{\cai} & 6 & 0.2 & \textbf{0.1 }& 6 & 0.3 & \textbf{0.2} & 7 & 0.9 & \textbf{0.6} & 7 & 19.3 & \textbf{9.8} & 7 & 22.9 & \textbf{6.6} & 7 & - & \textbf{14131.6} \\
& 9 & 0.1 & \textbf{0.0 }& 9 & 0.3 & \textbf{0.1} & 9 & 0.6 &\textbf{ 0.3} & 9 & 12.9 & \textbf{5.4} & 9 & 3.6 &\textbf{ 2.0} & 9 & - & \textbf{375.8} \\
& 12 & \textbf{0.0} & \textbf{0.0} & 12 & 0.1 & \textbf{0.0 }& 12 & \textbf{0.1} & \textbf{0.1} & 12 & 4.3 & \textbf{1.1} & 12 & 0.7 & \textbf{0.4} & 12 & 193.0 & \textbf{90.7} \\
\midrule
\multirow{3}{*}{\epi} & 5 & 48.2 & \textbf{36.0} & 5 & \textbf{62.6} & 66.3 & 7 & 1450.8 & \textbf{588.9} & 7 & - & \textbf{14507.9} & 10 & - & \textbf{3747.7} & 25 & - & \textbf{93791.8} \\
& 10 & 320.7 & \textbf{25.8} & 10 & 1813.7 & \textbf{109.6} & 12 & - & \textbf{267.9} & 12 & - & \textbf{26011.0} & 15 & - & \textbf{1137.0} & 30 & - & \textbf{733.7} \\
& 15 & 462.2 & \textbf{12.9} & 15 & 7951.5 & \textbf{56.9} & 17 & 6013.3 & \textbf{92.6} & 17 & - & \textbf{12645.6} & 20 & 1046.6 & \textbf{240.7} & 35 & 1.0 & \textbf{0.9 }\\
\midrule
\multirow{3}{*}{\ema} & 5 & 3.4 & \textbf{1.9} & 5 & 4.0 & \textbf{3.7} & 7 & 16.8 & \textbf{14.8} & 7 & 386.5 & \textbf{281.4} & 10 & 225.7 & \textbf{48.1} & 10 & - & \textbf{14966.8} \\
& 10 & 1.9 &\textbf{ 0.7 }& 10 & 11.4 & \textbf{1.7 }& 9 & 36.7 &\textbf{ 4.4} & 12 & 508.0 & \textbf{113.6} & 15 & 244.7 & \textbf{6.6} & 15 & - & \textbf{3577.9} \\
& 15 & \textbf{0.2} & \textbf{0.2} & 15 & 0.8 & \textbf{0.3} & 12 & 6.2 & \textbf{3.5} & 17 & 27.1 & \textbf{7.7 }& 20 & 118.8 & \textbf{0.4 } & 20 & - & \textbf{77.5} \\
\midrule
\multirow{3}{*}{\ama} & 6 & 2.8 & \textbf{1.0} & 6 & 4.2 & \textbf{2.7} & 7 & 6.0 & \textbf{1.7} & 7 & 39.5 & \textbf{18.3} & 7 & 53.1 &\textbf{ 10.5 }& 7 & - & \textbf{23008.3} \\
& 9 & 0.7 & \textbf{0.3} & 9 & 0.9 & \textbf{0.5} & 9 & 1.5 & \textbf{0.5} & 9 & 7.1 & \textbf{2.5} & 9 & 5.3 &\textbf{ 1.1} & 9 & 254.7 & \textbf{106.7} \\
& 12 & 0.3 & \textbf{0.1} & 12 & \textbf{0.1} & \textbf{0.1} & 12 & 0.3 & \textbf{0.1} & 12 & \textbf{0.2} & \textbf{0.2} & 12 & 0.4 & \textbf{0.2} & 12 & 2.6 & \textbf{1.0} \\
\midrule
\multirow{3}{*}{\dbl} & 5 & 4.9 & \textbf{0.5} & 5 & 6.1 & \textbf{0.9} & 7 & 1100.2 & \textbf{0.7} & 7 & 1518.9 & \textbf{11.8} & 7 & 1439.1 & \textbf{5.0} & 10 & - & \textbf{930.6} \\
& 7 & 1005.5 & \textbf{0.2} & 7 & 1192.8 & \textbf{0.3} & 12 & - & \textbf{0.2 }& 12 & - & \textbf{2.8} & 9 & 182028.8 & \textbf{1.8} & 15 & - & \textbf{550.6} \\
& 9 & 162301.4 & \textbf{0.1} & 9 & 193757.0 & \textbf{0.2} & 17 & - & \textbf{0.2 }& 17 & - & \textbf{2.3} & 12 & - & \textbf{1.3} & 20 & - & \textbf{459.3} \\
\midrule
\multirow{3}{*}{\pok} & 5 & 279.6 & \textbf{139.7} & 5 & \textbf{493.8} & 538.6 & 7 & 797.2 & \textbf{493.1} & 7 & 66895.4 & \textbf{8683.0} & 10 & - & \textbf{936.9} & 15 & - & \textbf{54975.3} \\
& 10 & 125.3 & \textbf{22.4} & 10 & 468.2 & \textbf{57.1} & 12 & 1209.2 & \textbf{67.6} & 12 & 16931.4 & \textbf{1584.5} & 20 & 2603.2 & \textbf{14.5} & 20 & - & \textbf{4817.3} \\
& 15 & 163.8 & \textbf{10.0} & 15 & 673.6 & \textbf{13.7} & 17 & 1075.8 & \textbf{14.8} & 17 & 14930.8 & \textbf{238.3} & 30 & 30.5 & \textbf{4.8} & 25 & - & \textbf{119.8} \\
\midrule
\multirow{3}{*}{\ski} & 10 & - & \textbf{4209.7} & 30 & - & \textbf{54752.0} & 30 & - & \textbf{25884.6} & 55 & - & \textbf{49106.6} & 40 & - & \textbf{67878.3} & 60 & - & \textbf{423542.2} \\
& 30 & - & \textbf{1849.2} & 40 & - & \textbf{13224.9} & 40 & - & \textbf{7784.1} & 60 & - & \textbf{1069.3} & 50 & - & \textbf{7355.9} & 65 & - &\textbf{ 2177.8 }\\
& 50 & - & \textbf{906.7} & 50 & - & \textbf{728.8} & 50 & - & \textbf{1201.2 }& 65 & - &\textbf{ 96.3} & 60 & - & \textbf{668.9} & 70 & - & \textbf{136.2} \\

		\bottomrule
	\end{tabular}
	
	\label{tab:running_time}
\end{table*}

In this section, we conduct extensive experiments to evaluate the performance of the proposed solutions. In addition, we also evaluate the effectiveness of our algorithms by presenting case studies and demonstrating their applications in Section~\ref{ssec:a1} and \ref{ssec:a2}.

\subsection{Experimental setup}

%\stitle{Algorithms.} We test four algorithms, \dlist, \plist, \dpivot and \ppivot. \dlist and \plist is the listing-based baseline on counting $s$-dclique and $s$-plex.  \dpivot and \ppivot is the pivot-based algorithm on counting $s$-dclique and $s$-plex. All of the four algorithms are equipped with the pruning techniques developed in Section~\ref{sec:baseline} in default.

%\stitle{Algorithms.} We evaluate 4 algorithms, namely \dlist, \plist, \dpivot, and \ppivot. \dlist and \plist are baseline algorithms that rely on listing-based approaches to count $s$-dclique and $s$-plex, respectively. On the other hand, \dpivot and \ppivot are pivot-based algorithms used to count $s$-dclique and $s$-plex, respectively. All algorithms are implemented in C++ and all of them are integrated with the pruning techniques developed in Section~\ref{sec:baseline} as their default setting. Since this is the first work to study the problem of \hcs counting, we do not consider other algorithms.

\stitle{Algorithms.} We evaluate 4 algorithms, namely \dlist, \plist, \dpivot, and \ppivot. \dlist and \plist are listing-based approaches to count $s$-dclique and $s$-plex respectively. On the other hand, \dpivot and \ppivot are pivot-based algorithms used to count $s$-dclique and $s$-plex respectively. All algorithms are implemented in C++ and all of them are integrated with the pruning techniques developed in Section~\ref{sec:baseline} as their default setting. Since this is the first work to study the problem of \hcs counting, we use the list-based algorithms \dlist and \plist as the baselines.

%\stitle{Datasets.} We choose $8$ real-world networks to test the performance. The details of the datasets are in Table~\ref{tab:datasets}. The columns of Table~\ref{tab:datasets} are the network name, the number of vertices, the number of edges, the degeneracy and the network type. All the datasets are from the SNAP project \cite{snapnets}.

\stitle{Datasets.} We selected 8 real-world networks to evaluate the performance of different algorithms. The details of the datasets can be found in Table~\ref{tab:datasets}. All the datasets used in our experiments are obtained from the SNAP project \cite{snapnets}.

All the experiments are conducted on a server with an AMD 3990X CPU, 256GB memory, and Linux CentOS 7 operating system. Similar to previous studies on enumerating maximal $s$-dcliques \cite{23sigmoddclique} and $s$-plexes \cite{plex_AAAI21, DAi_kplex,22wwwkplex}, we mainly evaluate different \hcs counting algorithms with $s\le 3$. This is because real-world applications often require that the subgraph pattern is cohesive and also very similar to clique \cite{plex_AAAI21, DAi_kplex,22wwwkplex}. When $s>3$, the \hcs may become sparse, making it more reasonable to consider small values of $s$, as previous studies have commonly done \cite{23sigmoddclique, plex_AAAI21, DAi_kplex,22wwwkplex}.

\comment{
\stitle{The choice of $s$.} Similar to previous studies on enumerating maximal $s$-dcliques \cite{23sigmoddclique} and $s$-plexes \cite{plex_AAAI21, DAi_kplex,22wwwkplex}, we mainly evaluate different \hcs counting algorithms with $s\le 3$. The reasons are as follows. First, real-world applications often require that the subgraph pattern is cohesive and also very similar to clique \cite{plex_AAAI21, DAi_kplex,22wwwkplex}. When $s>3$, the \hcs might be sparse, thus it is more reasonable to consider a small $s$ as done in previous work \cite{plex_AAAI21, DAi_kplex,22wwwkplex}. Second, for a large $s$, the problem becomes more challenging due to the exponential growth in the search space and the count as $s$ increases. In fact, even for maximal $s$-plex enumeration problem (often much easier than $s$-plex counting), existing solutions are confined to instances with $s\le3$ \cite{plex_AAAI21, DAi_kplex,22wwwkplex}. %These are the reasons why we only consider $s\le3$.
}

% pd-sdcc
% pl-pc
% sdcc-pc/pd-pl
% large q
%

\subsection{Performance studies}
%Table~\ref{tab:running_time} provides an overview of the performance of the algorithms. It shows the running time of \dlist, \dpivot, \plist, and \ppivot on all datasets with different values of $s$ and $q$.

\stitle{Comparing the listing and pivot-based algorithms.}  Comparing the performance of \dlist and \dpivot in Table~\ref{tab:running_time}, we observe that they exhibit similar performance when $q$ is small. For example, on \wik, \dlist takes $10.2$ seconds to count $(5,1)$-dclique, while \dpivot needs $9.62$ seconds. However, as $q$ becomes larger, \dpivot can be several orders of magnitude faster than  \dlist on many networks. For example, on \dbl, when $s=1$ and $q=9$, \dlist consumes 162301.4 seconds, while \dpivot takes only 0.1 seconds. We also find that on some networks, such as \cai and \ama, both \dlist and \dpivot perform similarly even for a large $q$. This is because the count of \hcs with size $q$ in each network differs. If the count is large, \dlist is often much slower than \dpivot, while if the count is not very large, \dlist performs comparably with \dpivot. For instance, \epi has $1.7\times 10^{9}$ $(15,1)$-dcliques, whereas \cai only has $3.2\times 10^4$ $(12,1)$-dcliques. Similarly, \ppivot is more efficient than \plist on complex networks when $q$ is large. These results demonstrate the high efficiency of the proposed pivot-based counting algorithms.

\stitle{Difference on counting $s$-dclique and $s$-plex.} From Table~\ref{tab:running_time}, we observe that \plist is slower than \dlist, and \ppivot is slower than \dpivot for a given $s$ and $q$. For instance, on \ski when $s=1$ and $q=30$, \dpivot is an order of magnitude faster than \ppivot. This result suggests that $s$-plex is a more complex structure than the $s$-dclique, which is often more difficult to count. %Specifically, $s$-dcliques allow at most $s$ missing edges in total, while $(s+1)$-plexes allow at most $s$ missing edges for each vertex.

\stitle{\dpivot and \ppivot with various $q$ and $s$.} In Table~\ref{tab:running_time}, we observe that the running time of \dpivot and \ppivot decreases as $q$ increases. However, the counter-intuitive thing is that the count of \hcs may not decrease as $q$ becomes larger. For instance, \ski has $9.3\times 10^{13}$ $(10,1)$-dcliques and $1.1\times 10^{21}$ $(30,1)$-dcliques, but \dpivot runs faster on counting $(30,1)$-dcliques than $(10,1)$-dcliques. This phenomenon occurs because larger values of $q$ can prune more candidate sets, thanks to the core-based and upper-bound based pruning techniques developed in Section~\ref{sec:baseline}. Note that the size of the search tree in Algorithm~\ref{alg:pivot} is almost not affected by the parameter $q$. Thus, no matter what the value of $q$ is, Algorithm~\ref{alg:pivot} will enumerate all the large \hcss. Therefore, the larger the value of $q$, the smaller the size of the candidate set, and the faster \dpivot and \ppivot are. %These results further confirm the high efficiency of our pivot-based counting algorithms.

In Table~\ref{tab:running_time}, we can also observe that the running time of \dpivot and \ppivot increases as the value of $s$ increases.  For instance, on the \wik network, \dpivot takes $9.62$ seconds for $(5,1)$-dclique and $134.9$ seconds for $(7,2)$-dclique, which is an increase of $14\times$. On the other hand, \ppivot takes $16.7$ seconds for $(5,1)$-plex and $3139.3$ seconds for $(7,2)$-plex, which is an increase of $188\times$. This is because the number of \hcss in a network increases significantly when the parameter $s$ increases.

%\stitle{The performance of \dpivot and \ppivot on various $s$.} In  Table~\ref{tab:running_time}, the running time of \dpivot and \ppivot increase by orders of magnitude with $s$, and the \ppivot increases even more.  On the \wik network, \dpivot spends $9.62$ seconds on $(1,5)$-dclique and $134.9$ seconds on $(2,7)$-dclique, where the increase is $14\times$. \ppivot spends $20.07$ seconds on $(1,5)$-plex and $4387.03$ seconds on $(2,7)$-plex, where the increase is $219\times$. This result means that $s$-plex is more sensitive to $s$ than $s$-dclique. It is intuitive and straightforward because $s$-plex will miss more edges than $s$-dclique as $s$ increases.

%\stitle{The performance of \dpivot and \ppivot on various $s$.} In Table~\ref{tab:running_time}, we can observe that the running time of \dpivot and \ppivot increases significantly as the value of $s$ increases. For instance, on the \wik network, \dpivot takes $9.62$ seconds for $(1,5)$-dclique and $134.9$ seconds for $(2,7)$-dclique, which is an increase of $14\times$. On the other hand, \ppivot takes $16.7$ seconds for $(2,5)$-plex and $3139.3$ seconds for $(3,7)$-plex, which is an increase of $188\times$. This result implies that $s$-plex is more sensitive to $s$ compared to $s$-dclique. It is intuitive as $s$-plex can miss more edges than $s$-dclique with the increase in $s$.

\begin{table}[t]  \vspace{-0.3cm}
%	\small
	\scriptsize
	\centering
	\caption{The reduction rate of candidate pruning technique.} \vspace*{-0.3cm}
	\begin{tabular}{c| c| c|c|  c|c|  c|c}
		\toprule
			\multirow{3}{*}{$\bm{q}$} &
		\multirow{3}{*}{\textbf{Networks}} & \multicolumn{6}{c}{\textbf{The reduced ratio of the candidate set}}  \\
		\cline{3-8}
	&	&   \multicolumn{2}{c|}{$\bm{s=1}$} &
		\multicolumn{2}{c|}{$\bm{s=2}$} &
		\multicolumn{2}{c}{$\bm{s=3}$} \\
		\cline{3-8}
%	&	& $1$-dclique &  $2$-plex
%		& $2$-dclique & $3$-plex
%		& $3$-dclique & $4$-plex \\
	&	& $s$-dclique &  $s$-plex
		& $s$-dclique & $s$-plex
		& $s$-dclique & $s$-plex \\
		\midrule

%\multirow{4}{*}{$8$}
%&\wik & 93.95\% & 94.02\% & 92.94\% & 91.54\% & 91.51\% & 85.18\% \\
%&\epi & 94.64\% & 94.92\% & 93.72\% & 92.89\% & 92.35\% & 83.48\% \\
%&\ama & 93.28\% & 93.36\% & 92.54\% & 92.11\% & 91.46\% & 85.43\% \\
%&\pok & 97.01\% & 97.29\% & 96.54\% & 95.90\% & 95.41\% & 88.67\% \\
%\cmidrule{1-8}
\multirow{4}{*}{$10$}
&\wik & 95.33\% & 95.39\% & 94.72\% & 94.02\% & 93.95\% & 91.52\% \\
&\epi & 95.81\% & 95.96\% & 95.30\% & 94.92\% & 94.64\% & 92.81\% \\
&\ama & 94.83\% & 93.94\% & 94.07\% & 93.41\% & 93.28\% & 92.06\% \\
&\pok & 97.66\% & 97.72\% & 97.36\% & 97.28\% & 97.01\% & 95.89\% \\
\cmidrule{1-8}
\multirow{4}{*}{$20$}
&\wik & 97.59\% & 97.67\% & 97.52\% & 97.44\% & 97.39\% & 97.19\% \\
&\epi & 97.62\% & 97.66\% & 97.55\% & 97.51\% & 97.45\% & 97.28\% \\
&\ama & 100\% & 100\% & 100\% & 100\% & 100\% & 100\% \\
&\pok & 99.80\% & 99.72\% & 99.71\% & 99.47\% & 99.60\% & 99.06\% \\

		\bottomrule
	\end{tabular}
	\vspace{-0.4cm}
	\label{tab:prune_technique_candidate}
\end{table}

\stitle{Evaluating the pruning technique: candidate reduction.} Table~\ref{tab:prune_technique_candidate} reports the reduction rate of the candidate pruning technique. $q$ is fixed to $10$ and $20$. The reduction rate is computed as $\frac{c_{pre}-c_{now}}{c_{pre}}$, where $c_{pre}$ is the total size of all the candidate set before pruning, i.e. $\sum_{v_i}{|\vec N(v_i) \cup \vec{N_2}(v_i)|}$ in line~3 of Algorithm~\ref{alg:hcslist} and $c_{now}$ is the total size after pruning. The table shows that the candidate reduction technique can remove over $90\%$ of unnecessary vertices, which is a significant speed-up. When $q$ goes larger, more vertices can be removed from the candidate set. For example, when $s=1$ and $q=10$, the total candidate set size of \wik after pruning on $s$-dclique counting is $3.5\times 10^5$. When $s=1$ and $q=20$, the size becomes $2.0\times 10^4$. Therefore, the candidate reduction based pruning technique is effective.%an effective way to improve the efficiency of \dpivot and \ppivot.

%\begin{table}[h]
%	\small
%	%\scriptsize
%	\centering
%	\caption{The running time with and without the upper bound.}
%	\begin{tabular}{c| c|c|  c|c}
%		\toprule
%		\multirow{3}{*}{\textbf{Networks}} & \multicolumn{4}{c}{\textbf{Running time (sec)}}  \\
%		\cline{2-5}
%		&   \multicolumn{2}{c|}{\dpivot} &
%		\multicolumn{2}{c|}{\ppivot} \\
%		\cline{2-5}
%		&  with &  without
%		& with & without\\
%		\midrule
%		
%		\wik &  &  &   &  \\
%		\epi &  &  &  &  \\
%		\ama &   &  &  &  \\
%		\pok & & & &  \\
%		\bottomrule
%	\end{tabular}
%	
%	\label{tab:upper}
%\end{table}

\begin{figure}[t] %\vspace*{-0.3cm}
	\subfigure[$s=1$ \label{sfig:upper_a}]{\includegraphics[width=0.48\linewidth]{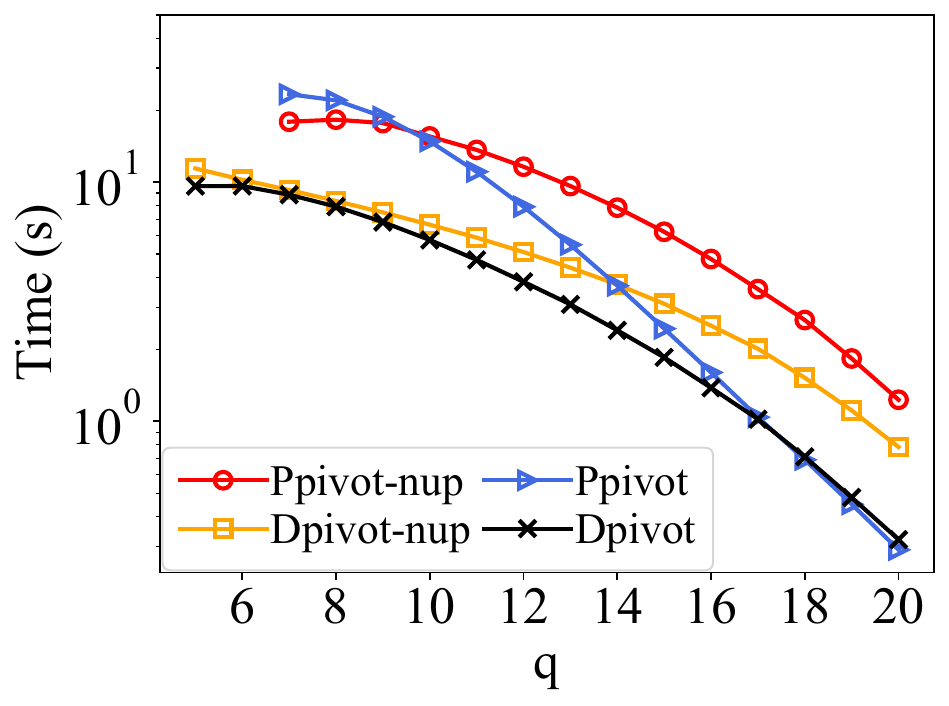}}
	\subfigure[$s=2$ \label{sfig:upper_b}]{\includegraphics[width=0.48\linewidth]{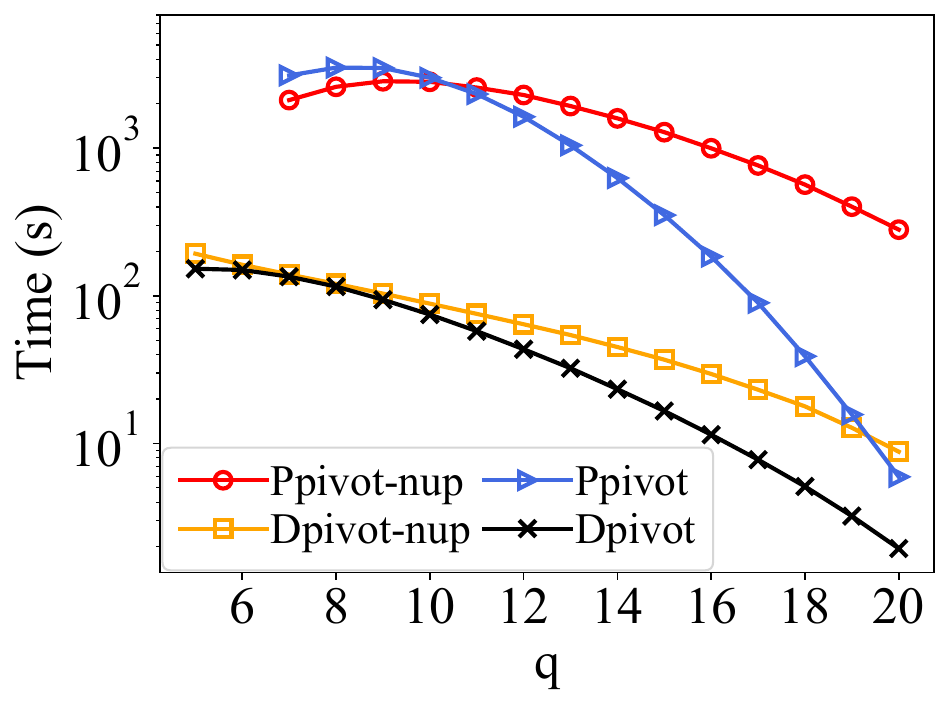} }
	\vspace{-0.4cm} \caption{Effectiveness of the upper bounds (\wik).}\vspace{-0.3cm}
	\label{fig:upper}
\end{figure}

\stitle{Evaluating the prune technique: upper bound.}  In Fig.~\ref{fig:upper}, we compare \ppivot against \nppivot, and compare \dpivot against \ndpivot on \wik, where \nppivot and \ndpivot are Algorithm~\ref{alg:pivot} without the prune techniques based on the upper bounds. Similar results can also be observed on the other datasets. As shown in Fig.~\ref{fig:upper}, the effectiveness of the upper bounds increases when $q$ becomes larger. The effect of $s$ on the effectiveness of the upper bounds also increases with increasing $s$. For instance, by fixing $q=20$, \nppivot is $4\times$ slower than \ppivot when $s=1$  while $47\times$ when $s=2$.  %Another case is that with $q=7$, \ppivot is slower than \nppivot when $s=1$ while faster when $s=2$.

\begin{table}[t!]
	%\small
	\scriptsize
	\centering
	\caption{The percentage of \hcss counted in combination} \vspace*{-0.3cm}
	\begin{tabular}{c|c| cc| cc}
		\toprule
		
	\multirow{2}{*}{$\mathbf{q}$}	&\multirow{2}{*}{\textbf{Networks}}
		&   \multicolumn{2}{c|}{$s$-dclique} &
		\multicolumn{2}{c}{$s$-plex} \\
		\cline{3-6}
	&	&  $s=1$  &$s=2$
		&$s=1$ & $s=2$\\
		
		\midrule

%\multirow{4}{*}{$6$}  & \wik & 83.0\%  & 75.4\%  & 65.7\%  & 39.7\%  \\
%& \epi & 78.7\%  & 69.8\%  & 62.4\%  & 35.0\%  \\
%& \ama & 100.0\%  & 99.8\%  & 99.6\%  & 93.7\%  \\
%& \pok & 93.2\%  & 88.4\%  & 81.9\%  & -  \\
%\midrule
%\multirow{4}{*}{$12$}  & \wik & 100.0\%  & 100.0\%  & 100.0\%  & 98.7\%  \\
%& \epi & 100.0\%  & 100.0\%  & 100.0\%  & 97.8\%  \\
%& \ama & 100.0\%  & 100.0\%  & 100.0\%  & 100.0\%  \\
%& \pok & 100.0\%  & 100.0\%  & 100.0\%  & 99.6\%  \\
		
\multirow{4}{*}{$7$}  & \wik & 95.7\%  & 91.8\%  & 87.6\%  & 57.4\%  \\
& \epi & 92.8\%  & 87.3\%  & 84.3\%  & 51.8\%  \\
& \ama & 100.0\%  & 100.0\%  & 100.0\%  & 98.4\%  \\
& \pok & 98.5\%  & 96.5\%  & 94.8\%  & 72.1\%  \\
\midrule
\multirow{4}{*}{$14$}  & \wik & 100.0\%  & 100.0\%  & 100.0\%  & 99.8\%  \\
& \epi & 100.0\%  & 100.0\%  & 100.0\%  & 99.6\%  \\
& \ama & 100.0\%  & 100.0\%  & 100.0\%  & 100.0\%  \\
& \pok & 100.0\%  & 100.0\%  & 100.0\%  & 100.0\%  \\
		\bottomrule
	\end{tabular}
	\vspace{-0.4cm}
	\label{tab:combi}
\end{table}

\begin{figure}[t] %\vspace*{-0.3cm}
	\subfigure[\ama  \label{sfig:mem_ama}]{\includegraphics[width=0.45\linewidth]{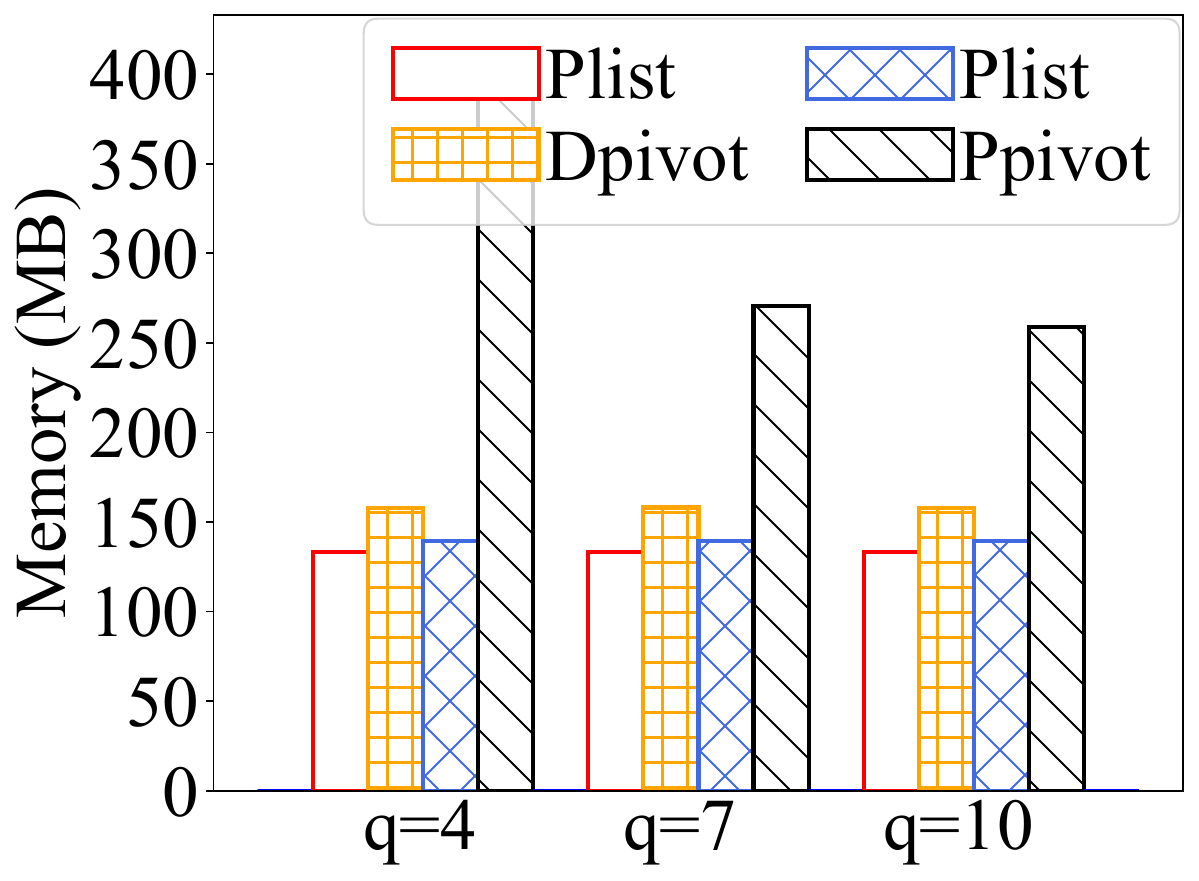}}
	\subfigure[\pok \label{sfig:mem_pok}]{\includegraphics[width=0.45\linewidth]{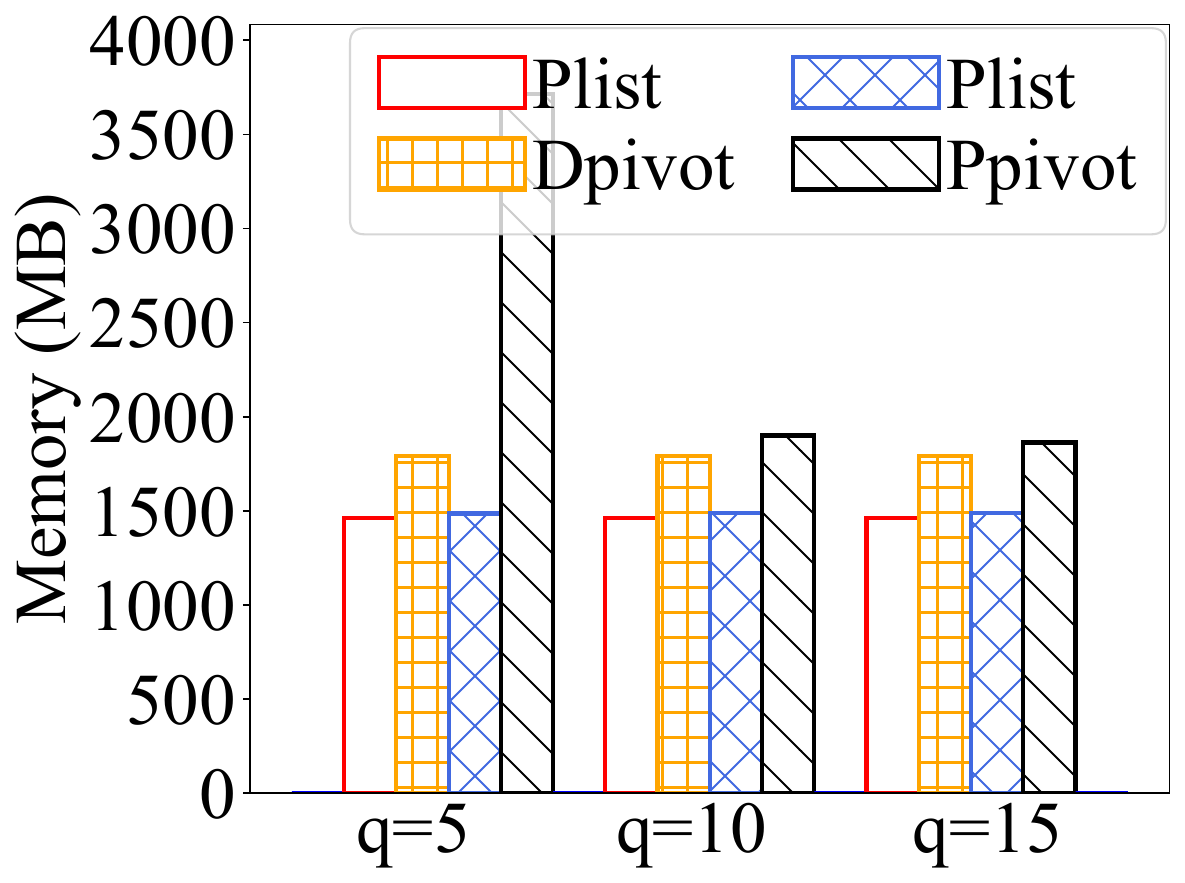} }
	\vspace{-0.4cm} \caption{Memory overheads of different algorithms}
	\label{fig:memory}
	\vspace{-0.4cm}
\end{figure}

\stitle{The effectiveness of the pivoting technique.} In Algorithm~\ref{alg:pivot}, the \hcss are counted by listing (line~3) or counted in a combinatoric manner (line~6). Clearly, the more \hcss are counted in a combinatoric manner, the better the acceleration of the pivot-vertex technique. Table~\ref{tab:combi} shows the percentage of \hcss counted in a combinatoric manner. In Table~\ref{tab:combi}, \dpivot has larger percentages than \ppivot. Moreover, \ppivot is more sensitive to the parameter $s$. This is because \dpivot always has a pivot vertex at each node of the recursion tree, while \ppivot depends on $R$, $C$ and $s$ as stated by Corollary~\ref{cor:plex_choose_pivot}. Table~\ref{tab:combi} also shows that the \hcss with larger size are more easy to count in a combinatoric way. With larger value of $q$, the condition $|R|=q-1$ (line~2 in Algorithm~\ref{alg:pivot}) is harder to meet. In general, our pivot-based algorithms enable a large number of \hcss to be counted in a combinatoric manner. These results further confirm the high efficiency of the proposed pivot-based solutions.

\stitle{Memory overheads.} Fig.~\ref{fig:memory} compares the memory costs of the proposed algorithms on \ama and \pok given that $s=1$. For other parameter settings and other datasets, the results are consistent. The pivot-based approaches require more memory compared to the listing-based solutions. This is because the recursive depth of the pivot-based algorithms is often deeper than the listing-based solutions, thus resulting in more space usages. However, we can also note that the memory overheads of the pivot-based solutions are still within the same order of magnitude as listing-based algorithms, due to the fact that both the listing-based and pivot-based methods are depth-first search algorithms which are typically space-efficient. For example, when $q=4$ , \plist consumes $140$ MB on \ama and \ppivot uses $394$ MB on the same datasets.

 %This is because the recursive depth of the pivot-based algorithms is often deeper
 %
% and also the recursive depth is
%
% although the pivot-based algorithms need more memory, their memory consumption is still within the same order of magnitude as listing-based algorithms. For example, when $q=4$ , \plist costs $140$ MB on \ama and \ppivot costs $394$ MB.

\stitle{Counting \hcs with size in $\mathbf{[q_l,q_r]}$ simultaneously.} As described in Section~\ref{ssec:discussion}, \dpivot and \ppivot can count \hcs with size in a range simultaneously. Table~\ref{tab:range} shows the running time, where the column $q_l$ means the running time of counting only \hcs with size $q_l$. The parameters are set as $s=1$, $q_l=5$ and $q_r=20$. As shown in Table~\ref{tab:range}, The running time of counting $[q_l,q_r]$ and counting only $q_l$ are within the same order of magnitude. The results with other parameters are similar. This is because counting $[q_l,q_r]$ and counting only $q_l$ has the same size of the search tree and only has difference in the leaves nodes (as described in Algorithm~\ref{alg:pivot} and Section~\ref{ssec:discussion}). These results further demonstrate the advantages of the pivot-based algorithms, compared to the list-based algorithms.

\begin{table}[t!]
%	\small
	\scriptsize
	\centering
	\vspace*{-0.1cm}
	\caption{The running time of counting \hcs with size in $\mathbf{[q_l,q_r]}$ simultaneously ($q_l=5, q_r=20$).} \vspace*{-0.3cm}
	\begin{tabular}{c| cc|  cc}
		\toprule
		\multirow{3}{*}{\textbf{Networks}} & \multicolumn{4}{c}{\textbf{Running time (sec)}}  \\
		\cline{2-5}
		&   \multicolumn{2}{c|}{\dpivot} &
		\multicolumn{2}{c}{\ppivot} \\
		\cline{2-5}
		&  ${q_l}$ &  ${[q_l,q_r]}$
		& ${q_l}$ & ${[q_l,q_r]}$\\
		\midrule
		
		\wik & 9.62 & 11.11 & 16.68  & 32.27 \\
		\epi & 35.98  & 47.595 & 66.31 &184.17  \\
		\ama & 1.02  & 1.86 & 4.72 & 4.74 \\
		\pok &139.73 &146.88 & 538.55 & 570.89 \\
		\bottomrule
	\end{tabular}
	\vspace{-0.25cm}
	\label{tab:range}
\end{table}

%\stitle{Local counting.} We compare the running time of global and local counting in Table~\ref{tab:local}. The parameters are set as $s=1, q=6$. Local counting includes counting \hcs in each vertex and edge. In Table~\ref{tab:local}, the local-vertex counting is slightly slower than the global counting. This is because the computation of local-vertex counting only need to scan the vertices as described in subsection~\ref{ssec:discussion}. Since the local-edge counting needs to scan the edges and the count of edges is larger than vertices, the running time of the local-edge is quite slower but still within the same order of magnitude.  Thus in Table~\ref{tab:local}, local-edge counting is at most $2.57 \times$ slower than local-vertex counting. These results implies that \dpivot and \ppivot are efficient on counting local \hcs.
\stitle{Local counting.} We compared the running time of global and local counting in Table~\ref{tab:local}, with parameters $s=1$. Local counting includes counting \hcss in each vertex and edge. In Table~\ref{tab:local}, we find that the local-vertex counting is slightly slower than the globally counting. This is because the computation of local-vertex counting only requires scanning the vertices, as described in Section~\ref{ssec:discussion}. However, the local-edge counting needs to scan the edges, and since the number of edges is larger than the number of vertices, the running time of local-edge counting is slower but still within the same order of magnitude. For example, in Table~\ref{tab:local} with $q=7$, local-edge counting is at most $2.42\times$ slower than global counting. These results indicate that \dpivot and \ppivot are very efficient in locally counting \hcss for each vertex or edge.

%\stitle{The count of \hcs.}

\begin{table}[t!]
	%	\small
	\scriptsize
	\centering
	\caption{The running time of locally counting} \vspace*{-0.3cm}
	\begin{tabular}{c|c| ccc|  ccc }
		\toprule
	\multirow{3}{*}{\textbf{q}} &\multirow{3}{*}{\textbf{Networks}} & \multicolumn{6}{c}{\textbf{Running time (sec)}}  \\
		\cline{3-8}
	&	&   \multicolumn{3}{c|}{\dpivot} &
		\multicolumn{3}{c}{\ppivot} \\
		\cline{3-8}
	& & global	&  vertex &  edge & global & vertex & edge\\
		\midrule
%\multirow{4}{*}{6}   & \wik   & 9.6  & 10.4  & 18.0  & 21.5  & 21.9  & 45.4  \\
%& \epi   & 35.5  & 38.1  & 67.1  & 78.3  & 79.3  & 187.1  \\
%& \ama   & 1.0  & 1.1  & 1.8  & 2.7  & 2.7  & 3.3  \\
%& \pok   & 80.1  & 84.3  & 130.6  & 291.5  & 293.9  & 375.8  \\
%\midrule
%\multirow{4}{*}{12}   & \wik   & 4.0  & 5.4  & 9.0  & 7.9  & 8.0  & 13.2  \\
%& \epi   & 20.8  & 26.8  & 68.8  & 92.6  & 93.7  & 210.6  \\
%& \ama   & 0.1  & 0.1  & 0.1  & 0.1  & 0.1  & 0.1  \\
%& \pok   & 15.4  & 18.0  & 29.4  & 30.2  & 30.3  & 44.5  \\

\multirow{4}{*}{7}   & \wik   & 11.3  & 11.9  & 21.7  & 29.6  & 30.1  & 58.4  \\
& \epi   & 43.3  & 45.5  & 94.5  & 119.6  & 121.8  & 288.9  \\
& \ama   & 0.8  & 0.8  & 1.2  & 1.8  & 1.8  & 2.2  \\
& \pok   & 60.6  & 65.2  & 109.1  & 188.5  & 190.3  & 268.4  \\
\midrule
\multirow{4}{*}{14}   & \wik   & 3.0  & 4.7  & 7.0  & 4.7  & 4.6  & 6.8  \\
& \epi   & 19.2  & 28.1  & 72.5  & 86.4  & 86.5  & 194.5  \\
& \ama   & 0.1  & 0.1  & 0.1  & 0.1  & 0.1  & 0.1  \\
& \pok   & 11.6  & 14.1  & 22.5  & 18.6  & 18.6  & 27.2  \\

		\bottomrule
	\end{tabular}
	\vspace{-0.3cm}
	\label{tab:local}
\end{table}

\subsection{Application 1 : \hcs-based graph clustering}\label{ssec:a1}

%We apply the \hcs counting algorithms to motif-based spectral clustering \cite{motifClustering, benson2016higher} and  motif-based network profiles computation \cite{motifProfile}.

%Motif-based spectral clustering is a classic spectral clustering algorithm that use the count of motif as the weight of edge. The experiment results show that \hcs-based spectral clustering has a great performance.

%Motif-based network profile can uniquely  characterize the networks. \cite{motifProfile} shows that the networks in the same domain have similar profiles.  One advantage of motif-based profile is that it is stable to the noise in the networks \cite{motifProfile}. We show that the profile computed by \hcs is more stable than clique.

%\stitle{\hcs-based graph clustering.} %The communities in networks are a set of vertices that the inter-relation is better than intra-relation.
In this section, we show the application of the \hcs counts for motif-based graph clustering. One of the most popular metrics in community detection is \textit{conductance} \cite{yang2012defining, schaeffer2007graph, leskovec2008statistical, fortunato2010community, motifClustering, benson2016higher}. The conductance is the ratio of the count of edges leaving the community and the count of edges in the community.  Similarly, the motif-based conductance \cite{motifClustering, benson2016higher} is the ratio of the count of motifs leaving the community and the count of motifs in the community, which is formulated as $
	\Phi(S) = \frac{C_{M}(S:\overline{S})}{min(C_{M}(S), C_{M}(\overline{S}))},
$
%\begin{equation}
%	\Phi(S) = \frac{C_{M}(S:\overline{S})}{min(C_{M}(S), C_{M}(\overline{S}))},
%\end{equation}
where $\overline{S}$ is the remainder vertices of $S$, $C_{M}(S)$ is the count of the motif $M$ in $G(S)$ and $C_{M}(S:\overline{S})$ is the count of the motif $M$ that contains both vertices in $S$ and $\overline{S}$.  $\Phi(S)$ measures how well a community $S$ preserves the occurrences of $M$ compared to its complement $\overline{S}$. A community with low motif-based conductance means that it preserves the occurrences of the motif $M$ well within the community.

%Since $C_{M}(S:\overline{S})$ is the count of $M$ in the graph cut $(S:\overline{S})$, $\Phi(S)$ can be seen as a normalized cut. Thus we can use the classic spectral clustering algorithm to minimize $\Phi(S)$ \cite{motifClustering, benson2016higher}. Specifically, define the weight matrix $W_M\in N^{|V|\times |V|}$ that $(W_M)_{ij}$ is the count of $M$ containing $i$ and $j$. Then the following steps is the same as the classic spectral clustering and we will get some clusters that the $\Phi(S)$ between the clusters are minimized.
To minimize the motif-based conductance $\Phi(S)$, we can use the classic spectral clustering algorithm, as suggested in previous studies \cite{motifClustering, benson2016higher}. Essentially, $\Phi(S)$ can be viewed as a normalized cut, where $C_{M}(S:\overline{S})$ represents the count of motifs in the graph cut $(S:\overline{S})$. To apply spectral clustering, we define a weight matrix $W_M\in N^{|V|\times |V|}$, where $(W_M)_{ij}$ is the count of motifs containing edge $(i, j)$. Using this weight matrix, we can follow the same steps as the classic spectral clustering algorithm to obtain clusters that minimize the motif-based conductance $\Phi(S)$ between them. This motif-based spectral clustering approach has been shown to be very effective in previous studies \cite{motifClustering, benson2016higher, lu2018community}, in which the clique was used as a motif. Similarly, we can use \hcs as a motif to devise an \hcs-based spectral clustering algorithm by the local counting property of \hcspivot as discussed in Section~\ref{ssec:discussion}. With our \hcspivot, we can get the weight matrix $W_M$ efficiently.

%Table~\ref{tab:cluster} compares the clustering performance of our \hcs-based spectral clustering algorithm against several famous algorithms. The tested network \kw{email}-\kw{Eu}-\kw{core} is downloaded from the SNAP project \cite{snapnets}, which has $42$ ground truth communities. The algorithms based on spectral clustering are all implemented by the famous scikit-learn python package \cite{scikit-learn}. Other famous algorithms are all the well-tested open source implementation \cite{python_louvain, scikit-learn, infomap,pSCAN}. Both in \cite{motifClustering, benson2016higher}, the authors find that the cliques have the best performance for clustering among the motifs. However, as shown in Table~\ref{tab:cluster}, we find that \hcs is better than cliques.
Table~\ref{tab:cluster} presents a comparison of the clustering performance of our proposed \hcs-based spectral clustering algorithm against several state-of-the-art graph clustering algorithms, including clique-based (3-clique and 4-clique) spectral clustering \cite{motifClustering, benson2016higher}, Louvain \cite{python_louvain}, pSCAN \cite{pSCAN}, label propagation algorithm (LPA) \cite{lpa}, and Infomap \cite{infomap,mapequation2023software}. We implement our spectral clustering algorithms using the commonly-used scikit-learn Python package \cite{scikit-learn}. For other popular algorithms, we use their open-source implementations that have been thoroughly tested \cite{python_louvain, scikit-learn, infomap, mapequation2023software,pSCAN}. We evaluate the algorithms on the \kw{email}-\kw{Eu}-\kw{core} network downloaded from the SNAP project \cite{snapnets}, which contains $42$ ground truth communities. Based on the ground truth communities, we can apply 4 widely-used metrics, including ARI \cite{vinh2009information}, Purity \cite{schutze2008introduction}, NMI \cite{witten2005practical}, and $F_1$ score, to evaluate the clustering performance of different algorithms. Due to the space limits, the detailed description of these metrics can be found in our full version \cite{fullversion}. As shown in Table~\ref{tab:cluster}, our $(3,1)$-dclique or $(3,1)$-plex model (when $q=3$ and $s=1$, $(3,1)$-dclique and $(3,1)$-plex are identical) achieves the best clustering performance with all 4 metrics. In general, both \hcs and clique-based spectral clustering algorithms (the first 5 rows) significantly outperform the other baselines to identify real-world communities, and our \hcs-based solutions can further outperform clique-based spectral clustering algorithms \cite{motifClustering, benson2016higher}. These results demonstrate the high effectiveness of the proposed solutions in community detection applications.

%Previous studies \cite{motifClustering, benson2016higher} have reported that cliques have the best performance for clustering among the motifs. However, our results, as shown in Table~\ref{tab:cluster}, demonstrate that \hcs outperforms cliques in this regard.

\begin{table}[t!]
	%	\small
	\scriptsize
	\centering
	\caption{The results of \hcs-based graph clustering} \vspace*{-0.3cm}
	\begin{tabular}{c| c|c|  c|c}
		\toprule
		\multirow{2}{*}{\textbf{Methods}} & \multicolumn{4}{c}{\textbf{Metrics}}  \\
		\cline{2-5}
		
		&  $\mathbf{ARI}$ &  $\mathbf{Purity}$ & $\mathbf{NMI}$ & $\mathbf{F_1}$ \\
		\midrule
		
		$(3,1)$-dclique/plex   & \textbf{ 0.47} & \textbf{0.67} &\textbf{ 0.68} & \textbf{0.49}\\
		(4,1)-dclique   & 0.34 & 0.62 & 0.62 & 0.38\\
		(4,1)-plex  & 0.33 & 0.63 & 0.63 & 0.36\\
		
		$3$-clique   & 0.30 & 0.64 & 0.64 & 0.33\\
		$4$-clique   & 0.23 & 0.59 & 0.59 & 0.27\\
		Louvain   & 0.33 & 0.45 & 0.58 & 0.38\\
	%	$2$-clique   & 0.27 & 0.64 & 0.63 & 0.30\\
		pSCAN  & 0.02 & 0.28 & 0.28 & 0.10\\
		LPA  & 0.10 & 0.52 & 0.49 & 0.13\\
		Infomap & 0.28 & 0.52 & 0.62 & 0.33\\
		\bottomrule
	\end{tabular}
	\vspace*{-0.3cm}
	\label{tab:cluster}
\end{table}

\comment{
\stitle{Case study on a word network.} We also conduct a case study that applies the \hcs-based spectral clustering algorithm on a word association network \cite{wordNetwork}. In this network, each vertex is a word and the words with strong associations are linked to each other. The network contains 5040 vertices and 55258 edges. We compare the results obtained by our \hcs-based spectral clustering algorithm and the state-of-the-art clique-based spectral clustering algorithm. %Fig.~\ref{sfig:a} and  Fig.~\ref{sfig:b} shows the cluster containing the word \kw{``HITCHHIKE"} founded by the 3-clique and $(3,1)$-dclique based methods respectively.
As shown in Fig.~\ref{sfig:a} and Fig.~\ref{sfig:b}, \kw{``HITCHHIKE"} connects two neighbors \kw{``STRANGER"} and \kw{``DANGER"} in the cluster identified by the 3-clique based solution. However, in the cluster identified by the (3,1)-dclique based method, \kw{``HITCHHIKE"} links \kw{``RIDE"} and \kw{``CAR"}. Intuitively, the word \kw{``HITCHHIKE"} is more relevant to \kw{``RIDE"} and \kw{``CAR"} than \kw{``STRANGER"} and \kw{``DANGER"}. Moreover, we can see that the cluster found by the 3-clique based approach only contains triangle structures, and the cluster identified by our solution includes richer and more relevant information than this approach. The reason could be that the clique based method is often too restrictive and our \hcs-based solution is more flexible and reasonable for graph clustering applications. These results further demonstrate the high effectiveness of our solutions.
%
%
%In Fig.~\ref{sfig:a}, \kw{``HITCHHIKE"} only has two neighbors, \kw{``STRANGER"} and \kw{``DANGER"}. In Fig.~\ref{sfig:b}, \kw{"HITCHHIKE"} also only has two neighbors, \kw{``RIDE"} and \kw{``CAR"}. From the perspective of clique, the word \kw{``HITCHHIKE"} in both clusters is only contained in a $3$-clique. But from the perspective of \hcs, \kw{``HITCHHIKE"} is not included in any \hcs in Fig.~\ref{sfig:a}, but in many \hcs in Fig.~\ref{sfig:b}. Therefore, the clustering results based on \hcs are more reasonable, which further confirm the high effectiveness of our solutions. %It is also quite intuitive that the relationship between \kw{"RIDE"} and \kw{"CAR"} in Fig.~\ref{sfig:b} is much more intimate than that between \kw{"STRANGER"} and \kw{"DANGER"} in Fig.~\ref{sfig:a}.

\begin{figure}[t!] \vspace*{-0.3cm}
	\subfigure[$3$-clique based method \label{sfig:a}]{\includegraphics[width=0.38\linewidth]{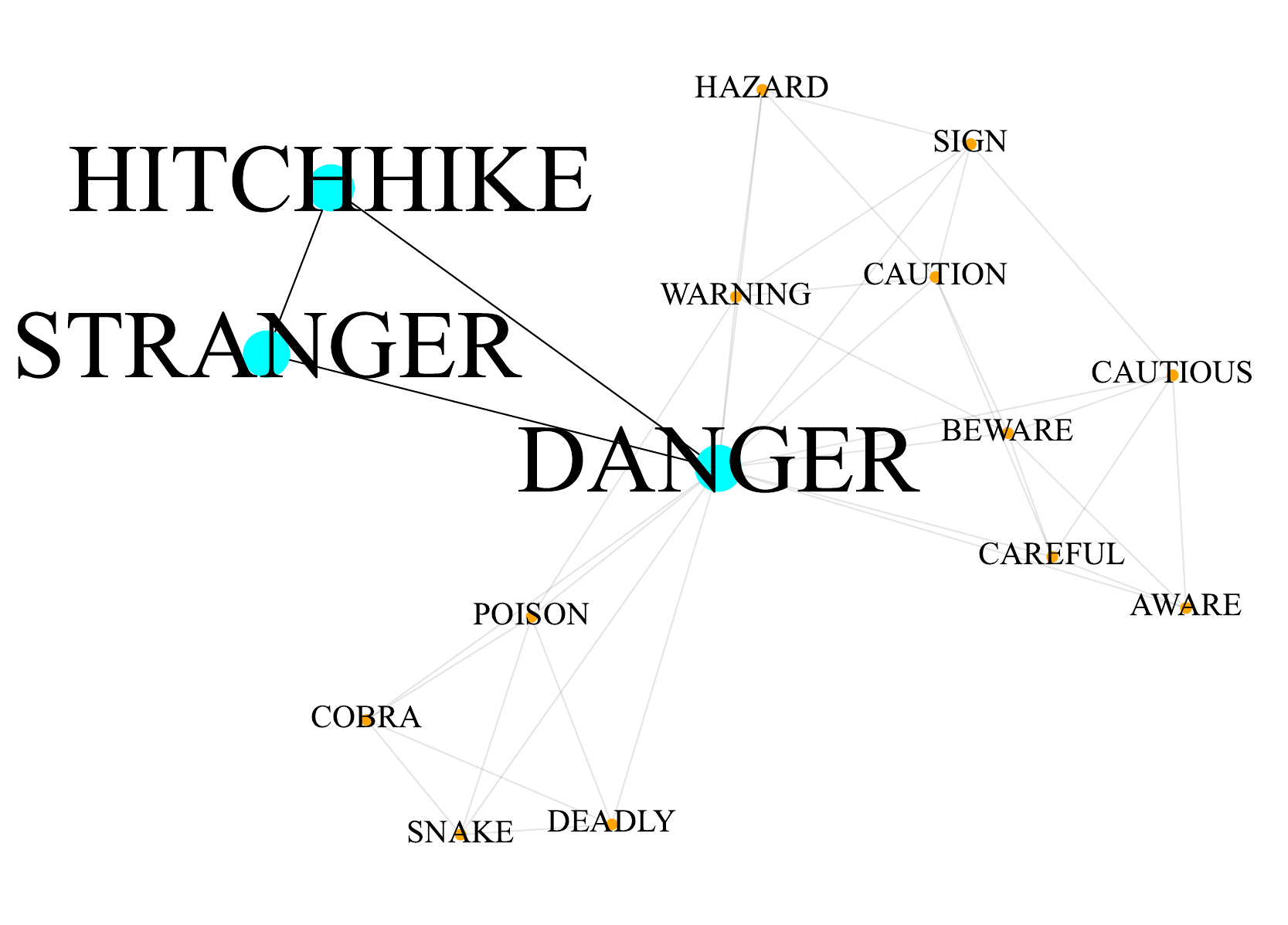}} \hspace{-0.3cm}
	\subfigure[$(3,1)$-dclique based method \label{sfig:b}]{\includegraphics[width=0.59\linewidth]{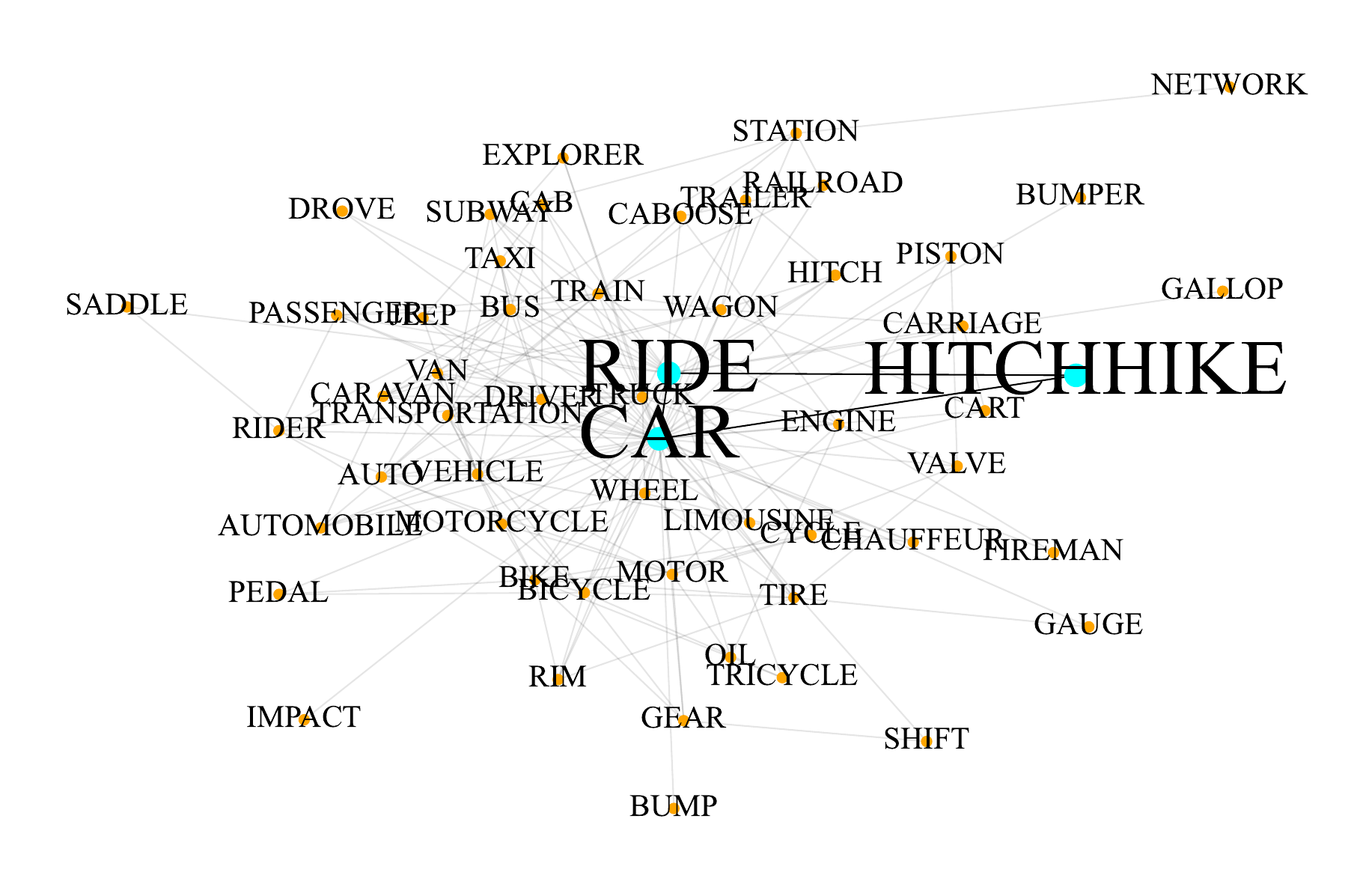} }
	\vspace{-0.4cm} \caption{The clusters containing \kw{"HITCHHIKE"} found by the clique-based and \hcs-based spectral clustering algorithms}
	\vspace{-0.15cm}
	\label{fig:clustering_case}
		\vspace{-0.3cm}
\end{figure}

}

\subsection{Application 2 : \hcs-based network analysis}\label{ssec:a2}
The graph profile (\gp) is a kind of characteristic vector of given networks, which has a lot of applications in network analysis \cite{guimera2007classes,alon2006introduction,mason2007graph,martinet2020robust, xu2008superfamily}. A nice property of \gp is that the networks in the same domain often have similar \gps \cite{motifProfile, 17prehighorder}.

%The most popular \gp is the subgraph ratio profile (\srp) \cite{motifProfile}. Given a sequence of subgraphs $\{M_1,M_2,...\}$, \srp is a vector where each item is the normalized significance ratio of the subgraph $M_i$. The significance ratio is obtained by subtracting the count of $M_i$ of a randomly generated null graph from the count of $M_i$ of the original graph, with some normalization \cite{motifProfile}. In our implementation, the null graph model is the algorithm introduced in \cite{coreNullModel} which can preserve the core number of origin graph. The sequence of subgraphs $\{M_1,M_2,...\}$ is set as six kinds of connected subgraphs with size $4$, which is the same as \cite{motifProfile}.
The subgraph ratio profile (\srp) is a well-known and widely used \gp  \cite{motifProfile}. The \srp is a vector that quantifies the relative significance of a sequence of subgraphs, denoted by $\{M_1,M_2,...\}$, where the $i_{th}$ entry in the vector corresponds to the normalized count of a specific subgraph $M_i$. This normalization involves subtracting the count of $M_i$ in a randomly generated null graph from that of the original graph, followed by a scaling \cite{motifProfile}. In our study, we use a null graph model based on the algorithm proposed in \cite{coreNullModel}, which preserves the core number of the original graph. The sequence of subgraphs $\{M_1,M_2,...\}$ is composed of six types of connected subgraphs with a size of four, following the methodology of \cite{motifProfile}.

%Based on the count of \hcs, we design a new graph profile called \hcs-based graph profile (\hgp). Given a sequence of size $\{q_1,q_2,...\}$, \hgp is a vector where the $i_{th}$ item is the ratio between the count of cliques and $\hcss$ with size $q_i$. Since the clique is a special kind of \hcs, the ratio is the probability of an \hcs being clique, which is the convergence trends of the graphs.
\stitle{\hcs-based graph profile.} We introduce a novel graph profile, termed the \hcs-based graph profile (\hgp), which is constructed based on the count of \hcs. The \hgp is a vector where the $i_{th}$ entry represents the ratio between the count of cliques and \hcss with size $q_i$. Since cliques are a special type of \hcs, this ratio can be interpreted as the probability of an \hcs being a clique, which characterizes the convergence behavior of the graphs. Since \hcspivot can count \hcss with size in a range $[q_l,q_r]$ simultaneously (Section~\ref{ssec:discussion}), we can compute \hgp efficiently.

\begin{figure}[t!] \vspace{-0.2cm}
	\subfigure[Amazon newtorks \label{sfig:pa}]{\includegraphics[width=0.45\linewidth]{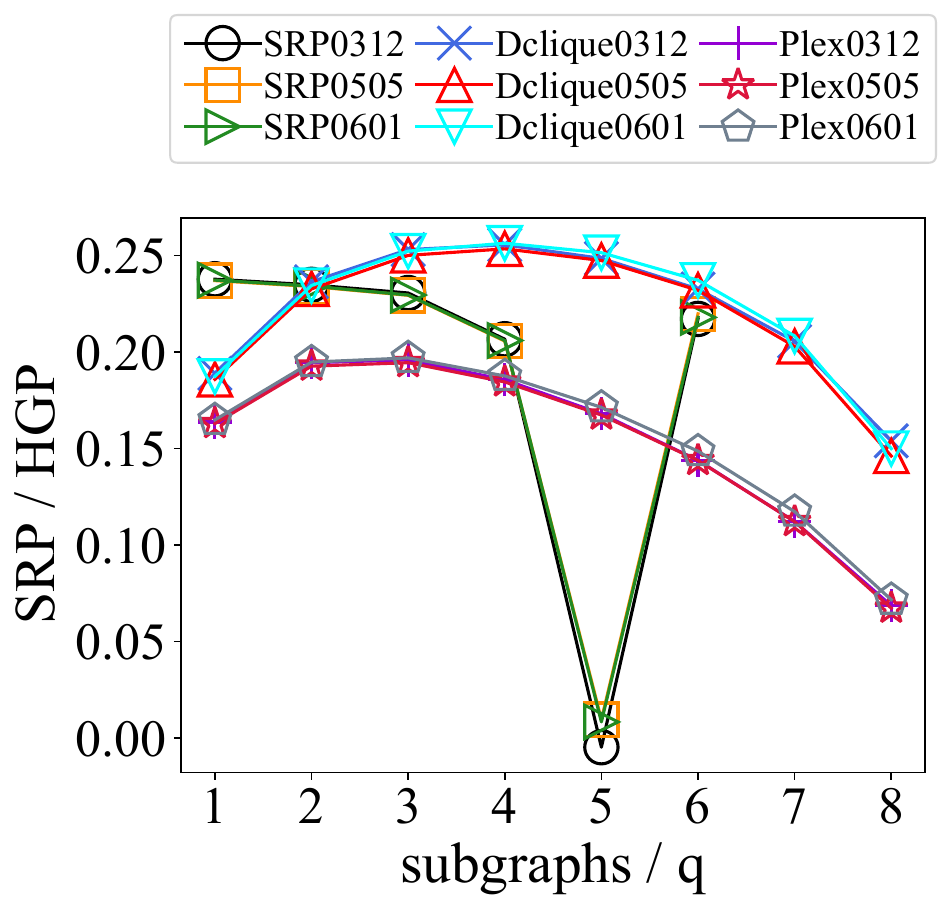}}\vspace{-0.2cm}
	\subfigure[\srps of the collaboration and social networks \label{sfig:pb}]{\includegraphics[width=0.45\linewidth]{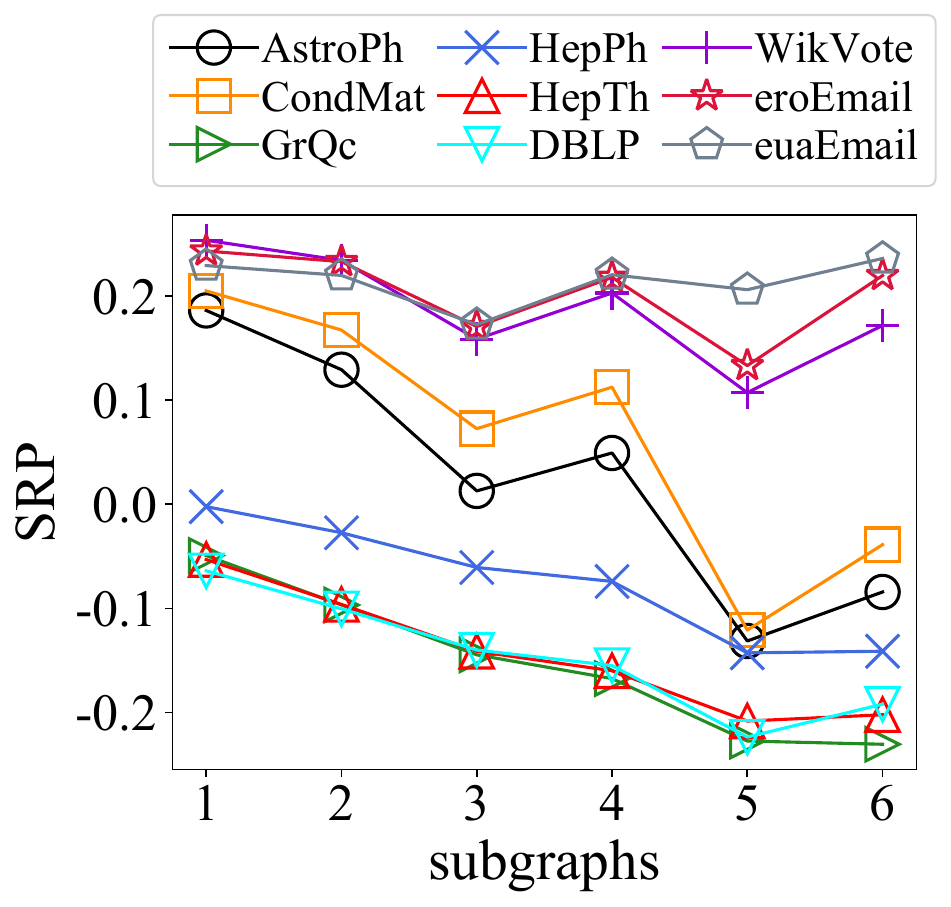} }
	\subfigure[dclique-based \hgps of collaboration and social networks \label{sfig:pc}]{\includegraphics[width=0.45\linewidth]{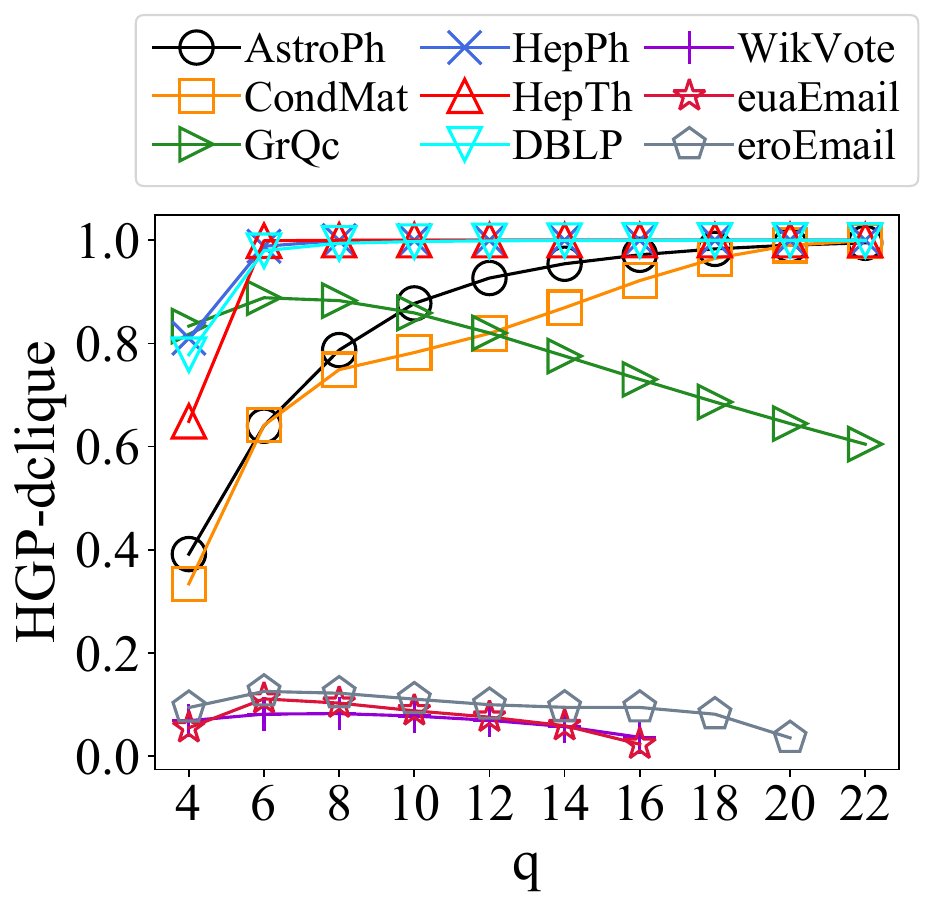}}%\vspace{-0.2cm}
	\subfigure[plex-based \hgps of collaboration and social networks \label{sfig:pd}]{\includegraphics[width=0.45\linewidth]{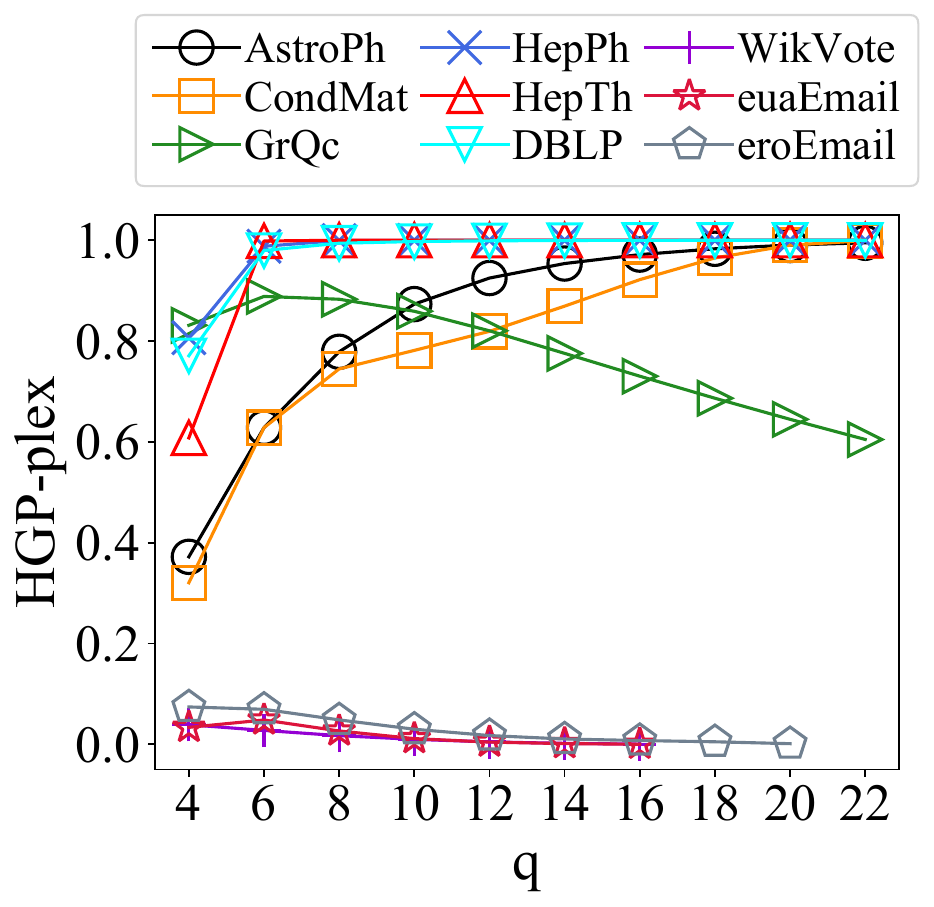} }
	
%	\vspace{-0.2cm}
	\caption{(1) Fig.~\ref{sfig:pa} are \srps and \hgps on the Amazon networks.  The Amazon0312, Amazon0505 and Amazon0601 have the same \srps and \hgps. (2) Fig.~\ref{sfig:pb}, Fig.~\ref{sfig:pc} and \ref{sfig:pd} are \srps and \hgps on 6 collaboration (from AstroPh, CondMat to DBLP) and 3 social (from WikVote to eroEmail) networks.  In Fig.~\ref{sfig:pb}, the networks in the two domains have similar \srps. In Fig.~\ref{sfig:pc} and Fig.~\ref{sfig:pd}, the networks in the same domain still have similar \hgps, but the \hgps of the networks in the two domains are different.  }
	\label{fig:profile}
	\vspace{-0.4cm}
\end{figure}

%In Fig.~\ref{fig:profile}, we compare the \srps and \hgps on the networks in three domains \cite{snapnets}. For representation simplicity, we use the count of $s$-dclique to compute the \hgps because the performance of the \hgps computed by the count of $s$-plex is similar. We set the sequence of size as $[4,5,...,20]$. Fig.~\ref{sfig:pa} and Fig.~\ref{sfig:pb} plot the \srps and \hgps on the Amazon networks. Fig.~\ref{sfig:pc} and Fig.~\ref{sfig:pd} plot the \srps and \hgps on six collaboration and three social networks.
In Fig.~\ref{fig:profile}, we compare the \srps and \hgps on the networks in three domains \cite{snapnets}. We set the sequence of size of \hgp as $\{4,5,...,20\}$. Fig.~\ref{sfig:pa} plots the \srps and \hgps on 3 Amazon networks. Fig.~\ref{sfig:pb},  Fig.~\ref{sfig:pc} and Fig.~\ref{sfig:pd} plot the \srps, dclique-based \hgps  and plex-base \hgps respectively on 6 collaboration and 3 social networks. Similar to \srps,  \hgps of the networks in the same domain have the same change tendency. For example, on the Amazon networks in Fig.~\ref{sfig:pa}, both Ama0312, Ama0505 and Ama0601 have exactly the same shapes of \hgps and \srps. The same performance of  \srp and \hgp implies that the proposed \hgp can characterize the network properties in the same domain as \srp. When the networks are in different domains, we find that \hgp performs better than \srp. Specifically, in Fig.~\ref{sfig:pb}, \srp cannot separate the collaboration and social networks. However, in Fig.~\ref{sfig:pc} and Fig.~\ref{sfig:pd}, \hgp separates the two kinds of networks clearly. As a result, \hgp is a better \gp than \srp when there exist networks in multiple domains. These results demonstrate the high effectiveness of the proposed \hgp to
characterize network properties.

%\begin{figure}[t!]
%	\subfigure[Amazon newtorks \label{sfig:pa}]{\includegraphics[width=0.45\linewidth]{./graphs/profile/srpAma.pdf}}
%	\subfigure[Amazon newtorks \label{sfig:pb}]{\includegraphics[width=0.45\linewidth]{./graphs/profile/hcsAma.pdf} }
%	\subfigure[Collaboration and social networks \label{sfig:pc}]{\includegraphics[width=0.45\linewidth]{./graphs/profile/srpCS.pdf}}
%	\subfigure[Collaboration and social networks \label{sfig:pd}]{\includegraphics[width=0.45\linewidth]{./graphs/profile/hcsCS.pdf} }
%	\caption{(1) Fig.~\ref{sfig:pa} and \ref{sfig:pb} are \srps and \hgps on the Amazon networks. Both \hgp and \srp extract the network Am00302 separately. (2) Fig.~\ref{sfig:pc} and \ref{sfig:pd} are \srps and \hgps on six collaboration (from AstroPh, CondMat to DBLP) and three social (from WikVote to eroEmail) networks.  In Fig.~\ref{sfig:pc}, the networks in the two domains have similar \srps. In Fig.~\ref{sfig:pd}, the networks in the same domain still have similar \hgps, but the \hgps of the networks in the two domains are different.  }
%	\label{fig:profile}
%\end{figure}

\section{Related work}
\stitle{Subgraph Counting.} Our work is related to the subgraph counting problem which includes general subgraph counting and specific subgraph counting \cite{surveyRibeiroPSAS21}. The recent exact algorithms for general subgraphs counting  \cite{17WWWmotif,OrbitsCounting,Ortmann,PGD} can compute the count efficiently, but they always have a small size constraint. To overcome the hardness, sampling-based approximate algorithms \cite{colorcodingTkdd, VLDB17PrioritySampling,pathsampling2018,pathsampling2015} for general subgraph counting are well studied. Recently, machine learning techniques are also applied for subgraph counting \cite{gnnZhaoYZLR21,nipsgnnChen0VB20,NeuralSubgraphCounting}. However, these method are often not very accurate and also cannot provide a theoretical guarantee of the estimated counts.

Despite the general algorithms, there are a lot of algorithms designed for counting specific important subgraphs, where the most representative one is $k$-clique. The first algorithm for $k$-clique counting was introduced by \cite{CN}, which is a listing-based solution. To improve the efficiency, some ordering based algorithms \cite{kClist, LiVldb} are designed. Instead of listing, Jain and Seshadhri \cite{PIVOTER} propose an elegant algorithm \kw{PIVOTER}, which can count all $k$-cliques without listing them, based on the classic maximal clique enumeration algorithm \cite{BK}. There also exist several approximate clique counting algorithms \cite{TuranShadow,ccpath}, which are often more efficient but cannot obtain the exact counts. All these algorithms are tailed to the problem of $k$-clique counting and cannot be applied to the general \hcs counting problem. In addition, we also note that the complexity of the hereditary subgraph counting problem was investigated in \cite{22stochcscountcomplex}. However, no specific counting algorithm was proposed in \cite{22stochcscountcomplex}.

\stitle{Maximal Hereditary Subgraph Enumeration.} Our work is also related to the problem of maximal hereditary subgraph enumeration, the goal of which is to enumerate all maximal hereditary subgraphs. The widely studied maximal hereditary subgraph model is the maximal clique. The most popular algorithm for maximal clique enumeration is the classic pivot-based Bron-Kerbosch algorithm \cite{BK, tomitaTime}. Since clique model is often too restrictive, several relaxed clique models include $s$-defective clique and $s$-plex are proposed \cite{Defective_bioinformatics_06,maximumDclique21, dcliqueAAAI2022,23sigmoddclique,Plex_seidman1978graph, plex_AAAI21, DAi_kplex}, which also satisfy the hereditary property. Recently, several pivot-based solutions for enumerating maximal $s$-defective cliques and maximal $s$-plex were proposed \cite{23sigmoddclique,DAi_kplex,22wwwkplex}, which generalize the classic pivoting technique proposed for maximal clique enumeration \cite{tomitaTime}. We also note that \cite{CohenKS08} developed a general algorithm to enumerate all maximal hereditary subgraph based on the reverse search framework \cite{96reverseserach}. However, this algorithm is often much slower than the pivot-based algorithms when processing real-world graphs. All the above mentioned algorithms are tailored to maximal hereditary subgraph enumeration, and they cannot be used for \hcs counting. This is because an \hcs can be located in many maximal hereditary subgraphs; and it is very difficult to reduce the repeated counts when using the maximal hereditary subgraphs to count \hcss.

%The maximal and maximum $s$-defective cliques are applied in biology and network analysis \cite{maximumDclique21, dcliqueAAAI2022, Defective_bioinformatics_06}. The maximal $s$-plex enumeration problem \cite{Plex_seidman1978graph, plex_AAAI21, DAi_kplex} involves finding all maximal $s$-plexes in a given graph. \cite{CohenKS08} aims to develop a general framework to list all maximal subgraphs with hereditary property, including the maximal  $s$-defective cliques and $k$-plex. However, their algorithm is not efficient compared to the specific designed algorithms. There are also other works on the clique-like models without hereditary property like $s$-clique \cite{luce1950connectivity, sCliqueBeharC18}, $\gamma$-quasi-cliques \cite{BrunatoHB07,quasiCliquePeiJZ05}, $k$-cores \cite{seidman1983network} and $k$-blocks \cite{moody2003structural}.

\section{Conclusion} \label{sec:conclusion}
In this work, we address a new problem of counting hereditary cohesive subgraphs (\hcss) in a graph. To tackle this problem, we first propose a listing-based framework with several carefully-designed pruning techniques to count the \hcss. To further improve the efficiency, we then develop a novel pivot-based framework which counts most \hcss in a combinatorial manner without the need for exhaustive listing. Based on our two frameworks, we devise several specific algorithms to count the $s$-defective cliques and $s$-plexes in a graph, which are two relaxed clique models and also satisfy the hereditary property. We conduct comprehensive experiments on 8 real-life networks to evaluate the efficiency of our algorithms. The results reveal that the pivot-based solutions exhibit significantly higher performance, being up to 7 orders of magnitude faster than the listing-based solution. In addition, we also evaluate the performance of our methods in graph clustering and network characterization applications, and the results demonstrate the high effectiveness of the proposed solutions.

%
%
%In this work, we introduce the problem of counting hereditary cohesive subgraphs. To the best of our knowledge, this is the first work to study the problem. We propose a listing-based framework, called \hcslist, and develop effective prune techniques to improve its performance. Additionally, we present an advanced algorithm that incorporates a new search strategy and a pivot vertex technique to further enhance the efficiency of the baseline. We also demonstrate the efficiency and effectiveness of our solutions through extensive experiments on 8 real-life graphs and two case studies. We hope that our work will inspire further research on subgraphs counting. %with the properties of hereditary.
%
%he specific design of the advanced framework on counting $s$-defective cliques and $k$-plexes are also described in details.

\bibliographystyle{ACM-Reference-Format}
\balance
\bibliography{reference}


\begin{thebibliography}{75}

%%% the Web address.

\ifx \showCODEN    \undefined \def \showCODEN     #1{\unskip}     \fi
\ifx \showDOI      \undefined \def \showDOI       #1{#1}\fi
\ifx \showISBNx    \undefined \def \showISBNx     #1{\unskip}     \fi
\ifx \showISBNxiii \undefined \def \showISBNxiii  #1{\unskip}     \fi
\ifx \showISSN     \undefined \def \showISSN      #1{\unskip}     \fi
\ifx \showLCCN     \undefined \def \showLCCN      #1{\unskip}     \fi
\ifx \shownote     \undefined \def \shownote      #1{#1}          \fi
\ifx \showarticletitle \undefined \def \showarticletitle #1{#1}   \fi
\ifx \showURL      \undefined \def \showURL       {\relax}        \fi
% invisible to TeX
\providecommand\bibfield[2]{#2}
\providecommand\bibinfo[2]{#2}
\providecommand\natexlab[1]{#1}
\providecommand\showeprint[2][]{arXiv:#2}

\bibitem[Ahmed et~al\mbox{.}(2017)]%
        {VLDB17PrioritySampling}
\bibfield{author}{\bibinfo{person}{Nesreen~K. Ahmed}, \bibinfo{person}{Nick~G.
  Duffield}, \bibinfo{person}{Theodore~L. Willke}, {and}
  \bibinfo{person}{Ryan~A. Rossi}.} \bibinfo{year}{2017}\natexlab{}.
\newblock \showarticletitle{On Sampling from Massive Graph Streams}.
\newblock \bibinfo{journal}{\emph{Proc. {VLDB} Endow.}} \bibinfo{volume}{10},
  \bibinfo{number}{11} (\bibinfo{year}{2017}), \bibinfo{pages}{1430--1441}.
\newblock


\bibitem[Ahmed et~al\mbox{.}(2015)]%
        {PGD}
\bibfield{author}{\bibinfo{person}{Nesreen~K. Ahmed}, \bibinfo{person}{Jennifer
  Neville}, \bibinfo{person}{Ryan~A. Rossi}, {and} \bibinfo{person}{Nick~G.
  Duffield}.} \bibinfo{year}{2015}\natexlab{}.
\newblock \showarticletitle{Efficient Graphlet Counting for Large Networks}. In
  \bibinfo{booktitle}{\emph{ICDM}}. \bibinfo{pages}{1--10}.
\newblock


\bibitem[Alon(2006)]%
        {alon2006introduction}
\bibfield{author}{\bibinfo{person}{Uri Alon}.} \bibinfo{year}{2006}\natexlab{}.
\newblock \bibinfo{booktitle}{\emph{An introduction to systems biology: design
  principles of biological circuits}}.
\newblock \bibinfo{publisher}{Chapman and Hall/CRC}.
\newblock


\bibitem[Avis and Fukuda(1996)]%
        {96reverseserach}
\bibfield{author}{\bibinfo{person}{David Avis} {and} \bibinfo{person}{Komei
  Fukuda}.} \bibinfo{year}{1996}\natexlab{}.
\newblock \showarticletitle{Reverse Search for Enumeration}.
\newblock \bibinfo{journal}{\emph{Discret. Appl. Math.}} \bibinfo{volume}{65},
  \bibinfo{number}{1-3} (\bibinfo{year}{1996}), \bibinfo{pages}{21--46}.
\newblock


\bibitem[Aynaud(2020)]%
        {python_louvain}
\bibfield{author}{\bibinfo{person}{Thomas Aynaud}.}
  \bibinfo{year}{2020}\natexlab{}.
\newblock \bibinfo{title}{python-louvain x.y: Louvain algorithm for community
  detection}.
\newblock
  \bibinfo{howpublished}{\href{https://github.com/taynaud/python-louvain}{\texttt{https://github.com/taynaud/python-louvain}}}.
\newblock


\bibitem[Batagelj and Zaversnik(2003)]%
        {corePeel}
\bibfield{author}{\bibinfo{person}{Vladimir Batagelj} {and}
  \bibinfo{person}{Matjaz Zaversnik}.} \bibinfo{year}{2003}\natexlab{}.
\newblock \showarticletitle{An O(m) Algorithm for Cores Decomposition of
  Networks}.
\newblock \bibinfo{journal}{\emph{CoRR}}  \bibinfo{volume}{cs.DS/0310049}
  (\bibinfo{year}{2003}).
\newblock


\bibitem[Benson et~al\mbox{.}(2016)]%
        {benson2016higher}
\bibfield{author}{\bibinfo{person}{Austin~R Benson}, \bibinfo{person}{David~F
  Gleich}, {and} \bibinfo{person}{Jure Leskovec}.}
  \bibinfo{year}{2016}\natexlab{}.
\newblock \showarticletitle{Higher-order organization of complex networks}.
\newblock \bibinfo{journal}{\emph{Science}} \bibinfo{volume}{353},
  \bibinfo{number}{6295} (\bibinfo{year}{2016}), \bibinfo{pages}{163--166}.
\newblock


\bibitem[Bressan et~al\mbox{.}(2018)]%
        {colorcodingTkdd}
\bibfield{author}{\bibinfo{person}{Marco Bressan}, \bibinfo{person}{Flavio
  Chierichetti}, \bibinfo{person}{Ravi Kumar}, \bibinfo{person}{Stefano
  Leucci}, {and} \bibinfo{person}{Alessandro Panconesi}.}
  \bibinfo{year}{2018}\natexlab{}.
\newblock \showarticletitle{Motif Counting Beyond Five Nodes}.
\newblock \bibinfo{journal}{\emph{{ACM} Trans. Knowl. Discov. Data}}
  \bibinfo{volume}{12}, \bibinfo{number}{4} (\bibinfo{year}{2018}),
  \bibinfo{pages}{48:1--48:25}.
\newblock


\bibitem[Bron and Kerbosch(1973)]%
        {BK}
\bibfield{author}{\bibinfo{person}{Coenraad Bron} {and} \bibinfo{person}{Joep
  Kerbosch}.} \bibinfo{year}{1973}\natexlab{}.
\newblock \showarticletitle{Finding All Cliques of an Undirected Graph
  (Algorithm 457)}.
\newblock \bibinfo{journal}{\emph{Commun. {ACM}}} \bibinfo{volume}{16},
  \bibinfo{number}{9} (\bibinfo{year}{1973}), \bibinfo{pages}{575--576}.
\newblock


\bibitem[Brunato et~al\mbox{.}(2007)]%
        {BrunatoHB07}
\bibfield{author}{\bibinfo{person}{Mauro Brunato}, \bibinfo{person}{Holger~H.
  Hoos}, {and} \bibinfo{person}{Roberto Battiti}.}
  \bibinfo{year}{2007}\natexlab{}.
\newblock \showarticletitle{On Effectively Finding Maximal Quasi-cliques in
  Graphs}. In \bibinfo{booktitle}{\emph{{LION}}}
  \emph{(\bibinfo{series}{Lecture Notes in Computer Science},
  Vol.~\bibinfo{volume}{5313})}. \bibinfo{pages}{41--55}.
\newblock


\bibitem[Chang et~al\mbox{.}(2017)]%
        {pSCAN}
\bibfield{author}{\bibinfo{person}{Lijun Chang}, \bibinfo{person}{Wei Li},
  \bibinfo{person}{Lu Qin}, \bibinfo{person}{Wenjie Zhang}, {and}
  \bibinfo{person}{Shiyu Yang}.} \bibinfo{year}{2017}\natexlab{}.
\newblock \showarticletitle{pSCAN: Fast and Exact Structural Graph Clustering}.
\newblock \bibinfo{journal}{\emph{{IEEE} Trans. Knowl. Data Eng.}}
  \bibinfo{volume}{29}, \bibinfo{number}{2} (\bibinfo{year}{2017}),
  \bibinfo{pages}{387--401}.
\newblock


\bibitem[Chang and Qin(2018)]%
        {subgraphcountingbook}
\bibfield{author}{\bibinfo{person}{Lijun Chang} {and} \bibinfo{person}{Lu
  Qin}.} \bibinfo{year}{2018}\natexlab{}.
\newblock \bibinfo{booktitle}{\emph{Cohesive Subgraph Computation Over Large
  Sparse Graphs}}.
\newblock \bibinfo{publisher}{Springer Cham}.
\newblock


\bibitem[Chang et~al\mbox{.}(2013)]%
        {13sigmodkecc}
\bibfield{author}{\bibinfo{person}{Lijun Chang}, \bibinfo{person}{Jeffrey~Xu
  Yu}, \bibinfo{person}{Lu Qin}, \bibinfo{person}{Xuemin Lin},
  \bibinfo{person}{Chengfei Liu}, {and} \bibinfo{person}{Weifa Liang}.}
  \bibinfo{year}{2013}\natexlab{}.
\newblock \showarticletitle{Efficiently computing k-edge connected components
  via graph decomposition}. In \bibinfo{booktitle}{\emph{{SIGMOD}}}.
  \bibinfo{pages}{205--216}.
\newblock


\bibitem[Chen et~al\mbox{.}(2021)]%
        {maximumDclique21}
\bibfield{author}{\bibinfo{person}{Xiaoyu Chen}, \bibinfo{person}{Yi Zhou},
  \bibinfo{person}{Jin{-}Kao Hao}, {and} \bibinfo{person}{Mingyu Xiao}.}
  \bibinfo{year}{2021}\natexlab{}.
\newblock \showarticletitle{Computing maximum \emph{k}-defective cliques in
  massive graphs}.
\newblock \bibinfo{journal}{\emph{Comput. Oper. Res.}}  \bibinfo{volume}{127}
  (\bibinfo{year}{2021}), \bibinfo{pages}{105131}.
\newblock


\bibitem[Chen et~al\mbox{.}(2020)]%
        {nipsgnnChen0VB20}
\bibfield{author}{\bibinfo{person}{Zhengdao Chen}, \bibinfo{person}{Lei Chen},
  \bibinfo{person}{Soledad Villar}, {and} \bibinfo{person}{Joan Bruna}.}
  \bibinfo{year}{2020}\natexlab{}.
\newblock \showarticletitle{Can Graph Neural Networks Count Substructures?}. In
  \bibinfo{booktitle}{\emph{NeurIPS}}.
\newblock


\bibitem[Chiba and Nishizeki(1985)]%
        {CN}
\bibfield{author}{\bibinfo{person}{Norishige Chiba} {and}
  \bibinfo{person}{Takao Nishizeki}.} \bibinfo{year}{1985}\natexlab{}.
\newblock \showarticletitle{Arboricity and Subgraph Listing Algorithms}.
\newblock \bibinfo{journal}{\emph{{SIAM} J. Comput.}} \bibinfo{volume}{14},
  \bibinfo{number}{1} (\bibinfo{year}{1985}), \bibinfo{pages}{210--223}.
\newblock


\bibitem[Cohen et~al\mbox{.}(2008)]%
        {CohenKS08}
\bibfield{author}{\bibinfo{person}{Sara Cohen}, \bibinfo{person}{Benny
  Kimelfeld}, {and} \bibinfo{person}{Yehoshua Sagiv}.}
  \bibinfo{year}{2008}\natexlab{}.
\newblock \showarticletitle{Generating all maximal induced subgraphs for
  hereditary and connected-hereditary graph properties}.
\newblock \bibinfo{journal}{\emph{J. Comput. Syst. Sci.}} \bibinfo{volume}{74},
  \bibinfo{number}{7} (\bibinfo{year}{2008}), \bibinfo{pages}{1147--1159}.
\newblock


\bibitem[Comandur and Tirthapura(2019)]%
        {tutoralSeshadhriT19}
\bibfield{author}{\bibinfo{person}{Seshadhri Comandur} {and}
  \bibinfo{person}{Srikanta Tirthapura}.} \bibinfo{year}{2019}\natexlab{}.
\newblock \showarticletitle{Scalable Subgraph Counting: The Methods Behind The
  Madness}. In \bibinfo{booktitle}{\emph{{WWW}}}. \bibinfo{pages}{1317--1318}.
\newblock


\bibitem[Conte et~al\mbox{.}(2018)]%
        {d2k_kdd18}
\bibfield{author}{\bibinfo{person}{Alessio Conte}, \bibinfo{person}{Tiziano~De
  Matteis}, \bibinfo{person}{Daniele~De Sensi}, \bibinfo{person}{Roberto
  Grossi}, \bibinfo{person}{Andrea Marino}, {and} \bibinfo{person}{Luca
  Versari}.} \bibinfo{year}{2018}\natexlab{}.
\newblock \showarticletitle{{D2K:} Scalable Community Detection in Massive
  Networks via Small-Diameter k-Plexes}. In \bibinfo{booktitle}{\emph{{KDD}}}.
  \bibinfo{pages}{1272--1281}.
\newblock


\bibitem[Dai et~al\mbox{.}(2023)]%
        {23sigmoddclique}
\bibfield{author}{\bibinfo{person}{Qiangqiang Dai}, \bibinfo{person}{Rong{-}Hua
  Li}, \bibinfo{person}{Meihao Liao}, {and} \bibinfo{person}{Guoren Wang}.}
  \bibinfo{year}{2023}\natexlab{}.
\newblock \showarticletitle{Maximal Defective Clique Enumeration}.
\newblock \bibinfo{journal}{\emph{Proc. {ACM} Manag. Data}}
  \bibinfo{volume}{1}, \bibinfo{number}{1} (\bibinfo{year}{2023}),
  \bibinfo{pages}{77:1--77:26}.
\newblock


\bibitem[Dai et~al\mbox{.}(2022)]%
        {DAi_kplex}
\bibfield{author}{\bibinfo{person}{Qiangqiang Dai}, \bibinfo{person}{Rong{-}Hua
  Li}, \bibinfo{person}{Hongchao Qin}, \bibinfo{person}{Meihao Liao}, {and}
  \bibinfo{person}{Guoren Wang}.} \bibinfo{year}{2022}\natexlab{}.
\newblock \showarticletitle{Scaling Up Maximal \emph{k}-plex Enumeration}. In
  \bibinfo{booktitle}{\emph{{CIKM}}}. \bibinfo{pages}{345--354}.
\newblock


\bibitem[Danisch et~al\mbox{.}(2018)]%
        {kClist}
\bibfield{author}{\bibinfo{person}{Maximilien Danisch}, \bibinfo{person}{Oana
  Balalau}, {and} \bibinfo{person}{Mauro Sozio}.}
  \bibinfo{year}{2018}\natexlab{}.
\newblock \showarticletitle{Listing k-cliques in Sparse Real-World Graphs}. In
  \bibinfo{booktitle}{\emph{{WWW}}}. \bibinfo{pages}{589--598}.
\newblock


\bibitem[Edler et~al\mbox{.}(2023)]%
        {mapequation2023software}
\bibfield{author}{\bibinfo{person}{Daniel Edler}, \bibinfo{person}{Anton
  Holmgren}, {and} \bibinfo{person}{Martin Rosvall}.}
  \bibinfo{year}{2023}\natexlab{}.
\newblock \bibinfo{title}{{The MapEquation software package}}.
\newblock \bibinfo{howpublished}{\url{https://mapequation.org}}.
\newblock


\bibitem[Epasto et~al\mbox{.}(2015)]%
        {15vldbegonet}
\bibfield{author}{\bibinfo{person}{Alessandro Epasto}, \bibinfo{person}{Silvio
  Lattanzi}, \bibinfo{person}{Vahab~S. Mirrokni}, \bibinfo{person}{Ismail
  Sebe}, \bibinfo{person}{Ahmed Taei}, {and} \bibinfo{person}{Sunita Verma}.}
  \bibinfo{year}{2015}\natexlab{}.
\newblock \showarticletitle{Ego-net Community Mining Applied to Friend
  Suggestion}.
\newblock \bibinfo{journal}{\emph{Proc. {VLDB} Endow.}} \bibinfo{volume}{9},
  \bibinfo{number}{4} (\bibinfo{year}{2015}), \bibinfo{pages}{324--335}.
\newblock


\bibitem[Eppstein et~al\mbox{.}(2013)]%
        {13maximalclique}
\bibfield{author}{\bibinfo{person}{David Eppstein}, \bibinfo{person}{Maarten
  L{\"{o}}ffler}, {and} \bibinfo{person}{Darren Strash}.}
  \bibinfo{year}{2013}\natexlab{}.
\newblock \showarticletitle{Listing All Maximal Cliques in Large Sparse
  Real-World Graphs}.
\newblock \bibinfo{journal}{\emph{{ACM} J. Exp. Algorithmics}}
  \bibinfo{volume}{18} (\bibinfo{year}{2013}).
\newblock


\bibitem[Focke and Roth(2022)]%
        {22stochcscountcomplex}
\bibfield{author}{\bibinfo{person}{Jacob Focke} {and} \bibinfo{person}{Marc
  Roth}.} \bibinfo{year}{2022}\natexlab{}.
\newblock \showarticletitle{Counting small induced subgraphs with hereditary
  properties}. In \bibinfo{booktitle}{\emph{{STOC}}}.
  \bibinfo{pages}{1543--1551}.
\newblock


\bibitem[Fortunato(2010)]%
        {fortunato2010community}
\bibfield{author}{\bibinfo{person}{Santo Fortunato}.}
  \bibinfo{year}{2010}\natexlab{}.
\newblock \showarticletitle{Community detection in graphs}.
\newblock \bibinfo{journal}{\emph{Physics reports}} \bibinfo{volume}{486},
  \bibinfo{number}{3-5} (\bibinfo{year}{2010}), \bibinfo{pages}{75--174}.
\newblock


\bibitem[Gao et~al\mbox{.}(2022)]%
        {dcliqueAAAI2022}
\bibfield{author}{\bibinfo{person}{Jian Gao}, \bibinfo{person}{Zhenghang Xu},
  \bibinfo{person}{Ruizhi Li}, {and} \bibinfo{person}{Minghao Yin}.}
  \bibinfo{year}{2022}\natexlab{}.
\newblock \showarticletitle{An Exact Algorithm with New Upper Bounds for the
  Maximum k-Defective Clique Problem in Massive Sparse Graphs}. In
  \bibinfo{booktitle}{\emph{{AAAI}}}. \bibinfo{pages}{10174--10183}.
\newblock


\bibitem[Guimera et~al\mbox{.}(2007)]%
        {guimera2007classes}
\bibfield{author}{\bibinfo{person}{Roger Guimera}, \bibinfo{person}{Marta
  Sales-Pardo}, {and} \bibinfo{person}{Luis~AN Amaral}.}
  \bibinfo{year}{2007}\natexlab{}.
\newblock \showarticletitle{Classes of complex networks defined by role-to-role
  connectivity profiles}.
\newblock \bibinfo{journal}{\emph{Nature physics}} \bibinfo{volume}{3},
  \bibinfo{number}{1} (\bibinfo{year}{2007}), \bibinfo{pages}{63--69}.
\newblock


\bibitem[Hocevar and Demsar(2014)]%
        {OrbitsCounting}
\bibfield{author}{\bibinfo{person}{Tomaz Hocevar} {and} \bibinfo{person}{Janez
  Demsar}.} \bibinfo{year}{2014}\natexlab{}.
\newblock \showarticletitle{A combinatorial approach to graphlet counting}.
\newblock \bibinfo{journal}{\emph{Bioinform.}} \bibinfo{volume}{30},
  \bibinfo{number}{4} (\bibinfo{year}{2014}), \bibinfo{pages}{559--565}.
\newblock


\bibitem[Huang et~al\mbox{.}(2014)]%
        {14sigmodtruss}
\bibfield{author}{\bibinfo{person}{Xin Huang}, \bibinfo{person}{Hong Cheng},
  \bibinfo{person}{Lu Qin}, \bibinfo{person}{Wentao Tian}, {and}
  \bibinfo{person}{Jeffrey~Xu Yu}.} \bibinfo{year}{2014}\natexlab{}.
\newblock \showarticletitle{Querying k-truss community in large and dynamic
  graphs}. In \bibinfo{booktitle}{\emph{{SIGMOD}}}.
  \bibinfo{pages}{1311--1322}.
\newblock


\bibitem[Jain and Seshadhri(2017)]%
        {TuranShadow}
\bibfield{author}{\bibinfo{person}{Shweta Jain} {and} \bibinfo{person}{C.
  Seshadhri}.} \bibinfo{year}{2017}\natexlab{}.
\newblock \showarticletitle{A Fast and Provable Method for Estimating Clique
  Counts Using Tur{\'{a}}n's Theorem}. In \bibinfo{booktitle}{\emph{WWW}}.
  \bibinfo{pages}{441--449}.
\newblock


\bibitem[Jain and Seshadhri(2020)]%
        {PIVOTER}
\bibfield{author}{\bibinfo{person}{Shweta Jain} {and} \bibinfo{person}{C.
  Seshadhri}.} \bibinfo{year}{2020}\natexlab{}.
\newblock \showarticletitle{The Power of Pivoting for Exact Clique Counting}.
  In \bibinfo{booktitle}{\emph{WSDM}}. \bibinfo{pages}{268--276}.
\newblock


\bibitem[Jha et~al\mbox{.}(2015)]%
        {pathsampling2015}
\bibfield{author}{\bibinfo{person}{Madhav Jha}, \bibinfo{person}{C. Seshadhri},
  {and} \bibinfo{person}{Ali Pinar}.} \bibinfo{year}{2015}\natexlab{}.
\newblock \showarticletitle{Path Sampling: {A} Fast and Provable Method for
  Estimating 4-Vertex Subgraph Counts}. In \bibinfo{booktitle}{\emph{WWW}}.
  \bibinfo{pages}{495--505}.
\newblock


\bibitem[Koevering et~al\mbox{.}(2021)]%
        {coreNullModel}
\bibfield{author}{\bibinfo{person}{Katherine~Van Koevering},
  \bibinfo{person}{Austin~R. Benson}, {and} \bibinfo{person}{Jon~M.
  Kleinberg}.} \bibinfo{year}{2021}\natexlab{}.
\newblock \showarticletitle{Random Graphs with Prescribed K-Core Sequences: {A}
  New Null Model for Network Analysis}. In \bibinfo{booktitle}{\emph{{WWW}
  '21}}, \bibfield{editor}{\bibinfo{person}{Jure Leskovec},
  \bibinfo{person}{Marko Grobelnik}, \bibinfo{person}{Marc Najork},
  \bibinfo{person}{Jie Tang}, {and} \bibinfo{person}{Leila Zia}} (Eds.).
  \bibinfo{pages}{367--378}.
\newblock


\bibitem[Leskovec and Krevl(2014)]%
        {snapnets}
\bibfield{author}{\bibinfo{person}{Jure Leskovec} {and} \bibinfo{person}{Andrej
  Krevl}.} \bibinfo{year}{2014}\natexlab{}.
\newblock \bibinfo{title}{{SNAP Datasets}: {Stanford} Large Network Dataset
  Collection}.
\newblock \bibinfo{howpublished}{\url{http://snap.stanford.edu/data}}.
\newblock


\bibitem[Leskovec et~al\mbox{.}(2008)]%
        {leskovec2008statistical}
\bibfield{author}{\bibinfo{person}{Jure Leskovec}, \bibinfo{person}{Kevin~J
  Lang}, \bibinfo{person}{Anirban Dasgupta}, {and} \bibinfo{person}{Michael~W
  Mahoney}.} \bibinfo{year}{2008}\natexlab{}.
\newblock \showarticletitle{Statistical properties of community structure in
  large social and information networks}. In \bibinfo{booktitle}{\emph{WWW}}.
  \bibinfo{pages}{695--704}.
\newblock


\bibitem[Li et~al\mbox{.}(2020)]%
        {LiVldb}
\bibfield{author}{\bibinfo{person}{Ronghua Li}, \bibinfo{person}{Sen Gao},
  \bibinfo{person}{Lu Qin}, \bibinfo{person}{Guoren Wang},
  \bibinfo{person}{Weihua Yang}, {and} \bibinfo{person}{Jeffrey~Xu Yu}.}
  \bibinfo{year}{2020}\natexlab{}.
\newblock \showarticletitle{Ordering Heuristics for k-clique Listing}.
\newblock \bibinfo{journal}{\emph{Proc. {VLDB} Endow.}} \bibinfo{volume}{13},
  \bibinfo{number}{11} (\bibinfo{year}{2020}), \bibinfo{pages}{2536--2548}.
\newblock


\bibitem[Li et~al\mbox{.}(2024)]%
        {fullversion}
\bibfield{author}{\bibinfo{person}{Rong{-}Hua Li}, \bibinfo{person}{Xiaowei
  Ye}, \bibinfo{person}{Fusheng Jin}, \bibinfo{person}{Yu{-}Ping Wang},
  \bibinfo{person}{Ye Yuan}, {and} \bibinfo{person}{Guoren Wang}.}
  \bibinfo{year}{2024}\natexlab{}.
\newblock \showarticletitle{Counting Cohesive Subgraphs with Hereditary
  Properties}. In \bibinfo{booktitle}{\emph{Full version:}}.
\newblock
\urldef\tempurl%
\url{https://github.com/LightWant/HCS}
\showURL{%
\tempurl}


\bibitem[Lu et~al\mbox{.}(2018)]%
        {lu2018community}
\bibfield{author}{\bibinfo{person}{Zhenqi Lu}, \bibinfo{person}{Johan
  Wahlstr{\"o}m}, {and} \bibinfo{person}{Arye Nehorai}.}
  \bibinfo{year}{2018}\natexlab{}.
\newblock \showarticletitle{Community detection in complex networks via clique
  conductance}.
\newblock \bibinfo{journal}{\emph{Scientific reports}} \bibinfo{volume}{8},
  \bibinfo{number}{1} (\bibinfo{year}{2018}), \bibinfo{pages}{5982}.
\newblock


\bibitem[Luce(1950)]%
        {luce1950connectivity}
\bibfield{author}{\bibinfo{person}{R~Duncan Luce}.}
  \bibinfo{year}{1950}\natexlab{}.
\newblock \showarticletitle{Connectivity and generalized cliques in sociometric
  group structure}.
\newblock \bibinfo{journal}{\emph{Psychometrika}} \bibinfo{volume}{15},
  \bibinfo{number}{2} (\bibinfo{year}{1950}), \bibinfo{pages}{169--190}.
\newblock


\bibitem[Martello et~al\mbox{.}(1999)]%
        {Knapsack}
\bibfield{author}{\bibinfo{person}{Silvano Martello}, \bibinfo{person}{David
  Pisinger}, {and} \bibinfo{person}{Paolo Toth}.}
  \bibinfo{year}{1999}\natexlab{}.
\newblock \showarticletitle{Dynamic Programming and Strong Bounds for the 0-1
  Knapsack Problem}.
\newblock \bibinfo{journal}{\emph{Management Science}} \bibinfo{volume}{45},
  \bibinfo{number}{3} (\bibinfo{year}{1999}), \bibinfo{pages}{414--424}.
\newblock


\bibitem[Martinet et~al\mbox{.}(2020)]%
        {martinet2020robust}
\bibfield{author}{\bibinfo{person}{L-E Martinet}, \bibinfo{person}{MA Kramer},
  \bibinfo{person}{W Viles}, \bibinfo{person}{LN Perkins}, \bibinfo{person}{E
  Spencer}, \bibinfo{person}{CJ Chu}, \bibinfo{person}{SS Cash}, {and}
  \bibinfo{person}{ED Kolaczyk}.} \bibinfo{year}{2020}\natexlab{}.
\newblock \showarticletitle{Robust dynamic community detection with
  applications to human brain functional networks}.
\newblock \bibinfo{journal}{\emph{Nature communications}} \bibinfo{volume}{11},
  \bibinfo{number}{1} (\bibinfo{year}{2020}), \bibinfo{pages}{2785}.
\newblock


\bibitem[Mason and Verwoerd(2007)]%
        {mason2007graph}
\bibfield{author}{\bibinfo{person}{Oliver Mason} {and} \bibinfo{person}{Mark
  Verwoerd}.} \bibinfo{year}{2007}\natexlab{}.
\newblock \showarticletitle{Graph theory and networks in biology}.
\newblock \bibinfo{journal}{\emph{IET systems biology}} \bibinfo{volume}{1},
  \bibinfo{number}{2} (\bibinfo{year}{2007}), \bibinfo{pages}{89--119}.
\newblock


\bibitem[Matula and Beck(1983)]%
        {degeneracyOrder}
\bibfield{author}{\bibinfo{person}{David~W. Matula} {and}
  \bibinfo{person}{Leland~L. Beck}.} \bibinfo{year}{1983}\natexlab{}.
\newblock \showarticletitle{Smallest-Last Ordering and clustering and Graph
  Coloring Algorithms}.
\newblock \bibinfo{journal}{\emph{J. {ACM}}} \bibinfo{volume}{30},
  \bibinfo{number}{3} (\bibinfo{year}{1983}), \bibinfo{pages}{417--427}.
\newblock


\bibitem[Milo et~al\mbox{.}(2004)]%
        {motifProfile}
\bibfield{author}{\bibinfo{person}{Ron Milo}, \bibinfo{person}{Shalev
  Itzkovitz}, \bibinfo{person}{Nadav Kashtan}, \bibinfo{person}{Reuven Levitt},
  \bibinfo{person}{Shai Shen-Orr}, \bibinfo{person}{Inbal Ayzenshtat},
  \bibinfo{person}{Michal Sheffer}, {and} \bibinfo{person}{Uri Alon}.}
  \bibinfo{year}{2004}\natexlab{}.
\newblock \showarticletitle{Superfamilies of evolved and designed networks}.
\newblock \bibinfo{journal}{\emph{Science}} \bibinfo{volume}{303},
  \bibinfo{number}{5663} (\bibinfo{year}{2004}), \bibinfo{pages}{1538--1542}.
\newblock


\bibitem[Ortmann and Brandes(2017)]%
        {Ortmann}
\bibfield{author}{\bibinfo{person}{Mark Ortmann} {and} \bibinfo{person}{Ulrik
  Brandes}.} \bibinfo{year}{2017}\natexlab{}.
\newblock \showarticletitle{Efficient orbit-aware triad and quad census in
  directed and undirected graphs}.
\newblock \bibinfo{journal}{\emph{Appl. Netw. Sci.}}  \bibinfo{volume}{2}
  (\bibinfo{year}{2017}), \bibinfo{pages}{13}.
\newblock


\bibitem[Pedregosa et~al\mbox{.}(2011)]%
        {scikit-learn}
\bibfield{author}{\bibinfo{person}{F. Pedregosa}, \bibinfo{person}{G.
  Varoquaux}, \bibinfo{person}{A. Gramfort}, \bibinfo{person}{V. Michel},
  \bibinfo{person}{B. Thirion}, \bibinfo{person}{O. Grisel},
  \bibinfo{person}{M. Blondel}, \bibinfo{person}{P. Prettenhofer},
  \bibinfo{person}{R. Weiss}, \bibinfo{person}{V. Dubourg}, \bibinfo{person}{J.
  Vanderplas}, \bibinfo{person}{A. Passos}, \bibinfo{person}{D. Cournapeau},
  \bibinfo{person}{M. Brucher}, \bibinfo{person}{M. Perrot}, {and}
  \bibinfo{person}{E. Duchesnay}.} \bibinfo{year}{2011}\natexlab{}.
\newblock \showarticletitle{Scikit-learn: Machine Learning in {P}ython}.
\newblock \bibinfo{journal}{\emph{Journal of Machine Learning Research}}
  \bibinfo{volume}{12} (\bibinfo{year}{2011}), \bibinfo{pages}{2825--2830}.
\newblock


\bibitem[Pei et~al\mbox{.}(2005)]%
        {quasiCliquePeiJZ05}
\bibfield{author}{\bibinfo{person}{Jian Pei}, \bibinfo{person}{Daxin Jiang},
  {and} \bibinfo{person}{Aidong Zhang}.} \bibinfo{year}{2005}\natexlab{}.
\newblock \showarticletitle{On mining cross-graph quasi-cliques}. In
  \bibinfo{booktitle}{\emph{KDD}}. \bibinfo{pages}{228--238}.
\newblock


\bibitem[Pinar et~al\mbox{.}(2017)]%
        {17WWWmotif}
\bibfield{author}{\bibinfo{person}{Ali Pinar}, \bibinfo{person}{C. Seshadhri},
  {and} \bibinfo{person}{Vaidyanathan Vishal}.}
  \bibinfo{year}{2017}\natexlab{}.
\newblock \showarticletitle{{ESCAPE:} Efficiently Counting All 5-Vertex
  Subgraphs}. In \bibinfo{booktitle}{\emph{WWW}}. \bibinfo{pages}{1431--1440}.
\newblock


\bibitem[Ribeiro et~al\mbox{.}(2022)]%
        {surveyRibeiroPSAS21}
\bibfield{author}{\bibinfo{person}{Pedro Ribeiro}, \bibinfo{person}{Pedro
  Paredes}, \bibinfo{person}{Miguel E.~P. Silva}, \bibinfo{person}{David
  Apar{\'{\i}}cio}, {and} \bibinfo{person}{Fernando M.~A. Silva}.}
  \bibinfo{year}{2022}\natexlab{}.
\newblock \showarticletitle{A Survey on Subgraph Counting: Concepts,
  Algorithms, and Applications to Network Motifs and Graphlets}.
\newblock \bibinfo{journal}{\emph{{ACM} Comput. Surv.}} \bibinfo{volume}{54},
  \bibinfo{number}{2} (\bibinfo{year}{2022}), \bibinfo{pages}{28:1--28:36}.
\newblock


\bibitem[Rosvall et~al\mbox{.}(2009)]%
        {infomap}
\bibfield{author}{\bibinfo{person}{Martin Rosvall}, \bibinfo{person}{Daniel
  Axelsson}, {and} \bibinfo{person}{Carl~T Bergstrom}.}
  \bibinfo{year}{2009}\natexlab{}.
\newblock \showarticletitle{The map equation}.
\newblock \bibinfo{journal}{\emph{The European Physical Journal Special
  Topics}} \bibinfo{volume}{178}, \bibinfo{number}{1} (\bibinfo{year}{2009}),
  \bibinfo{pages}{13--23}.
\newblock


\bibitem[Sariy{\"{u}}ce et~al\mbox{.}(2015)]%
        {15wwwnucleusdecomp}
\bibfield{author}{\bibinfo{person}{Ahmet~Erdem Sariy{\"{u}}ce},
  \bibinfo{person}{C. Seshadhri}, \bibinfo{person}{Ali Pinar}, {and}
  \bibinfo{person}{{\"{U}}mit~V. {\c{C}}ataly{\"{u}}rek}.}
  \bibinfo{year}{2015}\natexlab{}.
\newblock \showarticletitle{Finding the Hierarchy of Dense Subgraphs using
  Nucleus Decompositions}. In \bibinfo{booktitle}{\emph{WWW}}.
  \bibinfo{pages}{927--937}.
\newblock


\bibitem[Schaeffer(2007)]%
        {schaeffer2007graph}
\bibfield{author}{\bibinfo{person}{Satu~Elisa Schaeffer}.}
  \bibinfo{year}{2007}\natexlab{}.
\newblock \showarticletitle{Graph clustering}.
\newblock \bibinfo{journal}{\emph{Computer science review}}
  \bibinfo{volume}{1}, \bibinfo{number}{1} (\bibinfo{year}{2007}),
  \bibinfo{pages}{27--64}.
\newblock


\bibitem[Schutze et~al\mbox{.}(2008)]%
        {schutze2008introduction}
\bibfield{author}{\bibinfo{person}{Hinrich Schutze},
  \bibinfo{person}{Christopher~D Manning}, {and} \bibinfo{person}{Prabhakar
  Raghavan}.} \bibinfo{year}{2008}\natexlab{}.
\newblock \bibinfo{booktitle}{\emph{Introduction to information retrieval}}.
\newblock \bibinfo{publisher}{Cambridge University Press}.
\newblock


\bibitem[Seidman(1983)]%
        {seidman1983network}
\bibfield{author}{\bibinfo{person}{Stephen~B Seidman}.}
  \bibinfo{year}{1983}\natexlab{}.
\newblock \showarticletitle{Network structure and minimum degree}.
\newblock \bibinfo{journal}{\emph{Social networks}} \bibinfo{volume}{5},
  \bibinfo{number}{3} (\bibinfo{year}{1983}), \bibinfo{pages}{269--287}.
\newblock


\bibitem[Seidman and Foster(1978)]%
        {Plex_seidman1978graph}
\bibfield{author}{\bibinfo{person}{Stephen~B Seidman} {and}
  \bibinfo{person}{Brian~L Foster}.} \bibinfo{year}{1978}\natexlab{}.
\newblock \showarticletitle{A graph-theoretic generalization of the clique
  concept}.
\newblock \bibinfo{journal}{\emph{Journal of Mathematical sociology}}
  \bibinfo{volume}{6}, \bibinfo{number}{1} (\bibinfo{year}{1978}),
  \bibinfo{pages}{139--154}.
\newblock


\bibitem[Tomita et~al\mbox{.}(2006)]%
        {tomitaTime}
\bibfield{author}{\bibinfo{person}{Etsuji Tomita}, \bibinfo{person}{Akira
  Tanaka}, {and} \bibinfo{person}{Haruhisa Takahashi}.}
  \bibinfo{year}{2006}\natexlab{}.
\newblock \showarticletitle{The worst-case time complexity for generating all
  maximal cliques and computational experiments}.
\newblock \bibinfo{journal}{\emph{Theor. Comput. Sci.}} \bibinfo{volume}{363},
  \bibinfo{number}{1} (\bibinfo{year}{2006}), \bibinfo{pages}{28--42}.
\newblock


\bibitem[Tsourakakis(2015)]%
        {kdensWWW2015}
\bibfield{author}{\bibinfo{person}{Charalampos~E. Tsourakakis}.}
  \bibinfo{year}{2015}\natexlab{}.
\newblock \showarticletitle{The K-clique Densest Subgraph Problem}. In
  \bibinfo{booktitle}{\emph{WWW}}. \bibinfo{pages}{1122--1132}.
\newblock


\bibitem[Tsourakakis et~al\mbox{.}(2017)]%
        {motifClustering}
\bibfield{author}{\bibinfo{person}{Charalampos~E. Tsourakakis},
  \bibinfo{person}{Jakub Pachocki}, {and} \bibinfo{person}{Michael
  Mitzenmacher}.} \bibinfo{year}{2017}\natexlab{}.
\newblock \showarticletitle{Scalable Motif-aware Graph Clustering}. In
  \bibinfo{booktitle}{\emph{WWW}}. \bibinfo{pages}{1451--1460}.
\newblock


\bibitem[Usha Nandini~Raghavan and Kumara(2007)]%
        {lpa}
\bibfield{author}{\bibinfo{person}{Réka~Albert Usha Nandini~Raghavan} {and}
  \bibinfo{person}{Soundar Kumara}.} \bibinfo{year}{2007}\natexlab{}.
\newblock \showarticletitle{Near linear time algorithm to detect community
  structures in large-scale networks}.
\newblock \bibinfo{journal}{\emph{Phys. Rev. E}} \bibinfo{volume}{76},
  \bibinfo{number}{036106} (\bibinfo{year}{2007}).
\newblock


\bibitem[Vinh et~al\mbox{.}(2009)]%
        {vinh2009information}
\bibfield{author}{\bibinfo{person}{Nguyen~Xuan Vinh}, \bibinfo{person}{Julien
  Epps}, {and} \bibinfo{person}{James Bailey}.}
  \bibinfo{year}{2009}\natexlab{}.
\newblock \showarticletitle{Information theoretic measures for clusterings
  comparison: is a correction for chance necessary?}. In
  \bibinfo{booktitle}{\emph{Proceedings of the 26th annual international
  conference on machine learning}}. \bibinfo{pages}{1073--1080}.
\newblock


\bibitem[Wang et~al\mbox{.}(2022a)]%
        {NeuralSubgraphCounting}
\bibfield{author}{\bibinfo{person}{Hanchen Wang}, \bibinfo{person}{Rong Hu},
  \bibinfo{person}{Ying Zhang}, \bibinfo{person}{Lu Qin}, \bibinfo{person}{Wei
  Wang}, {and} \bibinfo{person}{Wenjie Zhang}.}
  \bibinfo{year}{2022}\natexlab{a}.
\newblock \showarticletitle{Neural Subgraph Counting with Wasserstein
  Estimator}. In \bibinfo{booktitle}{\emph{{SIGMOD}}}.
  \bibinfo{pages}{160--175}.
\newblock


\bibitem[Wang and Cheng(2012)]%
        {12vldbtruss}
\bibfield{author}{\bibinfo{person}{Jia Wang} {and} \bibinfo{person}{James
  Cheng}.} \bibinfo{year}{2012}\natexlab{}.
\newblock \showarticletitle{Truss Decomposition in Massive Networks}.
\newblock \bibinfo{journal}{\emph{Proc. {VLDB} Endow.}} \bibinfo{volume}{5},
  \bibinfo{number}{9} (\bibinfo{year}{2012}), \bibinfo{pages}{812--823}.
\newblock


\bibitem[Wang et~al\mbox{.}(2018)]%
        {pathsampling2018}
\bibfield{author}{\bibinfo{person}{Pinghui Wang}, \bibinfo{person}{Junzhou
  Zhao}, \bibinfo{person}{Xiangliang Zhang}, \bibinfo{person}{Zhenguo Li},
  \bibinfo{person}{Jiefeng Cheng}, \bibinfo{person}{John C.~S. Lui},
  \bibinfo{person}{Don Towsley}, \bibinfo{person}{Jing Tao}, {and}
  \bibinfo{person}{Xiaohong Guan}.} \bibinfo{year}{2018}\natexlab{}.
\newblock \showarticletitle{{MOSS-5:} {A} Fast Method of Approximating Counts
  of 5-Node Graphlets in Large Graphs}.
\newblock \bibinfo{journal}{\emph{{IEEE} Trans. Knowl. Data Eng.}}
  \bibinfo{volume}{30}, \bibinfo{number}{1} (\bibinfo{year}{2018}),
  \bibinfo{pages}{73--86}.
\newblock


\bibitem[Wang et~al\mbox{.}(2022b)]%
        {22wwwkplex}
\bibfield{author}{\bibinfo{person}{Zhengren Wang}, \bibinfo{person}{Yi Zhou},
  \bibinfo{person}{Mingyu Xiao}, {and} \bibinfo{person}{Bakhadyr Khoussainov}.}
  \bibinfo{year}{2022}\natexlab{b}.
\newblock \showarticletitle{Listing Maximal k-Plexes in Large Real-World
  Graphs}. In \bibinfo{booktitle}{\emph{WWW}}. \bibinfo{pages}{1517--1527}.
\newblock


\bibitem[Witten et~al\mbox{.}(2005)]%
        {witten2005practical}
\bibfield{author}{\bibinfo{person}{Ian~H Witten}, \bibinfo{person}{Eibe Frank},
  \bibinfo{person}{Mark~A Hall}, \bibinfo{person}{Christopher~J Pal}, {and}
  \bibinfo{person}{MINING DATA}.} \bibinfo{year}{2005}\natexlab{}.
\newblock \showarticletitle{Practical machine learning tools and techniques}.
  In \bibinfo{booktitle}{\emph{Data Mining}}, Vol.~\bibinfo{volume}{2}.
\newblock


\bibitem[Xu et~al\mbox{.}(2008)]%
        {xu2008superfamily}
\bibfield{author}{\bibinfo{person}{Xiaoke Xu}, \bibinfo{person}{Jie Zhang},
  {and} \bibinfo{person}{Michael Small}.} \bibinfo{year}{2008}\natexlab{}.
\newblock \showarticletitle{Superfamily phenomena and motifs of networks
  induced from time series}.
\newblock \bibinfo{journal}{\emph{Proceedings of the National Academy of
  Sciences}} \bibinfo{volume}{105}, \bibinfo{number}{50}
  (\bibinfo{year}{2008}), \bibinfo{pages}{19601--19605}.
\newblock


\bibitem[Yang and Leskovec(2012)]%
        {yang2012defining}
\bibfield{author}{\bibinfo{person}{Jaewon Yang} {and} \bibinfo{person}{Jure
  Leskovec}.} \bibinfo{year}{2012}\natexlab{}.
\newblock \showarticletitle{Defining and evaluating network communities based
  on ground-truth}. In \bibinfo{booktitle}{\emph{KDD}}. \bibinfo{pages}{1--8}.
\newblock


\bibitem[Ye et~al\mbox{.}(2022)]%
        {ccpath}
\bibfield{author}{\bibinfo{person}{Xiaowei Ye}, \bibinfo{person}{Rong{-}Hua
  Li}, \bibinfo{person}{Qiangqiang Dai}, \bibinfo{person}{Hongzhi Chen}, {and}
  \bibinfo{person}{Guoren Wang}.} \bibinfo{year}{2022}\natexlab{}.
\newblock \showarticletitle{Lightning Fast and Space Efficient k-clique
  Counting}. In \bibinfo{booktitle}{\emph{WWW}}. \bibinfo{pages}{1191--1202}.
\newblock


\bibitem[Yin et~al\mbox{.}(2017)]%
        {17prehighorder}
\bibfield{author}{\bibinfo{person}{Hao Yin}, \bibinfo{person}{Austin~R.
  Benson}, {and} \bibinfo{person}{Jure Leskovec}.}
  \bibinfo{year}{2017}\natexlab{}.
\newblock \showarticletitle{Higher-order clustering in networks}.
\newblock \bibinfo{journal}{\emph{Physical Review E}} \bibinfo{volume}{97},
  \bibinfo{number}{5} (\bibinfo{year}{2017}), \bibinfo{pages}{052306}.
\newblock


\bibitem[Yu et~al\mbox{.}(2006)]%
        {Defective_bioinformatics_06}
\bibfield{author}{\bibinfo{person}{Haiyuan Yu}, \bibinfo{person}{Alberto
  Paccanaro}, \bibinfo{person}{Valery Trifonov}, {and} \bibinfo{person}{Mark
  Gerstein}.} \bibinfo{year}{2006}\natexlab{}.
\newblock \showarticletitle{Predicting interactions in protein networks by
  completing defective cliques}.
\newblock \bibinfo{journal}{\emph{Bioinform.}} \bibinfo{volume}{22},
  \bibinfo{number}{7} (\bibinfo{year}{2006}), \bibinfo{pages}{823--829}.
\newblock


\bibitem[Zhao et~al\mbox{.}(2021)]%
        {gnnZhaoYZLR21}
\bibfield{author}{\bibinfo{person}{Kangfei Zhao}, \bibinfo{person}{Jeffrey~Xu
  Yu}, \bibinfo{person}{Hao Zhang}, \bibinfo{person}{Qiyan Li}, {and}
  \bibinfo{person}{Yu Rong}.} \bibinfo{year}{2021}\natexlab{}.
\newblock \showarticletitle{A Learned Sketch for Subgraph Counting}. In
  \bibinfo{booktitle}{\emph{{SIGMOD}}}. \bibinfo{pages}{2142--2155}.
\newblock


\bibitem[Zhou et~al\mbox{.}(2012)]%
        {12edbtkecc}
\bibfield{author}{\bibinfo{person}{Rui Zhou}, \bibinfo{person}{Chengfei Liu},
  \bibinfo{person}{Jeffrey~Xu Yu}, \bibinfo{person}{Weifa Liang},
  \bibinfo{person}{Baichen Chen}, {and} \bibinfo{person}{Jianxin Li}.}
  \bibinfo{year}{2012}\natexlab{}.
\newblock \showarticletitle{Finding maximal k-edge-connected subgraphs from a
  large graph}. In \bibinfo{booktitle}{\emph{{EDBT}}}.
  \bibinfo{pages}{480--491}.
\newblock


\bibitem[Zhou et~al\mbox{.}(2021)]%
        {plex_AAAI21}
\bibfield{author}{\bibinfo{person}{Yi Zhou}, \bibinfo{person}{Shan Hu},
  \bibinfo{person}{Mingyu Xiao}, {and} \bibinfo{person}{Zhang{-}Hua Fu}.}
  \bibinfo{year}{2021}\natexlab{}.
\newblock \showarticletitle{Improving Maximum k-plex Solver via Second-Order
  Reduction and Graph Color Bounding}. In \bibinfo{booktitle}{\emph{{AAAI}}}.
  \bibinfo{pages}{12453--12460}.
\newblock


\end{thebibliography}

%\newpage
%\input{appendix}
\end{document}